\documentclass[11pt,a4paper]{article}
\usepackage[T1]{fontenc}
\usepackage[utf8]{inputenc}
\usepackage{lmodern}
\usepackage[margin=25mm,headheight=14pt]{geometry}
\usepackage{amsmath,amssymb,amsthm,mathtools}
\usepackage{microtype}
\usepackage{enumitem,booktabs,array,longtable}
\usepackage{xcolor,fancyhdr,needspace}
\usepackage[unicode,colorlinks=true,linkcolor=blue!45!black,
 citecolor=blue!45!black,urlcolor=blue!45!black]{hyperref}
\numberwithin{equation}{section}
\newtheorem{theorem}{Theorem}[section]
\newtheorem{lemma}[theorem]{Lemma}
\newtheorem{proposition}[theorem]{Proposition}
\newtheorem{corollary}[theorem]{Corollary}

\theoremstyle{definition}

\theoremstyle{remark}

\newcommand{\R}{\mathbb R}
\newcommand{\Z}{\mathbb Z}

\newcommand{\eps}{\varepsilon}
\newcommand{\Ot}{\widetilde O}
\newcommand{\T}{\mathsf T}
\newcommand{\cB}{\mathcal B}
\newcommand{\cV}{\mathcal V}
\newcommand{\Bx}{\mathsf Q}
\newcommand{\va}{\mathbf a}\newcommand{\vb}{\mathbf b}\newcommand{\vc}{\mathbf c}
\newcommand{\ve}{\mathbf e}\newcommand{\vg}{\mathbf g}\newcommand{\vp}{\mathbf p}
\newcommand{\vt}{\mathbf t}\newcommand{\vu}{\mathbf u}
\newcommand{\vv}{\mathbf v}\newcommand{\vx}{\mathbf x}
\newcommand{\vy}{\mathbf y}\newcommand{\vz}{\mathbf z}
\newcommand{\vpi}{\boldsymbol\pi}\newcommand{\vlambda}{\boldsymbol\lambda}
\newcommand{\vell}{\boldsymbol\ell}\newcommand{\vzero}{\mathbf 0}
\newcommand{\vone}{\mathbf 1}
\newcommand{\vd}{\mathbf d}\newcommand{\vh}{\mathbf h}\newcommand{\vq}{\mathbf q}
\newcommand{\rhos}{\rho^{\ast}}
\DeclareMathOperator{\vol}{vol}
\DeclareMathOperator{\epi}{epi}
\DeclareMathOperator{\conv}{conv}
\DeclareMathOperator{\covol}{covol}

\DeclareMathOperator{\rank}{rank}
\DeclareMathOperator{\Span}{span}
\DeclareMathOperator{\poly}{poly}

\DeclareMathOperator{\inter}{int}

\allowdisplaybreaks[2]
\setlist{itemsep=2pt,topsep=4pt}
\ifdefined\pdfmapfile
\pdfmapfile{+lm.map}
\pdfmapfile{+cm.map}
\pdfmapfile{+symbols.map}
\fi

\begin{document}
\title{Even Faster Algorithms for Bounded Integer Programming}
\author{Lin Chen, Yuchen Mao, Guochuan Zhang\footnote{\{chenlin198662,maoyc,zgc\}@zju.edu.cn. Zhejiang University.}}
\date{\today}
\maketitle

\begin{abstract}
We study integer programs
$\min\{\vc^{\T}\vx: A\vx=\vb,\ \vell\le\vx\le\vu,\ \vx\in\Z^n\}$ with a
fixed number $d$ of equality constraints and constraint coefficients of
absolute value at most $N$. Chen, Mao, and Zhang recently solved such
programs in almost $N^{2d}$ time. We show that any polynomial saving for
$(\min,+)$-convolution carries over to this problem. If witnessed
$(\min,+)$-convolution of arrays of length $m$ can be computed in
$\Ot(m^{\rho})$ time with error probability $p$, then for every
$\eps>0$ bounded integer programming can be solved in
$\Ot\bigl(n+N^{(\max\{\rho,3/2\}+\eps)d}\bigr)$ time with the same error
probability $p$, up to polynomial factors in the encoding length and the
cost of one linear program. A deterministic convolution algorithm yields
a deterministic integer programming algorithm. Combined with recent
truly subquadratic algorithms for $(\min,+)$-convolution, this gives a
deterministic algorithm running in $\Ot(n+N^{1.99986\,d})$ time and a
randomized algorithm running in $\Ot(n+N^{1.9602\,d})$ time, up to the same
factors.
\end{abstract}

\section*{AI Statement} The same authors recently obtained an $O(N^{2d + \varepsilon})$-time algorithm for bounded integer programming~\cite{CMZ26b}.  Combining the integer programming technique and recent progress in (min,+) convolution, Zhou~\cite{Zhou26} showed that Knapsack can be solved in truly subquadratic time. We fed these two papers into GPT-6.2 Astra, and with almost no human input and after several runs, it obtained the result in this paper. The authors take no intellectual credit.

\newpage

\tableofcontents

\clearpage

\section{Introduction}
\label{det:intro}

We study the integer program
\begin{equation}\label{det:ip}
 \min\{\vc^{\T}\vx:\ A\vx=\vb,\ \vell\le\vx\le\vu,\ \vx\in\Z^n\},
 \qquad A\in\Z^{d\times n},\qquad
 N=\max\{2,\max_{i,j}|A_{ij}|\},
\end{equation}
in standard form with finite variable bounds, where all entries of
$\vb,\vc,\vell,\vu$ are integers. The parameter $d$ counts only the
equations $A\vx=\vb$; the variable bounds are finite, binary encoded,
and not counted. An algorithm must return an optimal integer vector or
correctly report infeasibility. Integer programming is NP-hard in
general~\cite{Karp72}, and already for $d=1$ when $N$ is part of the
input. The question is therefore how the running time depends on the
structural parameters $d$ and $N$, while the targets, bounds, and costs
may be arbitrarily large.

Chen, Mao, and Zhang~\cite{CMZ26b} recently solved~\eqref{det:ip} in
time $N^{2d+\eps}$ for every fixed $\eps>0$, up to the cost of reading
the input and solving one linear program. The exponent $2d$ matches
that of the algorithm of Jansen and Rohwedder for nonnegative
variables~\cite{JansenRohwedder19}. Jansen and Rohwedder also showed
that this exponent cannot be improved substantially unless
$(\min,+)$-convolution can be computed in truly subquadratic time. That
conditional barrier has now been removed: the deterministic real
$(\min,+)$-product algorithm of Chan, Itakura, Ma, and Yu~\cite{CIMY26}
yields a truly subquadratic $(\min,+)$-convolution
(Lemma~\ref{detmc:lemma}). Can the exponent $2d$ for bounded integer
programming be improved as well, and by how much?

We show that it can, and that the improvement is governed by the
exponent of $(\min,+)$-convolution. If convolution of arrays of length
$m$ takes time $m^{\rho}$, then bounded integer programming takes time
$N^{(\max\{\rho,3/2\}+\eps)d}$. The saving in the exponent is thus a
constant multiple of $d$, and the integer programming algorithm has the
same error probability as the convolution algorithm.

\subsection{Previous work}

\paragraph{Few constraints and small coefficients.}
For programs in standard form with $d$ equations and nonnegative
variables, Papadimitriou gave a pseudopolynomial algorithm for fixed
$d$~\cite{Papadimitriou81}. Eisenbrand and Weismantel used the Steinitz
lemma to obtain running time $(dN)^{O(d)}\|\vb\|_\infty^2$ for
nonnegative variables, and proved a proximity bound between optimal LP
and integer solutions that is independent of $n$, $\vb$, and the
objective~\cite{EW19}; proximity results of this kind go back to Cook,
Gerards, Schrijver, and Tardos~\cite{CGST86}. Jansen and Rohwedder
sharpened the dependence on $d$ and $N$ for nonnegative variables to
$O(\sqrt d\,N)^{2d}$ times lower-order factors, and related the exponent
$2d$ to $(\min,+)$-convolution~\cite{JansenRohwedder19}. Under the
Exponential Time Hypothesis, Fomin, Panolan, Ramanujan, and
Saurabh~\cite{FPRS23} and Knop, Pilipczuk, and Wrochna~\cite{KPW20}
showed that an exponent of order $d$ is necessary, up to lower-order
factors.

\paragraph{Bounded variables.}
Finite upper bounds make the problem harder to handle by these methods.
Eisenbrand and Weismantel obtained running time $n\cdot(dN)^{O(d^2)}$
for the bounded version~\cite{EW19}. Building on their near-optimal
algorithms for subset sum and for integer programming
feasibility~\cite{CMZ26}, Chen, Mao, and Zhang~\cite{CMZ26b} removed the
quadratic dependence on $d$ and obtained the deterministic running time
\[
 T_{\rm LP}+O_d(n)+O_{d,\eps}\bigl(N^{2d+\eps}\log^2(c_{\max}+2)\bigr),
 \qquad c_{\max}=\max_i|c_i|.
\]
Their algorithm answers strict decision queries by a recursion that
measures interior in the epigraph of the fractional cost function, and
it bounds the whole recursion by a single potential.

For $d=1$ the problem contains bounded knapsack. Combining the integer
programming technique of~\cite{CMZ26b} with recent progress on
$(\min,+)$-convolution, Zhou showed that bounded knapsack can be solved
in truly subquadratic time~\cite{Zhou26}. His algorithm uses affordable
fibres, a source-based accounting of the work, and convolution-based
tables of removed items. Part of the present paper carries this
accounting over to every fixed $d$
(Sections~\ref{fw:frame}--\ref{det:composition}). On its own, however,
this would only give the exponent $\bigl((4-\rho)/(3-\rho)+\eps\bigr)d$,
because a direct convolution-based reporter of affordable offsets is too
slow (Section~\ref{ov:interface}).

\paragraph{$(\min,+)$-convolution.}
Given arrays $F,G$ of length $m$, their $(\min,+)$-convolution is
$(F\oplus G)[k]=\min_{i+j=k}(F[i]+G[j])$. The naive algorithm takes
$O(m^2)$ time, and for a long time the best general bound was
$m^2/2^{\Omega(\sqrt{\log m})}$, obtained through the reduction to
$(\min,+)$-matrix product~\cite{BCDEHILPT14} and Williams' algorithm for
all-pairs shortest paths~\cite{Williams14}. The hypothesis that no
$O(m^{2-\eps})$ algorithm exists served as the basis for conditional
lower bounds by Cygan, Mucha, W\k{e}grzycki, and
W{\l}odarczyk~\cite{CMWW19} and by K\"unnemann, Paturi, and
Schneider~\cite{KPS17}. Among the problems shown to be equivalent to it
is $0$-$1$ knapsack parameterized by the capacity~\cite{CMWW19}. Chan,
Itakura, Ma, and Yu~\cite{CIMY26} recently gave a deterministic
algorithm for real $(\min,+)$-matrix product with exponent $2.9997$,
which yields deterministic witnessed $(\min,+)$-convolution in time
$\Ot(m^{1.99985})$ for integers of arbitrary bit length
(Lemma~\ref{detmc:lemma}).

\subsection{Our results}

By \emph{witnessed $(\min,+)$-convolution} we mean the following task:
given integer arrays $F,G$ of total length $m$, with entries $+\infty$
permitted and finite entries of at most $b$ bits, compute every value
$(F\oplus G)[k]=\min_{i+j=k}(F[i]+G[j])$ together with an attaining pair
$(i,j)$ for every finite value. An algorithm for it has \emph{error
probability~$p$} if, on every input, its entire output is correct with
probability at least $1-p$. Similarly, an algorithm \emph{solves} the
integer program~\eqref{det:ip} if it returns an optimal integer vector
or correctly reports infeasibility, and it has error probability~$p$ if
it does so on every input with probability at least $1-p$.

The parameter $\cB$ bounds the bit length of every input number, as
well as $\log(n+N+2)$. All intermediate quantities, including costs,
capacities, coordinates, and compressed profile data, then have
polynomially bounded bit length (Section~\ref{fw:arith}), and rational
intermediate quantities are handled exactly. The continuous relaxation
is solved once, deterministically, at cost $T_{\rm LP}$. Polynomial
degrees may depend on the fixed parameters, but never on $N$ or $n$. As in~\cite{CMZ26b}, vectors (points, targets,
prices, and linear functionals) are bold, prices are row vectors, and
$\T$ denotes transposition.

\begin{theorem}[Exponent and probability transfer]\label{det:main}
Let $1<\rho\le2$ and $0\le p<1$ be fixed. Suppose that witnessed
$(\min,+)$-convolution of arrays of total length $m$ can be solved in
worst-case time $\Ot(m^{\rho})\poly(b+\log m)$ with error
probability~$p$. Then for every fixed $\eps>0$, the integer
program~\eqref{det:ip} can be solved in worst-case time
\begin{equation}\label{det:main-bound}
 T_{\rm LP}+\Ot_{d,\rho,\eps,p}
 \Bigl(n+N^{\left(\max\{\rho,3/2\}+\eps\right)d}\Bigr)
 \poly_{d,\rho,\eps}(\cB)
\end{equation}
with error probability~$p$. If the convolution algorithm is
deterministic, so is the integer programming algorithm.
\end{theorem}

When $p>0$, the integer programming algorithm uses the constants hidden
in the running-time hypothesis, which are fixed together with the
convolution algorithm, to set its work budget and the time limits of
individual trials; the number of repetitions of each call is determined
by its index and by $p$ (Section~\ref{det:probability-section}). When $p=0$ it uses only the
convolution algorithm itself. Every call made by the algorithm has
entries of $b=\poly_{d,\rho,\eps}(\cB)$ bits (Section~\ref{fw:arith}), and
the degree of the factor $\poly_{d,\rho,\eps}(\cB)$
in~\eqref{det:main-bound} also depends on the degree of the polynomial
factor in the hypothesis.

Throughout the paper we write $\rhos=\max\{\rho,3/2\}$ for the
exponent in~\eqref{det:main-bound}. The term $\rho$ comes from the
signed removed-item tables (Section~\ref{det:transfer-section}), and
the term $3/2$ from the local LP atlas and the reporter of affordable
offsets (Section~\ref{det:section}). We make no optimality claim for
the exponent $3/2$. For $\rho=2$, which naive convolution attains, the
theorem recovers the exponent $(2+\eps)d$ of~\cite{CMZ26b}, with a
polynomial factor in $\cB$ in place of $\log^2(c_{\max}+2)$. The hidden
constant depends on $p$ only as $p\to1$: for fixed running-time
constants of the convolution algorithm, the amplification overhead is
uniform over $0\le p\le1/2$ (Section~\ref{det:probability-section}).

Instantiating Theorem~\ref{det:main} with the deterministic convolution
algorithm derived from~\cite{CIMY26}, and with the randomized
witnessed convolution algorithm of Zhou~\cite[Lemma~3.6]{Zhou26}, gives
the following explicit bounds.

\begin{corollary}[Explicit running times]\label{det:explicit}
The integer program~\eqref{det:ip} can be solved deterministically in
time
\begin{equation}\label{det:explicit-det}
 T_{\rm LP}+\Ot_d\bigl(n+N^{1.99986\,d}\bigr)\poly_d(\cB),
\end{equation}
and by a randomized algorithm with error probability at most $1/3$ in
time
\begin{equation}\label{det:explicit-rand}
 T_{\rm LP}+\Ot_d\bigl(n+N^{1.9602\,d}\bigr)\poly_d(\cB).
\end{equation}
\end{corollary}

The randomized convolution algorithm combines the weighted-cache 3SUM
decision algorithm of Wang~\cite{Wang26}, of exponent
$\alpha=1.920279408185$ on polynomially bounded integers, with Zhou's
reduction to witnessed convolution of arbitrary bit length, which has
exponent $1+\alpha/2=1.9601397\ldots$~\cite[Lemma~3.6 and
Appendix~B]{Zhou26}.

The savings $2-1.99986$ and $2-1.9602$ in the exponent of $N$ are
absolute constants per equation, so the improvement over $N^{2d}$ grows
with $d$. For $d=1,2,3$ the randomized exponents are $1.9602$, $3.9204$,
and $5.8806$. In both cases the bottleneck is the convolution exponent
$\rho$; the exponent $3/2$ would become the bottleneck only for a
convolution algorithm with exponent below $3/2$. For comparison, a
blocked-convolution reporter of affordable offsets with the same
convolution primitive would give the exponent
$(4-\rho)/(3-\rho)\approx1.9617$ per equation instead of $1.9602$. The
corollary is derived in Section~\ref{det:instantiation}.

Theorem~\ref{det:main} complements the conditional lower bound of
Jansen and Rohwedder~\cite{JansenRohwedder19}. There,
$(\min,+)$-convolution was identified as the obstacle to improving the
exponent $2d$; here we show that every polynomial improvement for
$(\min,+)$-convolution does translate into an improvement for bounded
integer programming, with a saving proportional to $d$, until the
exponent reaches $3d/2$.

\paragraph{Our approach.}
We use the structural recursion of~\cite{CMZ26b}, with Zhou's
source-based accounting~\cite{Zhou26} extended to an arbitrary fixed
rank. Its running time consists of structural work of order $N^d$, up
to slack factors $N^{O(\delta)}$, plus three kinds of local computation,
each repeated over many \emph{sources} or recursive \emph{states}:
(i)~exact tables of the fractional cost function on a window (the
\emph{local LP atlas}); (ii)~lists of affordable symmetric offsets
around a target (the \emph{reporter of affordable offsets}); and
(iii)~minimum-cost tables of removed items in a box (the \emph{signed
removed-item tables}). Only the last task needs $(\min,+)$-convolution. We compute the
removed-item tables by a deterministic reduction to convolution, at cost
$\Ot(M^{\rho})$ on a box of $M$ cells. We build the atlas
deterministically in time $\Ot(V^{3/2})$ for a window of normalized
volume $V$. We report affordable offsets output-sensitively and without
convolution, using only the convexity of the fractional cost function.
Up to slack factors, the accounting bounds the total volume of all
sources by $N^{d}$ and the $\rho$-th moments of the state masses by
$N^{\rho d}$; together these give the exponent $\max\{\rho,3/2\}d$. A
randomized convolution primitive is the only source of randomness, and
its errors are amplified away by repetition with a growing number of
repetitions per call.
Section~\ref{det:overview} explains these ingredients informally.

\paragraph{Provenance.}
The structural framework for bounded integer programming is that of
Chen, Mao, and Zhang~\cite{CMZ26b,CMZ26}. The affordable-fibre
arguments, source accounting, phase coalescence, and macro potential are
inherited or adapted from Zhou~\cite{Zhou26}; we claim none of these
global ideas as new. We extend them to an arbitrary fixed rank in
Sections~\ref{fw:frame}--\ref{det:composition}, which state precisely
the forms of the structural results of~\cite{CMZ26b,CMZ26,Zhou26} that
we use. Our contributions are this extension and the implementations of
four computational interfaces.

The canonical-subset and halver technique is due to
Chan~\cite{Chan26}; we supply the version needed here, with repeated
columns, witnesses, and signed costs. The deterministic numerical
primitive is the real-input result of Chan, Itakura, Ma, and
Yu~\cite{CIMY26}. The reporter of affordable offsets
(Section~\ref{det:reporter-section}) avoids convolution altogether:
it uses only the convexity of the fractional cost
function and the supporting prices stored in the atlas. This is the
change that improves the exponent from $(4-\rho)/(3-\rho)$ to
$\max\{\rho,3/2\}$.

The randomized numerical primitive is Zhou's witnessed convolution
algorithm~\cite[Lemma~3.6]{Zhou26}, built on the 3SUM decision algorithm
of Wang~\cite{Wang26}; we use it as a black box. The deterministic bound
uses a weaker convolution exponent.

\subsection{Further related work}

\paragraph{Knapsack.}
Bounded knapsack is the case $d=1$ of~\eqref{det:ip} with an inequality
constraint and nonnegative weights. For $0$-$1$ knapsack, running times
in terms of the largest weight $w_{\max}$ were studied by Polak,
Rohwedder, and W\k{e}grzycki~\cite{PolakRW21}, and Jin and Bringmann
independently obtained $\Ot(n+w_{\max}^2)$~\cite{Jin24,Bringmann24};
Chen, Lian, Mao, and Zhang gave faster algorithms for bounded knapsack
via fine-grained proximity results~\cite{CLMZ24}. Since $0$-$1$
knapsack is equivalent to $(\min,+)$-convolution~\cite{CMWW19},
quadratic bounds of this kind were conditionally optimal under the
$(\min,+)$-convolution hypothesis.
Zhou's truly subquadratic algorithm for bounded knapsack~\cite{Zhou26}
converts the new convolution algorithms into a polynomial saving for
$d=1$; Theorem~\ref{det:main} gives a saving of the same kind for every
fixed number of constraints.

\paragraph{Subset sum and feasibility.}
Without costs the picture is different: subset sum with one constraint
can be solved in time $O(n+w_{\max}^{1+o(1)})$~\cite{JinMao26,CMZ26},
and Chen, Mao, and Zhang extended their method to integer programming
feasibility~\cite{CMZ26}. Costs add real difficulty beyond
feasibility. In~\cite{CMZ26b} and here, this difficulty shows up in two
places: depth must be measured in the epigraph of the fractional cost
function, and the costs of removed items must be compared by
convolution-type tables.

\paragraph{Organization.}
Section~\ref{det:overview} gives an informal overview of the proof.
Sections~\ref{det:section}--\ref{detmc:top} implement the local
computations. Section~\ref{det:section} constructs the local LP atlas
deterministically, pays for its slower column-body queries, and gives
the output-sensitive reporter of affordable offsets.
Section~\ref{det:transfer-section} constructs deterministic signed
removed-item tables. Section~\ref{detmc:top} supplies the deterministic
and randomized instantiations of the convolution premise for arbitrary
bit lengths. Sections~\ref{fw:frame}--\ref{fw:charge}
develop the recursion: the compressed root and the common objective;
affordable geometry on one rational grid; states, sources, and the
local reductions; and the charging of sources and arithmetic phases.
Finally, Section~\ref{det:composition} proves the global moment and
source bounds, Theorem~\ref{det:main} (including the amplification of a
randomized primitive), and Corollary~\ref{det:explicit}.

\section{Technical overview}
\label{det:overview}

The main idea is simple: we treat the structural recursion for bounded
integer programming as a black box and change only how its local
computations are carried out, and hence what they cost. The recursion
of~\cite{CMZ26b}, with the source accounting of~\cite{Zhou26} extended
to every fixed rank (Sections~\ref{fw:frame}--\ref{det:composition}),
reduces the problem to three kinds of local tasks on pools of items with
small normalized volume. Its running time is a sum of terms, one for
each kind of task, and each term is a power of $N^d$; the exponent of
the whole algorithm is the largest of these powers. We keep the
recursion, its pools, windows, and witnesses unchanged. For each local
task we give an implementation whose cost is a power of the local size
with exponent at most $\max\{\rho,3/2\}$. The recursion itself is
sketched only in Section~\ref{ov:framework}.

\paragraph{Notation.}
As in~\cite{CMZ26b}, for a pool $I$ of labelled items with columns
$\va_i$ and costs $c_i$ we write
\[
 Z(I)=\sum_{i\in I}[\vzero,\va_i],\qquad
 P_I=\conv\{\pm\va_i:i\in I\},\qquad
 \Psi(I)=\frac{r!}{2^r}\,\frac{\vol_r(P_I)}{\covol(\Lambda_I)},
\]
where $\Lambda_I$ is the current lattice of the pool and $r=\rank I$;
we also write $T_I$ for the number of types (classes of items with equal
column). The set $Z(I)$ is the zonotope of fractional sums, $P_I$ is the
column hull, and $\Psi(I)$ is its normalized volume. Up to a factor
depending on $r$, the normalized volume counts the lattice points in
$P_I$, so it is the natural measure of the size of a pool. The fractional cost
function of the pool is
\[
 f_I(\vt)=\min\bigl\{\vc_I^{\T}\vx:\ A_I\vx=\vt,\ \vzero\le\vx\le\vone\bigr\},
\]
with $f_I=+\infty$ outside $Z(I)$; it is convex and piecewise linear.
The symbols $\delta$, $\nu=\lceil N^\delta\rceil$, and $D$ (the
stored-rank bound) play the same roles as in~\cite{CMZ26b}; their values
are fixed in Section~\ref{fw:scales}. The
convolution exponent is $\rho$ and the target slack is $\eps$; $p$
denotes the error probability of the convolution primitive and $\ell$
the number of repetitions used to amplify it. Tildes suppress
polylogarithmic factors, and in this overview we also suppress the
polynomial factors in~$\cB$.

\paragraph{Roadmap.}
Section~\ref{ov:interface} recalls where the recursion spends its time,
and why the exponent is $\max\{\rho,3/2\}d$ once every local task stays
within its budget. The remaining subsections explain how each budget is
met.
\begin{itemize}
\item \textbf{Build the LP atlas in time $V^{3/2}$}
(Section~\ref{ov:atlas}). Traverse the linearity tiles of $f_I$ by
pivoting in price space; answer each pivot by a support query on a
truncated hierarchy of small LP dictionaries; and use a lattice
counting bound to show that each tile involves few types.
\item \textbf{Report affordable offsets without convolution}
(Section~\ref{ov:reporter}). The set of affordable offsets around a
target is convex, and the atlas supplies cutting planes for it. A
centerpoint search lists it output-sensitively.
\item \textbf{Compute removed-item tables in time $M^{\rho}$}
(Section~\ref{ov:transfer}). Encode a box of vectors as scalars, and
compute all scalar capacities by a deterministic recursion that halves
cardinalities with canonical subsets and a rational halver.
\item \textbf{Run the recursion} (Section~\ref{ov:framework}). Anchors,
discrete-rank dispatch, deep reductions, and a phase-coalescing
potential bound the source volumes and the moments of the state masses.
\item \textbf{Instantiate and amplify the convolution primitive}
(Sections~\ref{ov:conv} and~\ref{ov:composition}). A deterministic
primitive from real $(\min,+)$-products gives $\rho=1.99985$; Zhou's
randomized primitive gives $\rho=1.9601397\ldots$; a randomized
primitive is amplified by repetition.
\end{itemize}
The following table summarizes the four computational interfaces and
how we implement them.
\begin{center}
\begin{tabular}{@{}>{\raggedright\arraybackslash}p{.24\textwidth}>{\raggedright\arraybackslash}p{.45\textwidth}>{\raggedright\arraybackslash}p{.23\textwidth}@{}}
\toprule
Component & Deterministic implementation & Resulting charge\\
\midrule
Local hull dictionaries & Direct LP dictionaries inside spatial blocks
of at most $\lceil\sqrt M\,\rceil$ lattice cells &
$\Ot(V^{3/2})$ per atlas\\[3pt]
Reporter of affordable offsets & Convex cutting-plane enumeration with
LP supporting prices; no convolution &
$\Ot(V^{3/2}+(Q+O)\sqrt V)$ per source\\[3pt]
Signed removed-item tables & Canonical subsets on labelled index trees
and rational conditional-expectation halvers &
$\Ot(M^{\rho})$ per table\\[3pt]
Convolution primitive & Deterministic real $(\min,+)$-products, square
blocks, and integer witness tags; randomized: Zhou's reduction from
3SUM decision~\cite{Zhou26} &
$\Ot(m^{1.99985})\poly(\cB)$, resp.\ $\Ot(m^{1.96014})\poly(\cB)$\\
\bottomrule
\end{tabular}
\end{center}

\subsection{Where the time goes}
\label{ov:interface}

We recall the recursion only to the extent needed to see its cost. As
in~\cite{CMZ26b}, one LP and the Steinitz lemma reduce~\eqref{det:ip} to
a binary problem with few labelled items. Optimization is then reduced
to strict decision queries of the form ``is there a binary selection of
weight $\vt$ and cost below $s$?''. A query is answered by a recursion
that compares the target with the fractional cost function. A target
with little \emph{affordable} room leads to subproblems on thinner
pools. A target with ample room is either solved by rounding and
repairing, or leads to a proper sublattice or a new thin coordinate.

Following Zhou~\cite{Zhou26}, we organize the work of this recursion
around immutable \emph{sources}. A source $S$ has a pool $I_S$ and a
window $K_S$ of targets, and it answers queries about \emph{affordable
offsets}: for a midpoint $\vt$, an allowance $s$, and a radius $H$, it
lists the lattice points of
\[
 S(\vt,s,H)=\{\vz:\ f(\vt+\vz)+f(\vt-\vz)\le2s\}\cap HP,
\]
where $f=f_{I_S}$ and $P=P_{I_S}$. These are the symmetric moves around
$\vt$ that remain within the cost allowance. In addition, each processed
state $u$ of the recursion needs tables of minimum-cost subsets of
\emph{removed items}, with prescribed vector sums in a box. The total
size of these tables is proportional to the buffered lattice mass $M_u$
of the state.

The accounting (Theorem~\ref{fw:global}) bounds all these quantities in
terms of
\[
 \Xi=(\nu N)^d,\qquad \nu=\lceil N^\delta\rceil,
\]
which is $N^d$ up to the slack factor $N^{d\delta}$, and a further slack
$N^\beta$. Let $V_S$ be the normalized volume of the window of a source,
$Q_S$ its number of queries, and $O_S$ its number of reported offsets.
Then
\[
 \cV=1+\sum_S(V_S+Q_S+O_S)=\Ot(\Xi N^\beta)
 \qquad\text{and}\qquad
 \sum_uM_u^{\theta}=\Ot(\Xi^\theta N^\beta)\quad(1\le\theta\le2),
\]
and all remaining structural work costs $\Ot(\nu\,\Xi N^\beta)$. Thus
the total work is
\[
 \sum_S\bigl(\text{atlas}+\text{reporter}\bigr)(S)
 +\sum_u\text{tables}(u)
 +\Ot(\nu\,\Xi N^\beta).
\]
If the atlas and the reporter of a source cost
$\Ot(V_S^{3/2}+(Q_S+O_S)\sqrt{V_S})$, then, since $V_S\le\cV$, the first
sum is $\Ot(\cV^{3/2})=\Ot(\Xi^{3/2}N^{3\beta/2})$. If a removed-item
table of mass $\mu$ costs $\Ot(\mu^{\rho})$, then by the moment bound
with $\theta=\rho$ the second sum is $\Ot(\Xi^{\rho}N^\beta)$. With
$\delta$ and $\beta$ small compared with $\eps$, the three terms are at
most $N^{(3/2+\eps)d}$, $N^{(\rho+\eps)d}$, and $N^{(1+\eps)d}$. This is
Theorem~\ref{det:main} for exact convolution replies.

With naive quadratic convolution the table term is $\Xi^2$, of order
$N^{2d}$, which is the exponent of~\cite{CMZ26b}. To profit from a
faster convolution, the other two kinds of local work must stay below
$N^{\rho d}$. The natural approach fails for the reporter. Reporting
the offsets $\vz$ with $f(\vt+\vz)+f(\vt-\vz)\le2s$ for many midpoints
is a convolution-type task, since the two arguments have the fixed sum
$2\vt$. Blocked convolution of length $L$ costs
$\Ot(V^2/L^{2-\rho}+QV/L+OL)$, and balancing the three terms gives
$\cV^{(4-\rho)/(3-\rho)}$, which is larger than $\cV^{\rho}$ for every
$\rho<2$. Two further obstacles concern determinism: the natural atlas
construction is randomized incremental hull construction, and the
removed-item tables of~\cite[Lemma~6.4]{Zhou26} use color coding. We
avoid all three problems.

\subsection{A deterministic local LP atlas}
\label{ov:atlas}

The \emph{atlas} of a source is the exact table of $f_I$ on the grid
points of its window $K$, with a supporting price and a basic witness
at every entry. Its size is governed by
\[
 V=V(K,I)=1+\frac{\vol_r(K+6P_I)}{\covol(\Lambda_I)},
\]
and we construct it deterministically in time $\Ot(V^{3/2})$
(Theorem~\ref{det:atlas}).

\paragraph{Tiles and pivots.}
Since $f_I$ is the value of a parametric LP, the target space is tiled
by the regions on which one basis is optimal. For a generic objective,
an optimal price $\vpi$ determines the items with $c_i<\vpi\va_i$ (taken
fully), $r$ tight basic items (taken fractionally), and the rest (not
taken). The corresponding region is a parallelepiped
\[
 T=\vz_T+B_T[0,1]^r,
\]
where $\vz_T$ is the sum of the strictly selected columns and $B_T$ is
the basis matrix. These tiles are the projected lower faces of the
lifted zonotope, and they tile $Z(I)$ face to face. We find all tiles
meeting the window by a breadth-first search, starting from the tile at
the origin. Crossing a facet of a tile releases one basic equality: the
price moves along a ray $\vpi+\lambda\vv$, and the next tile is
determined by the first item whose reduced cost $c-\vpi\va$ reaches
zero, from either side. Thus each step of the traversal is one \emph{event query}.

\paragraph{Event queries as support queries.}
Items with equal columns form a type with a nondecreasing profile of
marginal costs. For a type that is not currently \emph{active} (with no
selected copy and no basic copy), only its first marginal cost
$c_{\va,1}$ can create the next event. Viewing the columns $\va$ as
sites with heights $c_{\va,1}$, the question whether the ray is still
feasible at a given $\lambda$ is a support query
$\min_{\va}(c_{\va,1}-\vpi'\va)\ge0$ over the inactive sites, and the
exact first event follows by a binary search over rational ratios of
bounded bit length (Lemma~\ref{det:event}). The at most $s$ active
types are handled directly. So we need a structure for support queries
over integer sites in a box of $M$ lattice points, \emph{with
exclusions} of $s$ sites.

\paragraph{Dictionaries without hulls.}
For sites in a small box of $U$ lattice points we store, at every lattice
point $\vz$ of the box, the value of the lower convex envelope of the
heights at $\vz$, together with a supporting plane, or a separator of
$\vz$ from the hull of the sites. This takes one fixed-dimensional LP
per box point, i.e.\ time $\Ot(tU)$ for $t$ sites, and no convex hull is
ever constructed. A deterministic linear-time LP solver in fixed
dimension that returns an optimal basis gives the value, a globally
valid supporting plane by duality, and a witness on at most $r+1$ sites,
including at boundary points and for lower-dimensional hulls
(Lemma~\ref{det:small-dictionary}). A support
query in a direction $\vpi$ is then answered by a cutting-plane search
over the integer points of the box. We query a deterministic integer
centerpoint of Basu and Oertel~\cite{BasuOertel17}; the stored
subgradient or separator gives a cut through it; and each cut removes a
constant fraction of the candidates, so $O(\log U)$ cuts suffice
(Lemma~\ref{det:support}).

To handle a large box we bisect it into $O(M/b)$ \emph{block roots} of
at most $b$ lattice points and store dictionaries only at and below the
block roots. Preprocessing costs $\Ot(Tb+M)$, and a support query with
$s$ exclusions is answered by the $O(M/b+s\log M)$ maximal subtrees that
avoid the excluded sites (Lemma~\ref{det:forest}). With
$b=\lceil\sqrt M\,\rceil$, preprocessing is $\Ot(M^{3/2})$ and a query
costs $\Ot(\sqrt M+s)$. A lattice-adapted box around $P_I$ has
$M=O(\Psi(I))=O(V)$ points.

\paragraph{Few types per tile.}
The traversal visits $O(V)$ tiles, because their normalized
determinants are positive integers and they lie in a bounded
neighbourhood of the window. It remains to bound the number $s$ of
active types per tile. After the costs are normalized to be
nonnegative at a reference price (Lemma~\ref{det:window}), the corner
$\vz_T$ is a positive integral
combination of \emph{distinct} lattice vectors that all lie in the open
halfspace $\{\vpi\vx>0\}$. In dimension one, $k$ distinct positive
integers have sum at least $k^2/2$, so a sum bounded by $X$ forces
$k\le\sqrt{2X}$. The pointed distinct-type bound
(Lemma~\ref{det:pointed}) is the $r$-dimensional version: if such a
combination lies in $HP$, the number of distinct vectors is
$O(\sqrt{H\Psi})$. Applied to the tiles of the window this gives
$s=O(\sqrt V)$ (Lemma~\ref{det:window}). Hence each of the $O(V)$
pivots costs $\Ot(\sqrt V)$, and the traversal matches the
preprocessing: both are $\Ot(V^{3/2})$. This balance is the source of
the exponent $3/2$.

\paragraph{Slower body queries.}
The same hierarchy with zero heights answers support, membership,
separation, and gauge queries for the column hull $P_I$. Each such
query takes time $\Ot(\sqrt{\Psi(I)})$, whereas a near-linear hull
dictionary would answer it in time polynomial in $\cB$ alone. Every
source makes $\Ot(Q_S+O_S+V_S)$ such calls, so their total cost is
$\sum_S(Q_S+O_S+V_S)\sqrt{V_S}\le\cV^{3/2}$
(Proposition~\ref{det:source-charge}).

\subsection{Reporting affordable offsets by convexity}
\label{ov:reporter}

The only structural fact we use about the reporting problem is that $f$
is convex. Hence
\[
 g_{\vt}(\vz)=f(\vt+\vz)+f(\vt-\vz)
\]
is convex in $\vz$, and the set $S(\vt,s,H)$ of affordable offsets is a
convex set, intersected with a dilate of the column hull. We list its
grid points by a cutting-plane search, using the atlas as an oracle
(Lemma~\ref{det:reporter}).

\paragraph{Cuts from the atlas.}
Let $\vz$ be a candidate grid point. If an endpoint $\vt\pm\vz$ is not
a finite entry of the atlas, a precomputed separator of that endpoint
from the convex hull of the finite entries in its bucket cuts $\vz$
off. If
$\vz\notin HP$, a body query gives a separator. Otherwise both endpoints
carry exact values and supporting prices $\vp_\pm$, and the subgradient
inequalities at $\vt+\vz$ and $\vt-\vz$ give, for every affordable
offset $\vu$,
\[
 (\vp_+-\vp_-)(\vu-\vz)\le g_{\vt}(\vu)-g_{\vt}(\vz)\le2s-g_{\vt}(\vz).
\]
If $g_{\vt}(\vz)\le2s$ then $\vz$ is an output. Otherwise, if
$\vp_+\ne\vp_-$ the left-hand side is a cut through $\vz$ that keeps
every output, and if $\vp_+=\vp_-$ the inequality certifies that there
is no output at all.

\paragraph{Output-sensitive enumeration.}
With these cuts, deterministic integer centerpoints again find an output
in a search box, or certify that none exists, after $O(\log V)$ cuts.
Bisecting the search box and pruning empty parts lists $k$ outputs with
$O((1+k)\log^2V)$ such tests. Each test costs $\Ot(\sqrt V)$ because of
the slower deterministic separators. Including the bucket dictionaries,
a source with $Q$ queries and $O$ outputs costs
$\Ot(V^{3/2}+(Q+O)\sqrt V)$, and the total over all sources is
$\Ot(\cV^{3/2})$, the same order as the atlases. Since no convolution is
used, this change improves the exponent from $(4-\rho)/(3-\rho)$ to
$\max\{\rho,3/2\}$.

\subsection{Removed-item tables from convolution}
\label{ov:transfer}

The removed-item tables are the only place where
$(\min,+)$-convolution is needed. For a pool $I$ of labelled items with
integer columns and costs of arbitrary sign, and a box
$\Bx=\prod_j[-\Delta_j,\Delta_j]$ with $M$ lattice points, we need a
table $\mathsf D$ on $\Bx$ whose finite entries are witnessed by actual
subsets of labels and which is sandwiched as
\[
 \min_{Y\subseteq I,\ \sum_Y\va_i=\vz}c(Y)
 \ \le\ \mathsf D(\vz)\ \le\
 \min_{Y\in\mathcal F_{\Bx}(I),\ \sum_Y\va_i=\vz}c(Y),
\]
where $\mathcal F_{\Bx}(I)$ consists of the \emph{hereditarily clipped}
subsets, all of whose subsums stay in $\Bx$. Zhou~\cite{Zhou26} builds
such tables with color coding. We give a deterministic reduction to
witnessed convolution with running time $\Ot(M^{\rho})$
(Theorem~\ref{det:transfer}).

\paragraph{From vectors to scalars.}
Split the columns into orthants and reflect each orthant into
$\prod_j[0,\Delta_j]$. A mixed-radix encoding
$E(\va)=\sum_ja_j\prod_{i<j}(\Delta_i+1)$ maps this box injectively to
scalars below $M$. Different vectors may have the same scalar sum
because of carries, but every carry decreases the digit sum
$\sum_ja_j$. Adding a large multiple of the digit sum to every cost makes
the digit sum the primary criterion, so the scalar optimum at $E(\vz)$
is attained by a subset with vector sum exactly $\vz$, also along every
partial sum that the clipping requires. The orthant tables are then
merged by carry-free signed convolutions, clipping to $\Bx$ after every
merge.

\paragraph{All scalar capacities.}
For scalar weights we need minimum costs at all exact weights up to a
capacity $t$, in time $\Ot(t^{\rho})$
(Proposition~\ref{det:transfer-scalar}). We follow the derandomization
of Chan~\cite{Chan26}. Split the items into dyadic weight layers
$[a,2a)$. A solution of weight at most $t$ uses at most $k=t/a$ items of
one layer, so it suffices to compute, for items of weight at most
$u<2a$, a witnessed table on $[0,ku]$ that dominates every subset of at
most $k$ items. To do so, split the pool into two halves such that every
relevant subset is balanced, solve the two halves recursively with
cardinality about $k/2$, and convolve the results. The convolution
lengths shrink geometrically. A node with cardinality bound $k_v$ costs
about $(k_vu)^{\rho}$, and the cardinality bound shrinks by a factor
$\varsigma$ slightly above $1/2$, so the total is
$(ku)^{\rho}\sum_j(2\varsigma^{\rho})^j=O((ku)^\rho)$ because $\rho>1$. This
is where we use the assumption $\rho>1$.

\paragraph{Balanced splits deterministically.}
A random split is balanced for a fixed subset, but we need one split
that works for every subset at once. As in~\cite{Chan26}, we represent
every subset, up to an equally good replacement with the same weight
and cardinality, as a union of few \emph{canonical subsets} of size at
most $b$, and we balance only the canonical subsets. The canonical
subsets come from exact-cardinality tables on a balanced tree, and the
split is found by the method of conditional expectations applied to an
exponential potential, with exact rational arithmetic
(Lemma~\ref{det:transfer-halver}). One point needs care for integer
programs: \cite{Chan26} keeps a single best item per weight, which is
not valid for labelled $0$--$1$ items, since two labels with the same
column may both be selected. We therefore build the tree on sorted
label indices rather than on weights; the canonical tables then remain
exact for repeated columns and for costs of arbitrary sign
(Lemma~\ref{det:transfer-canonical}).

\subsection{Convolution primitives}
\label{ov:conv}

\paragraph{Deterministic.}
Chan, Itakura, Ma, and Yu~\cite{CIMY26} give a deterministic algorithm
for real $(\min,+)$-products of $s\times s$ matrices in time
$\Ot(s^{2.9997})$, and its comparisons involve only sums of a constant
number of inputs, so it runs with polynomial bit work on integers of
arbitrary length. The standard blocking reduction writes a
$(\min,+)$-convolution of length $m$ as $O(\sqrt m)$ products of
$\sqrt m\times\sqrt m$ matrices, which costs
$\Ot(\sqrt m\cdot m^{2.9997/2})=\Ot(m^{1.99985})$. Witnesses come for
free by encoding the index of the first argument into the low-order
digits of the entries (Lemma~\ref{detmc:lemma}).

\paragraph{Randomized.}
Wang~\cite{Wang26} gives a deterministic 3SUM decision algorithm with
exponent $\alpha=1.920279408185$ on polynomially bounded integers.
Zhou~\cite[Appendix~B]{Zhou26} turns it into witnessed convolution of
arbitrary bit length with exponent $1+\alpha/2=1.9601397\ldots$: block
pairs of length $\sqrt m$ are tested for indexed equalities by 3SUM
decision, a random prime reduces large entries to polynomial size, and
the output is recovered one bit per round. We use this result as a
black box (Lemma~\ref{detmc:random}).

\subsection{The recursion and its accounting}
\label{ov:framework}

Sections~\ref{fw:frame}--\ref{det:composition} develop the recursion in
which these local tasks arise. One LP and the Steinitz lemma reduce the
problem to a compressed pool of $O_d(N^{2d})$ signed binary units
(Lemma~\ref{fw:kernel}). A single integer perturbation of the costs,
the \emph{common objective}, makes every set of tight labels independent
in every pool that will ever occur. Hence all local LPs, phase
comparisons, and source tables use one objective
(Section~\ref{fw:objective}).

A state of the recursion is a pool with a batch of strict queries. Its
queries are scanned in decreasing order of their gap above the
fractional optimum, and each unmarked query becomes an \emph{anchor}.
The source of an anchor reports the affordable offsets $S_v$. A
discrete rounding of these offsets gives their span $\mathcal L_v$ and a
parallelotope $R_v$ comparable with $S_v$, and the anchor then marks the
nearby queries (Section~\ref{fw:geom}). The windows of different anchors are disjoint,
so the masses $\mu_v$ of the anchors sum to at most the mass of the state. The
span $\mathcal L_v$ decides the dispatch: if it is proper, the light
items lose rank; if it is full but $S_v$ does not contain a large dilate
of the column hull, the light items form a thinner pool; otherwise the
anchor is \emph{deep} (Theorem~\ref{fw:dispatch}). Boundary targets are
automatically of the first kind, so no minimal face is ever computed. A
deep anchor uses the canonical cores, common progressions, and
affordable fibres of~\cite{CMZ26b,CMZ26,Zhou26}: it either finds a
solution, restricts to a proper sublattice, or lifts the pool to rank
$r+1$ with a volume loss of order $\nu$ (Section~\ref{fw:deep}). Every
transfer to a child computes a removal table and an allowance merge, and
builds the child source window before any LP filter, so that failed
branches are paid (Section~\ref{fw:charge}).

Restrictions to sublattices are coalesced into \emph{phases} with one
root pool, in which an injective endpoint extension and a common-price
argument bound the total mass by a logarithmic factor
(Lemma~\ref{fw:phase-moments}). A potential, which multiplies the
moment $M^\theta$ by a small power of $\Psi$ and by polylogarithmic
factors rewarding rank and the unused lift budget, decreases
geometrically along every other kind of edge
(Section~\ref{fw:global-sec}); this gives the moment and
source bounds of Theorem~\ref{fw:global}. These are exactly the
quantities used in Section~\ref{ov:interface}.

\subsection{Composition and amplification}
\label{ov:composition}

With exact convolution replies, Section~\ref{ov:interface} already gives
the running time. Correctness follows from the structural recursion
(Section~\ref{fw:correct}) together with three local guarantees: every
atlas entry is exact; every required affordable offset is reported; and
every removed-item table dominates every hereditarily clipped assignment
and is witnessed by actual labels. All other routines are
deterministic: integer normal forms, LLL, fixed-dimensional integer
programming and linear programming, integer centerpoints, and the
progression construction of~\cite{CMZ26}. So the convolution primitive
is the only possible source of randomness.

For a randomized primitive with error probability $p$, each call is
repeated $\ell$ times. Every returned value is checked against its
witness, and the minimum valid candidate is kept at each index. Every
valid candidate is an attained value and hence an upper bound on the
true minimum, so a single fully correct trial makes the combined reply
exact. The combined call therefore fails with probability at most
$p^\ell$, even conditionally on the history. The subtle point is that
the number of calls is adaptive. We give the $c$th call
$\ell_c=O(\log(c/p)/\log(1/p))$ repetitions, so the failure
probabilities sum to less than $p$ with no bound on the number of calls
fixed in advance. Only the worst-case running time needs a guard.
Each call is charged its hypothesised cost $\tau_c$, computed from its
size and the primitive's running-time constants, and each trial is
stopped after $\tau_c$ steps. If the logical work plus the total charge
would exceed an envelope $W_\star$, valid for every execution with
exact replies, we stop with a failure report. Since $\tau_c\ge1$, at
most $W_\star$ calls are made, so
$\ell_c=O(1+(\log N+\log\cB)/\log(1/p))$ throughout
(Section~\ref{det:probability-section}).

\section{A deterministic construction of the local LP atlas}
\label{det:section}

This section builds the local LP atlas deterministically. A
near-linear dictionary of integer sites can be obtained by randomized
incremental hull construction, but we do not attempt to derandomize it.
Instead we trade preprocessing against query time, which yields a
deterministic local LP atlas with exponent $3/2$. All algorithms in this
section are deterministic. Constants and polynomial degrees may depend
on the fixed rank $r$, but the exponent $3/2$ does not.

We first fix the local model and state the main result,
Theorem~\ref{det:atlas}. We then build exact dictionaries on small
boxes, one LP per box point, and assemble them into a truncated
hierarchy that balances preprocessing against query time. This
hierarchy drives a deterministic traversal of the linearity tiles of
the fractional cost function, which proves the theorem. Finally, we
bound the total charge of the slower column-body queries and give the
reporter of affordable offsets
(Section~\ref{det:reporter-section}).

\subsection{The local model and the statement}

Let $I$ be a finite labelled pool of nonzero columns $\va_i$ with
rational costs $c_i$ and variables $0\le x_i\le1$. Let $\Lambda_I$ be
its current lattice; it has rank $r\ge1$ and contains all columns.
Items with equal signed columns are grouped into $T_I$ types. A type may
have a binary-encoded multiplicity and a nondecreasing marginal-cost
profile $c_{\va,1}\le c_{\va,2}\le\cdots$. The profile interface
supplies exact marginal access, identifiers of actual labels, prefix
counts and costs, threshold counts and costs, and predecessor and
successor queries. Each operation takes polynomial bit work and depends
only logarithmically on the multiplicity. Reversed, restricted, and
affinely re-costed views support the same operations.

We fix a rational objective that is generic in the following sense:
\begin{equation}\label{det:generic}
 \{i:\ c_i=\vpi\va_i\}\ \text{is independent and has at most $r$
 elements, for every row vector }\vpi.
\end{equation}
In words, at every price the items with zero reduced cost are linearly
independent, so the objective admits no degenerate ties. This objective may be the single common,
finitely perturbed objective of the whole computation; no new
perturbation is made for an individual source. Write
\[
 f_I(\vt)=\min\bigl\{\vc_I^{\T}\vx:\ A_I\vx=\vt,\ \vzero\le\vx\le\vone\bigr\},
\]
with $f_I(\vt)=+\infty$ outside $Z(I)$, and let $P_I$ and $\Psi(I)$ be as
in Section~\ref{det:overview}. For a compact rational polytope $K$ of
constant description complexity put
\begin{equation}\label{det:volume}
 V(K,I)=1+\frac{\vol_r(K+6P_I)}{\covol(\Lambda_I)}.
\end{equation}
Thus $V(K,I)$ measures, in lattice cells, the window $K$ enlarged by a
constant multiple of the column hull.

We use the following conventions on input size and arithmetic. Before
its access structures are prepared, the explicit compressed profile
representation has size $n$. Once immutable profiles have been
prepared, a \emph{pool scan} means a scan of the $T_I$ type records. The
bit parameter $\cB$ covers all coordinates, costs, capacities,
identifiers, profile sums, and source data. As in the integer-cost
model, affine normalizations use common denominators of bounded bit
length; no operation takes a least common multiple over unrelated table
entries. Exact fixed-dimensional rational linear programming and
separation--optimization are used in their bit-complexity
models~\cite{GLS88}.

\begin{theorem}[Deterministic local LP atlas]\label{det:atlas}
Fix the rank $r$ and a grid denominator $q$. The exact table of $f_I$ on
$K\cap(\Lambda_I/q)$, together with all fine LP tiles needed there, can
be constructed deterministically in time
\begin{equation}\label{det:atlas-time}
 \Ot_{r,q}\bigl(n+V(K,I)^{3/2}\bigr)\poly_r(\cB).
\end{equation}
Every finite entry comes with an exact supporting price and a compressed
basic witness for the common objective. The aggregate target and cost of
a witness are available in $O_r(1)$ arithmetic operations, its at most
$r$ basic labels are explicit, and the selected prefix of any requested
original type can be accessed without scanning the other types.
Infeasible entries are identified exactly. With prepared immutable
profiles, the term $n$ may be replaced by a scan of the $T_I$ type
records.
\end{theorem}

As a by-product, the construction provides exact support, membership,
separation, and gauge queries for $P_I$, with
$\Ot_r(\Psi(I)^{3/2})\poly_r(\cB)$ preprocessing and
$\Ot_r(\sqrt{\Psi(I)})\poly_r(\cB)$ time per query. These queries are
slower than those of the randomized near-linear dictionary;
Proposition~\ref{det:source-charge} pays for the difference explicitly.

\subsection{Deterministic dictionaries on small boxes}

The basic building block is a dictionary that stores, at every lattice
point of a small box, the value of the lower convex hull of the given
sites together with a supporting plane and a witness. It is built by
solving one fixed-dimensional LP per box point, so its cost is the
number of sites times the number of box points. A support query is then
answered by a centerpoint search over the stored points.

\begin{lemma}[Direct integer-site dictionary]\label{det:small-dictionary}
Let $A=\{\va_1,\ldots,\va_t\}\subset\Z^r$ lie in an integer coordinate
box $\Bx$ containing $U$ lattice points, and let $h_1,\ldots,h_t$ be
rational heights. Define
\[
 D_A=\conv A,\qquad
 F(\vx)=\min\Bigl\{\sum_j\lambda_jh_j:\
      \sum_j\lambda_j\va_j=\vx,\ \sum_j\lambda_j=1,\ \lambda_j\ge0\Bigr\}.
\]
In deterministic time $\Ot_r((t+1)U)\poly_r(\cB)$ one can store, at
every $\vz\in\Bx\cap\Z^r$, either
\begin{enumerate}
\item $F(\vz)$, an affine function that is a supporting plane of $F$ valid
      on all of $D_A$, and a convex-combination witness on at most $r+1$
      actual sites; or
\item a strict rational separator of $\vz$ from $D_A$.
\end{enumerate}
This covers lower-dimensional $D_A$, boundary points, and heights of
arbitrary sign. The dictionary uses $O_r(U)\poly_r(\cB)$ space, and
$O_r(t+U)\poly_r(\cB)$ space including its input. No convex hull is
constructed.
\end{lemma}

\begin{proof}
We solve two fixed-dimensional linear programs at every box point: one
decides membership in $D_A$ and returns a separator, and the other
returns the value, a supporting plane, and a sparse witness.

Duplicate sites may be replaced by a representative of minimum height.
Both programs below have $r+1$ variables, $t$ constraints, one per
site, and $2(r+1)$ box constraints on the variables, so each feasible
region is a polytope. We solve them with a deterministic linear-time algorithm for linear
programming in fixed dimension~\cite{Megiddo84,ChazelleMatousek96}, in
its LP-type form with a fixed lexicographic tie rule. Over a nonempty
polytope the lexicographically greatest optimal point exists and is a
unique vertex, so the tie rule is well defined. The algorithm uses $O_r(t+1)$
violation tests and basis computations on $O_r(1)$ constraints, so every
number it compares is the solution of a fixed-size subsystem and has
$\poly_r(\cB)$ bits. It returns this optimal vertex together with an optimal
\emph{basis}: a set of at most $r+1$ constraints (sites or box
constraints) on which the restricted program has the same optimum.

\emph{Membership and separation.} For an integer $\vz\in\Bx$ solve
\[
 \max_{\beta,\vv}\ \vv\vz-\beta
 \quad\text{subject to}\quad
 \vv\va_j\le\beta\ \text{for all }j,\qquad
 \|\vv\|_\infty\le1,\ |\beta|\le r2^{\cB}.
\]
The box constraints only bound the program; they are added to the
constraint list. By Farkas' lemma the optimum is positive exactly when
$\vz\notin D_A$, and then the optimal $(\vv,\beta)$ is a strict rational
separator: $\vv\vx\le\beta<\vv\vz$ for all $\vx\in D_A$.

\emph{Value, plane, and witness.} If $\vz\in D_A$, consider the dual pair
\begin{equation}\label{det:site-lp}
 F(\vz)=\min\Bigl\{\sum_j\lambda_jh_j:\ \sum_j\lambda_j(\va_j,1)=(\vz,1),\
 \boldsymbol\lambda\ge\vzero\Bigr\}
 =\max_{\alpha,\vpi}\bigl\{\alpha+\vpi\vz:\
 \alpha+\vpi\va_j\le h_j\ \text{for all }j\bigr\},
\end{equation}
which we solve, in the bounded form~\eqref{det:site-lp-boxed} below, by
the fixed-dimensional algorithm.
The primal is feasible, so the dual optimum is finite and is attained by
LP duality. The optimal face of the right-hand program can be unbounded,
however. It has a nonzero lineality space whenever $A$ does not affinely
span $\R^r$; for a single site at $\vzero$, the optimal intercept is
$\alpha=h_1$ and the slope $\vpi$ is unrestricted. It also has nonzero
recession directions whenever $\vz$ lies on the boundary of $D_A$. A
lexicographic rule applied directly to the dual coordinates can then be
unbounded, so we optimize over a bounded representative instead.

\emph{Bounded representative.} The optimal face
\[
 \Omega_{\vz}=\bigl\{(\alpha,\vpi):\ \alpha+\vpi\va_j\le h_j\ \text{for all }j,\
 \alpha+\vpi\vz=F(\vz)\bigr\}
\]
is a nonempty polyhedron in $\R^{r+1}$. Its data have $\poly_r(\cB)$
bits, because $F(\vz)$ is the value of a basic primal solution. A
minimal face of $\Omega_{\vz}$ is an affine subspace
$\{\vy:\ A'\vy=\vb'\}$ cut out by a subsystem of these constraints
taken with equality. Solving a maximal nonsingular square subsystem of
$A'\vy=\vb'$ by Cramer's rule, with the remaining coordinates set to
zero, gives a point $\vy^\circ\in\Omega_{\vz}$ of encoding length at
most $\varphi$, where $\varphi=\poly_r(\cB)$ is an explicit bound depending only on $r$ and
$\cB$. Put $R=2^{\varphi}+1$. Every coordinate of $\vy^\circ$ then has
absolute value less than $R$. Run the fixed-dimensional algorithm on the
boxed program
\begin{equation}\label{det:site-lp-boxed}
 \max_{\alpha,\vpi}\bigl\{\alpha+\vpi\vz:\
 \alpha+\vpi\va_j\le h_j\ \text{for all }j,\
 |\alpha|\le R,\ |\pi_i|\le R\ (1\le i\le r)\bigr\}.
\end{equation}
Its feasible region is a polytope, so the tie rule selects a unique
optimal vertex $(\alpha,\vpi)$ whose coordinates have $\poly_r(\cB)$
bits. Since $\vy^\circ$ is feasible for~\eqref{det:site-lp-boxed}, the
boxed optimum is $F(\vz)$. The box only removes points, so $(\alpha,\vpi)$
is also an optimal solution of the right-hand program
of~\eqref{det:site-lp}.

\emph{Supporting plane.} The affine function
$\vx\mapsto\alpha+\vpi\vx$ supports $F$ on all of $D_A$: for every convex
combination $\vx=\sum_j\mu_j\va_j$, dual feasibility gives
$\alpha+\vpi\vx\le\sum_j\mu_jh_j$, and minimizing over the representations
of $\vx$ gives $\alpha+\vpi\vx\le F(\vx)$, with equality at $\vz$. Its
coefficients have $\poly_r(\cB)$ bits, a fact used by the support queries
below.

\emph{Witness on actual sites.} Let $B$ be the returned basis, with
$|B|\le r+1$; it may contain box constraints. The LP dual
of~\eqref{det:site-lp-boxed} is
\[
 \min\ \sum_j\lambda_jh_j+R\Bigl(\nu^++\nu^-+\sum_{i=1}^r(\mu_i^++\mu_i^-)\Bigr)
 \ \ \text{s.t.}\ \
 \sum_j\lambda_j\va_j+\boldsymbol\mu^+-\boldsymbol\mu^-=\vz,\
 \sum_j\lambda_j+\nu^+-\nu^-=1,
\]
over nonnegative variables, in which each box constraint has become an
artificial column of cost $R$. The program~\eqref{det:site-lp-boxed}
restricted to $B$ has the same finite optimum $F(\vz)$, so by LP duality
the dual above, restricted to the columns in $B$, is feasible with value
$F(\vz)$. Such a solution on at most $r+1$ columns is found by a
fixed-size computation. Extended by zeros, it is optimal for the full
dual above, whose optimum is also $F(\vz)$. Complementary slackness
against the optimal point $\vy^\circ$ of~\eqref{det:site-lp-boxed},
which satisfies every box constraint strictly, forces every artificial
variable to be zero. Hence the solution is a convex combination
$\sum_{j\in B}\lambda_j\va_j=\vz$ of at most $r+1$ actual sites with
$\sum_{j\in B}\lambda_jh_j=F(\vz)$, which is the required witness.

Neither program assumes that $D_A$ is full-dimensional, that $\vz$ is an
interior point, or that the heights have a sign. Each box point costs
$O_r(t+1)\poly_r(\cB)$ time, and a record of constant size is stored.
Repeating this at the $U$ box points proves the bounds.
\end{proof}

Both the support queries below and the reporter of
Section~\ref{det:reporter-section} search the integer points of a box
with cutting planes. We isolate this search once.

\begin{lemma}[Centerpoint search]\label{det:centerpoint}
Let $E_0\subset\R^r$ be a rational polytope of constant description
complexity with $U$ integer points, given by $O_r(1)$ inequalities of
total encoding length at most $L_0$. Maintain a rational polytope $E$,
initially $E_0$, and repeat the following step: if $E\cap\Z^r=\emptyset$,
stop; otherwise pick an integer point $\vz\in E$ as below and call an
oracle, which either stops the search or returns a nonzero integral row
vector $\vg$, after which $E$ is replaced by
$E\cap\{\vu:\ \vg\vu\le\vg\vz-1\}$. Then the search stops after
$O_r(\log(U+1))$ oracle calls. If every returned cut has at most $L$
bits, all other work is $\poly_r(L_0+L,\log(U+1))$, deterministically,
and every queried point has $\poly_r(L_0+L,\log(U+1))$ bits.
\end{lemma}

The term $L_0$ cannot be dropped: $E_0=[2^{\cB},2^{\cB}+1]$ has only two
integer points, and writing either of them takes $\Omega(\cB)$ bits. In
all our applications $E_0$ is a box, or a subbox of one, with
$\poly_r(\cB)$-bit corners, so $L_0=\poly_r(\cB)$ is already covered by
the factor $\poly_r(\cB)$.

\begin{proof}
Integer feasibility of $E$ and, when needed, the affine span of
$E\cap\Z^r$ are computed in fixed dimension~\cite{Lenstra83,GLS88}; the
latter uses at most $r$ rank increases, each found by integer linear
optimization. For the centerpoint we use~\cite[Theorem~30]{BasuOertel17}
with $n=r$ integer and $d=0$ continuous coordinates. Its measure
$\mu_{{\rm mixed},E}$, defined in~\cite[(7)]{BasuOertel17}, is then the
uniform probability measure on $E\cap\Z^r$, and its depth
$f_\mu(\vz)$ is the infimum of $\mu(\{\vy:\vu(\vy-\vz)\ge0\})$ over unit
vectors $\vu$~\cite[(3)]{BasuOertel17}. The theorem returns, in time
polynomial in the input size of $E$ for fixed $r$, an integer point
$\vz\in E$ with $f_\mu(\vz)\ge2^{-r^2}(d+1)^{-(r+1)}=2^{-r^2}$. Every
closed halfspace containing $\vz$ contains one of the halfspaces in this
infimum, so it contains at least a fraction $\alpha_r=2^{-r^2}$ of
$E\cap\Z^r$. The algorithm is deterministic, and it is separate from the
Monte Carlo algorithm of~\cite[Theorem~33]{BasuOertel17}, which we do
not use. With $d=0$ it consists of the following steps.
\begin{itemize}
\item For large lattice width it rounds a centerpoint of the uniform
measure on $E$, which is computed by their Theorem~27.
\item Otherwise it finds a flatness direction. It then recurses on the
$O_r(1)$ integer slices, weights each recursive point by the number of
integer points of its slice, and takes a centerpoint of these $O_r(1)$
weighted points by brute force.
\end{itemize}
The slice counts, which their proof does not discuss, are computed
deterministically in polynomial time in fixed
dimension~\cite{Barvinok94}. For $r=1$ we do not rely on their
argument, whose large-width branch gives depth only close to $1/2$:
a median of $E\cap\Z$, found by counting, has depth $1/2$ directly.
Only some constant $\alpha_r>0$ depending on $r$ matters below. An integer point $\vu$
removed by the cut satisfies $\vg\vu>\vg\vz-1$, hence
$\vg\vu\ge\vg\vz$, so the cut removes the integer points of a closed
halfspace containing $\vz$. Each step therefore removes at least an
$\alpha_r$ fraction of $E\cap\Z^r$, and after $O_r(\log(U+1))$ steps
$E\cap\Z^r$ is empty. The polytope $E$ is described by the $O_r(1)$
inequalities of $E_0$, of total length $L_0$, and $O_r(\log(U+1))$ cuts
of at most $L$ bits each. The fixed-dimensional integer feasibility,
affine-span, and centerpoint computations run in time polynomial in this
description length, and every integer point they return has
polynomially many bits; this gives both bounds.
\end{proof}

\begin{lemma}[Support from a small dictionary]\label{det:support}
After the preprocessing of Lemma~\ref{det:small-dictionary}, a query
with a rational row vector $\vpi$ returns
\[
 \min_{\va_j\in A}(h_j-\vpi\va_j)
\]
and an attaining actual site in deterministic time
$\poly_r(\cB,\log(U+1))$.
\end{lemma}

\begin{proof}
The minimum in question is also the minimum of $F(\vz)-\vpi\vz$ over
$D_A\cap\Z^r$. Indeed, actual sites are feasible integer points, and
every other feasible value is a convex combination of values at actual
sites. Run the search of Lemma~\ref{det:centerpoint} on $E_0=\Bx$, whose
corners have $\poly_r(\cB)$ bits, so that $L_0=\poly_r(\cB)$, keeping the best value $F(\vz)-\vpi\vz$ found so far. At a queried
point $\vz$ the oracle reads the dictionary. If $\vz\notin D_A$, it
returns the stored separator, scaled to be integral; this keeps all of
$D_A$. If $\vz\in D_A$, it updates the incumbent and lets $\vg$ be the
stored subgradient minus $\vpi$. If $\vg=\vzero$, the incumbent is
optimal and the search stops. Otherwise it returns $\vg$ scaled to a
nonzero integral row vector; by the subgradient inequality every
strictly improving integer point survives the cut. Hence, when the search
stops, the incumbent is optimal.

All queried points are integer points of the original box, where the
dictionary is stored, and every cut is read from the dictionary, so its
bit length is $\poly_r(\cB)$. Finally, an optimal dictionary witness is
a convex combination of values at actual sites, all of which are at
least the optimum; hence every site of positive weight is an attaining
site. No random choice is made.
\end{proof}

\subsection{A truncated spatial hierarchy}

A single dictionary on the whole box would cost the number of sites
times the number of box points, which can be quadratic in the box
size $M$. We therefore cut
the box into blocks of about $\sqrt M$ points and store dictionaries
only inside blocks. This makes preprocessing $M^{3/2}$, at the price of
queries that visit about $\sqrt M$ blocks.

First we need a box of the right size. Both here and in
Lemma~\ref{fw:rounding} we use the following greedy basis.

\begin{lemma}[Greedy well-shaped basis]\label{det:greedy-basis}
Let $X$ be a finite set of vectors spanning a $k$-dimensional space
$\mathcal L$, put $P_X=\conv(X\cup-X)$, and let $B=(\vb_1,\dots,\vb_k)$
be independent vectors of $X$. Repeat the following scan: if some
$\vx\in X$ has a coordinate of absolute value greater than two in the
basis $B$, replace the corresponding basis vector by $\vx$. Each
replacement multiplies $|\det B|$, measured in $\mathcal L$, by more than
two, and always $|\det B|\le k!\vol_k(P_X)/2^k$. Hence there are at most
$\log_2\bigl(k!\vol_k(P_X)/(2^k|\det B_{\rm init}|)\bigr)$ replacements,
and at termination
\[
 P_X\subseteq B[-2,2]^k\subseteq2kP_X .
\]
Each scan uses $O_k(|X|)$ fixed-size linear-algebra operations.
\end{lemma}

\begin{proof}
By Cramer's rule, replacing $\vb_l$ by $\vx$ multiplies the determinant
by the absolute value of the $l$th coordinate of $\vx$. The
crosspolytope $\conv(\pm\vb_1,\dots,\pm\vb_k)$ lies in $P_X$ and has
volume $2^k|\det B|/k!$, which gives the upper bound and hence the count.
At termination every point of $X$ lies in the symmetric convex set
$B[-2,2]^k$, and so does $P_X$. Conversely,
$B[-1,1]^k\subseteq k\conv(\pm\vb_1,\dots,\pm\vb_k)\subseteq kP_X$.
\end{proof}

We work in a genuine basis of
$\Lambda_I$ and use the following constructive lattice-box fact:
\begin{equation}\label{det:box}
 P_I\subseteq\Bx_P=\{F\vz:\ |z_j|\le L_j\}
      \subseteq\kappa_rP_I,
 \qquad L_j\in\Z_{\ge1},\qquad
 M:=\prod_j(2L_j+1)=O_r(\Psi(I)).
\end{equation}
The basis $F$ and the widths $L_j$ can be found deterministically in
$\Ot_r(T_I+1)\poly_r(\cB)$ time. For completeness we sketch the
construction. Apply Lemma~\ref{det:greedy-basis} to the type columns,
starting from independent columns $C$. Since $|\det C|\ge\covol(\Lambda_I)$
initially, at most $\log_2\Psi(I)$ replacements occur. In the final
coordinates the body contains the unit
crosspolytope and lies in $[-2,2]^r$, and the lattice $C^{-1}\Lambda_I$
contains $\Z^r$ and has covolume $\covol(\Lambda_I)/|\det C|$. An
LLL-reduced basis $\vb_1,\dots,\vb_r$ of this lattice, with Gram--Schmidt
lengths $l_i$, satisfies $l_i=O_r(1)$, $\|\vb_i\|=O_r(l_i)$, and
$\prod_il_i=\covol(\Lambda_I)/|\det C|$. The bound $l_i=O_r(1)$
follows by projecting a unit vector nontrivially onto the last Gram--Schmidt
direction and using the Lov\'asz inequalities. Back substitution in the
size-reduced basis bounds the coordinates of points of the body by
$O_r(1/l_i)$. Integral widths of these orders give
$M=O_r(|\det C|/\covol(\Lambda_I))=O_r(\Psi(I))$, while the geometric
radius of the box stays bounded by a constant depending on $r$. This
proves~\eqref{det:box}; rational LLL and integer normal forms are
deterministic~\cite{LLL82}.

\begin{lemma}[Truncated hierarchy tradeoff]\label{det:forest}
Let $T$ distinct integer sites with rational heights lie in an integer
box of cardinality $M$. For every integer $1\le b\le M$ there is a
deterministic data structure with
\begin{align*}
 \text{preprocessing time}
   &\le \Ot_r(Tb+M)\poly_r(\cB),\\
 \text{space}
   &\le \Ot_r(M)\poly_r(\cB).
\end{align*}
After $s$ sites are specified as excluded, a support query over the
remaining sites takes
\begin{equation}\label{det:forest-query}
 \Ot_r(M/b+s+1)\poly_r(\cB)
\end{equation}
time and returns an attaining site that is not excluded.
\end{lemma}

\begin{proof}
Bisect integer coordinate intervals until every terminal box has at most
$b$ lattice points. A box can always be split along an interval
$[\beta^-,\beta^+]$ with at least two integer points: put
$h=\lfloor(\beta^-+\beta^+)/2\rfloor$ and use $[\beta^-,h]$ and
$[h+1,\beta^+]$. No integer site belongs to two children or is lost at a
boundary, and each child has between one third and two thirds of the
parent's points. Consequently, unless the initial box is already
terminal, every terminal box has more than $b/3$ points. There are
$O(M/b)$ terminal boxes, called \emph{block roots}, and the depth is
$O(\log(M+1))$.

Inside each block root, continue the same spatial hierarchy down to
individual sites, but store no dictionary strictly above the block
roots. At
every nonempty node at or below a block root, with $t_v$ input sites and
$U_v\le b$ grid points, build the direct dictionary of
Lemma~\ref{det:small-dictionary}; empty nodes are omitted. At each depth
of the forest the boxes are disjoint, so $\sum_vt_v\le T$ and
$\sum_vU_v\le M$. Since a stored node has $t_v\ge1$, its cost is
$O_r(t_vU_v)\poly_r(\cB)$, and the total cost at one depth is at most
$O_r(bT)\poly_r(\cB)$. Summing over the logarithmically many depths
proves the preprocessing and space bounds. Only $\Ot_r(M)$ point
certificates of constant size and actual-site witnesses are stored; the
node input lists add $\Ot_r(T)$ references, and $T\le M$.

For a query with exclusions, mark the paths from the $s$ excluded leaves
to their block roots. The maximal unmarked subtrees together with the
unmarked block roots partition the remaining sites. Their number is
\[
 O\bigl(M/b+s\log(M+1)\bigr),
\]
since there are $O(M/b)$ block roots and each marked path contributes at
most one unmarked sibling per level. The bound is additive: the number of
blocks is not multiplied by the number of exclusions. Apply
Lemma~\ref{det:support} once to each of these canonical nodes and take
the minimum. The paths and nodes are listed within the same bound using
the stored site coordinates, and no dictionary is rebuilt.
\end{proof}

Taking $b=\lceil\sqrt M\,\rceil$ gives preprocessing time
$\Ot_r(M^{3/2})\poly_r(\cB)$ and query time
$\Ot_r(\sqrt M+s+1)\poly_r(\cB)$ for support with exclusions. This is
the tradeoff used by the atlas. We do not claim to match the
$\Ot_r(M)$ preprocessing of the randomized dictionary deterministically.

\subsection{Exact events and deterministic traversal}

The atlas is computed by walking from tile to tile of the fractional
cost function. Each step moves the price along a ray and asks which
label becomes tight first. We answer this with the hierarchy of
Lemma~\ref{det:forest}, treating the few types that are already
selected as exclusions. A lattice counting bound
(Lemmas~\ref{det:pointed} and~\ref{det:window}) shows that there are
few such types, which keeps each step cheap.

At a fine LP tile with price $\vpi$, call a type \emph{active} if its
strict selected prefix is nonempty or it contains an actual basic label.
Suppose that at most $s$ types are active. In the hierarchy of
Lemma~\ref{det:forest}, we use the first marginal cost of each type as
its height and exclude exactly the active types. Every remaining type
then satisfies $\vpi\va<c_{\va,1}$.

\begin{lemma}[Exact next event]\label{det:event}
Consider the ray $\vpi+\lambda\vv$, $\lambda>0$, that releases one basic
equality and keeps the other $r-1$. The first crossing of an actual
label along this ray, or the fact that there is none, can be computed
deterministically in time
\[
 \Ot_r(M/b+s+1)\poly_r(\cB)
\]
after the preprocessing of Lemma~\ref{det:forest}.
\end{lemma}

\begin{proof}
We treat the non-excluded types by binary search with support queries,
and the at most $s$ excluded types directly.

For a type that is not excluded, the only possible first event is its
first marginal, and only if $\vv\va>0$. The ray starts inside every
halfspace $\vpi'\va\le c_{\va,1}$. Membership in their intersection is
decided exactly by a support query with exclusions, and it is monotone
along the ray. Every finite exit ratio
\[
 \frac{c_{\va,1}-\vpi\va}{\vv\va}
\]
has numerator and denominator of absolute value at most $2^L$, for a
known $L=\poly_r(\cB)$. A membership test at a sufficiently large power
of two detects an infinite exit. Otherwise binary search yields a bracket
of width less than $2^{-2L}$ whose lower end is feasible and whose upper
end is infeasible. Any constraint violated at the upper end has its exit
ratio in this bracket, as does the minimum ratio. Distinct candidate
ratios differ by at least $2^{-2L}$, so the returned constraint attains
the minimum exactly. The number of support queries is polynomial
in~$\cB$.

For each excluded type, take the least marginal strictly above $\vpi\va$
if $\vv\va>0$, and the greatest marginal strictly below it if
$\vv\va<0$. An empty side of a finite profile contributes no event, and a
currently tight marginal is skipped. Compare these at most $s$ ratios
with the answer for the non-excluded types. This handles increasing and
decreasing prefixes and finite upper bounds without expanding
multiplicities.
\end{proof}

Three volume comparisons below (Lemmas~\ref{det:pointed}
and~\ref{det:window} and Theorem~\ref{fw:dispatch}) use one elementary
fact about central sections.

\begin{lemma}[Central sections]\label{det:central}
Let $P\subset\R^r$ be a centrally symmetric convex body, let
$\vlambda\ne\vzero$ be a row vector, put
$A_0=\vol_{r-1}(P\cap\ker\vlambda)$ and
$w=h_P(\vlambda)/\|\vlambda\|$. Then:
\begin{enumerate}
\item $\frac2r\,wA_0\le\vol_r(P)\le2wA_0$;
\item for $0<\epsilon\le1$, the slab
$\{\vx\in P:\ |\vlambda\vx|\le\epsilon h_P(\vlambda)\}$ has volume at
most $2\epsilon wA_0\le r\epsilon\vol_r(P)$;
\item if a convex set $C$ contains $P\cap\ker\vlambda$ and a point $\vz$
with $\vlambda\vz=Hh_P(\vlambda)$, $H>0$, then
$H\vol_r(P)\le2r\vol_r(C)$.
\end{enumerate}
\end{lemma}

\begin{proof}
By Brunn's theorem the $(r-1)$-volumes of the sections of $P$ parallel to
$\ker\vlambda$ have a concave $(r-1)$st root; by symmetry it is even, so
the central section is the largest. Integrating over the width $2w$
gives the upper bound in item~1 and item~2. The two cones over the
central section with apexes at points of $P$ where $\vlambda$ equals
$\pm h_P(\vlambda)$ lie in $P$ and have total volume $2wA_0/r$, which is
the lower bound in item~1. In item~3 the cone from $\vz$ over the central
section lies in $C$ and has volume $HwA_0/r\ge H\vol_r(P)/(2r)$.
\end{proof}

The traversal needs the following geometric sparsity bound. It says
that a lattice point at gauge distance $H$ that is written as a
positive combination of distinct vectors in an open halfspace uses
only about $\sqrt{H\Psi}$ distinct vectors.

\begin{lemma}[Pointed distinct-type bound]\label{det:pointed}
Let $P$ be a symmetric convex body containing $r$ independent vectors of
a rank-$r$ lattice $\Lambda$, and let $\Psi=\Psi_r(P,\Lambda)=
r!\vol_r(P)/(2^r\covol\Lambda)$. Let $A\subset P\cap\Lambda$ be a set of
distinct vectors lying in an open halfspace through $\vzero$, and let
$\vt=\sum_{\va\in A}n_{\va}\va\in HP$ with positive integers $n_{\va}$
and $H\ge1$. Then
\[
 |A|\le C_r\sqrt{H\Psi}.
\]
\end{lemma}

\begin{proof}
We use two elementary lattice estimates. First, in rank $k$, a
centrally symmetric convex body containing $k$ independent lattice
vectors contains $O_k(\vol/\covol\Lambda)$ lattice points. (Symmetry is
needed: a thin neighbourhood of a long lattice segment can contain two
independent lattice vectors and arbitrarily many lattice points with
arbitrarily small area.) To see this, note that the convex hull of its
lattice points contains $\vzero$ and $k$ independent lattice vectors,
hence is full-dimensional, and triangulate it using all of these lattice
points as vertices;
each full simplex has volume at least $\covol\Lambda/k!$, and every
point is a vertex of some simplex (compare~\cite[Lemma~5.2]{CMZ26b}).
Second, for a functional $\vlambda$ that is primitive in the dual
lattice of $\Lambda$, with $P_0=P\cap\ker\vlambda$,
\[
 \Psi_{r-1}(P_0,\Lambda\cap\ker\vlambda)\le\Psi_r(P,\Lambda)/h_P(\vlambda).
\]
This is the lower bound of Lemma~\ref{det:central}(1) together with
$\covol(\Lambda\cap\ker\vlambda)=\|\vlambda\|\covol\Lambda$. Extending
independent lattice vectors in a smaller span to independent lattice
vectors in $P$ gives a flag in which every omitted width is at least one;
thus passing to the span of $A$ can only decrease the normalized section
volume.

We now induct on the rank of the span of $A$ and may assume it is full.
The idea is to slice by a functional that is nonnegative on $A$: few
vectors can have positive height, and the vectors at height zero are
handled by induction. The cone generated by $A$ is pointed. A primitive extreme ray $\vlambda$ in the dual lattice
of $\Lambda$ of its dual cone is nonnegative on $A$, and $\ker\vlambda$ contains $r-1$
independent vectors of $A$. Put $W=h_P(\vlambda)\ge1$ and
$\Psi_0=\Psi_{r-1}(P_0,\Lambda\cap\ker\vlambda)\le\Psi/W$. (The support
value $W$ need not be integral; only the heights of lattice vectors
are.) Each layer of $P$ at a positive integer height contains at most
$C_r\Psi_0$ lattice points, because differences within a layer lie in
$2P_0$. If $m_+$ of the distinct vectors have positive height, their
height sum is at least $m_+^2/(2C_r\Psi_0)$, while their
multiplicity-weighted height sum is at most $HW$. Hence
$m_+=O_r(\sqrt{H\Psi})$. Their total multiplicity is also at most $HW$,
so the sum of the vectors at height zero lies in $H(1+W)P_0$. By
induction there are at most
$C_{r-1}\sqrt{H(1+W)\Psi_0}=O_r(\sqrt{H\Psi})$ such vectors. Rank zero is
empty, which completes the induction.
\end{proof}

\begin{lemma}[Window sparsity and tile mass]
\label{det:window}
After one exact LP normalization, a convex guard of constant description
complexity can be chosen for the source window $K$ such that its retained
full tiles have total normalized determinant mass $O_r(V(K,I))$, and
every retained tile strictly selects at most $O_r(\sqrt{V(K,I)})$
distinct types.
\end{lemma}

\begin{proof}
We find an optimal price for a point of the window, normalize the costs
there, and enlarge the window by a constant multiple of the column hull
to obtain the guard. The tile mass bound is then a volume comparison,
and the sparsity bound follows from Lemma~\ref{det:pointed}.

\emph{An optimal price.} Find $\vy\in Z(I)\cap K$, or decide that the
intersection is empty, with a deterministic fixed-dimensional rational LP
oracle. One type scan computes the support function
$h_{Z(I)}(\vv)=\sum_{\va}u_{\va}\max(0,\vv\va)$, where $u_{\va}$ is the
multiplicity of type $\va$, together with a witnessing endpoint
assignment. Exact separation--optimization and primal recovery
therefore use $O_r(T_I)\poly_r(\cB)$ work. At a feasible $\vy$, an
optimal price $\vpi_0$ is found within the same bound, because the
value and a subgradient of
\[
 f_I^*(\vpi)=\sum_{i\in I}\max(0,\vpi\va_i-c_i)
\]
are obtained by one scan of the compressed profiles. Optimal tile prices
solve $r$ equalities for actual labels and hence have polynomially
bounded bit length, so exact fixed-dimensional optimization applies.
Condition~\eqref{det:generic} makes the resulting primal assignment
basic.

\emph{Normalization.} Normalize at $\vpi_0$ as
in~\cite[(5.1)]{CMZ26b}: $e_i=[\vpi_0\va_i>c_i]$, $\sigma_i=1-2e_i$,
$\bar\va_i=\sigma_i\va_i$, $\bar c_i=|c_i-\vpi_0\va_i|\ge0$, and put
$\vb_0=\sum_ie_i\va_i$. The substitution $x_i=e_i+\sigma_iz_i$ turns
$\vy$ into a normalized target and leaves $P_I$ and $\Lambda_I$
unchanged. The normalized target satisfies $\vy-\vb_0\in rP_I$, since
it is a sum of at most $r$ basic columns with coefficients in $[0,1]$. Each new signed type combines at most a reversed
prefix and a suffix of the two old orientations. Threshold operations are
sums of two interval operations, and selection in their sorted union
uses a logarithmic search. Actual identifiers and orientations are
retained, so no multiplicity is expanded.

\emph{The guard and the tile mass.} Take
\[
 W=K-\vb_0+(r+1)\Bx_P,\qquad C=W+rP_I.
\]
Since $\vy-\vb_0\in(K-\vb_0)\cap rP_I$, the set $W$ contains $\Bx_P$.
It also has a strict margin around $K-\vb_0$ and constant description
complexity, and it satisfies $\vol(C)/\covol(\Lambda_I)=O_r(V(K,I))$. At an
arrangement vertex $\vpi$, the $r$ tight actual columns form an
invertible basis matrix $B_T$, and the optimal target set is the tile
\[
 T=\vz_T+B_T[0,1]^r,\qquad
 \vz_T=\sum_{\bar c_i<\vpi\bar\va_i}\bar\va_i.
\]
These projected lower faces of the lifted zonotope form a face-to-face
tiling with disjoint interiors, and their masses
$\Delta_T=|\det B_T|/\covol(\Lambda_I)$ are positive integers. If $T$ has
an interior point in $W$, then $T\subseteq W+rP_I=C$. Hence the retained
mass and the number of retained tiles are $O_r(V(K,I))$.

\emph{Sparsity.} For a convex $C\supseteq P_I$ let
$H=\max_{\vz\in C}\|\vz\|_{P_I}$. Suppose $\vlambda$ supports $P_I$ at
value one and $\vlambda\vz=H$. Since $C\supseteq P_I$,
Lemma~\ref{det:central}(3) gives
\[
 H\vol(P_I)\le2r\vol(C),
 \qquad H\Psi(I)=O_r(V(K,I)).
\]
The strict corner $\vz_T$ is a positive integral combination of distinct
selected types lying in the open halfspace
$\{\vx:\vpi\vx>0\}$, because $\vpi\bar\va>\bar c\ge0$ for each of them.
Lemma~\ref{det:pointed} therefore bounds their number by
$O_r(\sqrt{V(K,I)})$.
\end{proof}

\begin{proof}[Proof of Theorem~\ref{det:atlas}]
We start at the tile containing the origin, cross facets one at a time
with Lemma~\ref{det:event}, and explore the retained tiles by
breadth-first search. Afterwards we enumerate grid points tile by tile
and attach witnesses.

The normalization and the box construction above are deterministic. An
empty feasible intersection is handled immediately. Zero columns, if
present before the model above is set up, have independent optimal
prefixes, and rank zero has only the target $\vzero$. Otherwise prepare
the forest of first-cost dictionaries with $b=\lceil\sqrt M\,\rceil$.
Since $M=O_r(\Psi(I))=O_r(V)$, where $V=V(K,I)$, the deterministic
preprocessing cost is $\Ot_r(V^{3/2})\poly_r(\cB)$.

\emph{The origin tile.} Maximize the sum of $r$ independent column
functionals over $\{\vpi:\ \vpi\va\le c_{\va,1}\ \text{for all types}\}$.
This polyhedron has nonempty interior: the at most $r$ columns of zero
normalized cost are independent, so a small vector that is negative on
them leaves all other inequalities strict. The polyhedron is pointed,
because it contains the constraints of $r$ independent columns. The
maximizing face is bounded: a recession direction $\vd$ satisfies
$\vd\va\le0$ for each of the $r$ objective columns, and it does not
change the objective, so the sum of these $r$ nonpositive numbers is zero
and each of them vanishes; independence gives $\vd=\vzero$. The bounded
face is the convex hull of its vertices, each of $\poly_r(\cB)$ bits, so
adding the box $\|\vpi\|_\infty\le R$ with $R=2^{\varphi}+1$ for a
suitable explicit $\varphi=\poly_r(\cB)$ leaves the maximizing face
unchanged and slack in every box constraint. We solve this
fixed-dimensional LP by a type scan with the algorithm and lexicographic
tie rule of Lemma~\ref{det:small-dictionary}, which returns an optimal
\emph{vertex} $\vpi$ of the bounded maximizing face. At a vertex of a
pointed polyhedron in $\R^r$ the tight constraints have rank $r$; no box
constraint is tight, so at least $r$ first labels are tight; by~\eqref{det:generic} at most $r$
are, and they are independent. Hence exactly $r$ first labels are
tight, they form an invertible basis, and no label is strictly
selected because $\vpi\va\le c_{\va,1}$ for every type, so the strict
corner is $\vzero$.

\emph{Crossing a facet.} For $T=\vz+B[0,1]^r$, the facet on which
coordinate $j$ is fixed to $\gamma\in\{0,1\}$ has outgoing price
direction $\vv=(2\gamma-1)\ve_j^{\T}B^{-1}$. Use Lemma~\ref{det:event}
to find its first positive crossing. If label $h$ arrives, replace basis
label $j$ by $h$ and update the strict endpoint assignment by
\[
 \ve'=\ve+\gamma\vone_j-[\vv\va_h<0]\,\vone_h .
\]
The other $r-1$ equalities remain tight, and genericity permits exactly
one arrival and yields an independent new basis. The two tiles share the
facet from opposite sides. If no label arrives, the facet lies on the
boundary of the zonotope. These formulas also cover consecutive copies
of the same type and both endpoint directions.

\emph{Traversal.} Retain exactly the full tiles that have an interior
point in $\inter W$, using a fixed-dimensional strict intersection test.
Their adjacency graph is connected. Indeed, any two interior points can
be joined through the convex set $\inter(Z(I)-\vb_0)\cap\inter W$, and the path
can be perturbed to avoid faces of codimension two. The origin tile belongs to this graph. Every tile
meeting the requested compact window lies in it, even at a boundary
contact, because the window has a strict guard margin. Breadth-first
search therefore reaches all required tiles. The sorted tuple of basic
labels is a unique key for a tile, and rejected neighbours are charged to
the retained facets that examine them.

By Lemma~\ref{det:window} there are $O_r(V)$ retained tiles, each
selecting at most $S=O_r(\sqrt V)$ types. A facet query excludes at most
$S+r$ types and hence takes $\Ot_r(\sqrt M+S+1)\poly_r(\cB)$ time. The
total traversal cost is
\[
 \Ot_r\bigl(M^{3/2}+V(\sqrt M+\sqrt V+1)\bigr)\poly_r(\cB)
 =\Ot_r(V^{3/2})\poly_r(\cB).
\]
A sparse map of the selected prefixes, of size $O_r(S+1)$ and with its
aggregate target and cost, may be copied for each retained tile and
updated by the two label changes above within the same bound. Store the
signed costs $\sigma_ic_i$ of the common objective as well as the
normalized costs; their difference is the affine term
$\vpi_0\bar\va_i$, so the same profile counts and sums give both
aggregates.

\emph{Grid points.} Enumerate the closed $\Lambda_I/q$-grid points of
each tile using Hermite or Smith normal form. A half-open fundamental
parallelotope contains exactly its index many points, and closing it
adds at most a factor depending on $r$. The total number of incidences is
therefore $O_r(q^rV)$, paid for by the tile mass. Filter to $K-\vb_0$ and
undo the translation. The requested grid itself can be enumerated by
recursive coordinate bisection and fixed-dimensional integer
feasibility, with polynomial bit work per reported point; thus no
unadapted bounding box is allocated. The grid has $O_r(q^rV)$ points,
since
translates of a fundamental cell around distinct grid points fit in a
constant dilate of the buffered window. Entries left unfilled have value
$+\infty$.

\emph{Witnesses.} At each entry solve the fixed-size basis system for the
basic fractions. The strict prefix assignment, the basic fractions, the
normalization baseline, and the orientation bits give the exact witness
and price for the common objective, and its aggregate target and cost are
stored. For a requested original type $\va$, query its immutable profile
at the returned common price $\vpi$: its strict prefix has
$\#\{j:\ c_{\va,j}<\vpi\va\}$ copies, and the at most $r$ actual basic
labels give the boundary and fraction corrections. This remains valid
after opposite signed types have been merged internally. In aggregate
form, if $\bar\ve_T$ is the strict assignment in normalized coordinates,
the strict assignment for the common objective is
\[
 \ve_T^{\rm com}=\ve+(\sigma_i\bar e_{T,i})_i
                  -\sum_{i\in B_T:\ e_i=1}\vone_i.
\]
The correction removes complemented tight labels from the strict
prefixes and costs $O_r(1)$. These representations meet the stated
witness access bounds. Including the initial profile preparation
proves~\eqref{det:atlas-time}.
\end{proof}

\subsection{Column-body queries and their total charge}

The deterministic column-body queries cost $\sqrt{\Psi(I)}$ each
rather than polylogarithmic time. We show that this loss is absorbed by
the same budget $\cV^{3/2}$ that pays for the atlases.

\begin{corollary}[Deterministic column-body oracles]
\label{det:body-oracles}
After $\Ot_r(\Psi(I)^{3/2})\poly_r(\cB)$ deterministic preprocessing,
exact support queries with an attaining signed column, membership
queries, strict separation queries, and gauge queries for $P_I$ take
$\Ot_r(\sqrt{\Psi(I)})\poly_r(\cB)$ time each.
\end{corollary}

\begin{proof}
Support queries come directly from the hierarchy; the other three
kinds of queries reduce to support by polarity. Apply the forest
construction, without exclusions, to the integer sites
$\{\pm\va_i\}$ with zero heights in the box~\eqref{det:box}. Minimizing
$-\vv\va$ gives $h_{P_I}(\vv)$ and an attaining column in the stated
time. This support oracle is a strong separation oracle for the bounded
polar body $P_I^\circ=\{\vv:\ \vv\va\le1\text{ for all signed
columns}\}$. Exact fixed-dimensional separation--optimization maximizes
$\vv\vx$ over $P_I^\circ$ with polynomially many oracle calls and
returns $\|\vx\|_{P_I}$; a value above one yields a strict separator.
Vertices of the polar body and a bounding box for it have polynomially
bounded bit length because the columns span the lattice.
\end{proof}

The accessible volume estimate
\[
 \widehat V(K,I)=1+\vol(K+6\Bx_P)/\covol(\Lambda_I)
\]
can be computed deterministically by fixed-dimensional rational geometry,
since $K+6\Bx_P$ has constant description complexity. It is within a
factor depending only on $r$ of $V(K,I)$, and it does not require the
whole column hull.

\begin{proposition}[Charging the slower support queries]
\label{det:source-charge}
Let the immutable sources $S$ have charged volumes $V_S$, external query
counts $Q_S$, and counts $O_S$ of reported offsets, all counted before
exact clipping, and put
\[
 \cV=1+\sum_S(V_S+Q_S+O_S).
\]
Suppose that every source makes at most $\Ot_r(Q_S+O_S+V_S)$ column-body
oracle calls, up to a polynomial bit factor. Then deterministic atlas
construction, column-body preprocessing, and all these oracle calls
cost
\[
 \Ot_r(\cV^{3/2})\poly_r(\cB),
\]
in addition to the initial preparation of explicit profiles.
\end{proposition}

\begin{proof}
Only sources with a nonempty window $K_S$ build these structures or
make these calls. Indeed, the source constructor
(Section~\ref{fw:constructor}) stops at its third step, before any
atlas, column-body structure, or reporter is built, when no candidate
remains; such a source makes no oracle calls and contributes nothing
here. Otherwise the window contains every candidate midpoint, because
$K_S\supseteq C+H_b\Bx_P$ (Section~\ref{det:reporter-section}), so it is
nonempty. A rank-zero source has $\Psi=1\le V_S$. For a source with
nonempty window, $\Psi(I_S)=O_r(V_S)$, because $K_S+6P_{I_S}$ contains a
translate of $6P_{I_S}$. Without this, the bound would fail: for
$K_S=\emptyset$ we have $V_S=1$ while $\Psi(I_S)$ may be large.
Consequently, summing over the sources that build structures,
\[
 \sum_S\Psi(I_S)^{3/2}+\sum_SV_S^{3/2}=O_r(\cV^{3/2})
\]
and
\[
 \sum_S(Q_S+O_S+V_S)\sqrt{\Psi(I_S)}
 \le O_r(\sqrt{\cV})\sum_S(Q_S+O_S+V_S)
 =O_r(\cV^{3/2}).
\]
The hypothesis covers the concrete geometric uses in adaptive reporting
of affordable bodies. These are the polylogarithmically many membership
tests per query and per emitted offset in Lemma~\ref{det:reporter}
below, including offsets later clipped away, and a constant number of
support calls per anchor for the facet tests of rounded bodies. Section
rounding, span computations, and marking use explicit reported lists or
polytopes of constant complexity, so they introduce no uncharged family
of support queries. Rejected sources and raw outputs before clipping
remain included in $\cV$.
\end{proof}

\subsection{Reporting affordable offsets without convolution}
\label{det:reporter-section}

One could list the affordable offsets of a source by blocked
$(\min,+)$-convolution of its atlas with itself. For a block length $L$
this costs $\Ot(V^2/L^{2-\rho}+QV/L+OL)$, and balancing the three terms
gives $\cV^{(4-\rho)/(3-\rho)}$ globally, which would be the
bottleneck. Instead, we use an output-sensitive enumeration that relies
only on convexity.

Consider one immutable source of positive rank $r$, with pool $I$,
lattice $\Lambda=\Lambda_I$, column hull $P=P_I$, fractional domain
$Z=Z(I)$, and fractional cost function $f=f_I$ (with $f=+\infty$ off
$Z$). Fix the refinement $q$ of Theorem~\ref{det:atlas} and put
$G=\Lambda/q$. The source has a window $K$, a candidate crop $C$, and a
birth halo $H_b\ge1$. The window contains both endpoints of every
queried offset (Section~\ref{fw:states}); concretely,
$K\supseteq C+H_b\Bx_P$ with $\Bx_P$ the box of~\eqref{det:box}. A
query consists of a midpoint
$\vt\in C\cap\Lambda$, an allowance $s$, and a radius $0<H\le H_b$, and
asks for the set
\[
 S(\vt,s,H)\cap G,\qquad
 S(\vt,s,H)=\{\vz:\ f(\vt+\vz)+f(\vt-\vz)\le2s\}\cap HP .
\]
Since $f$ is convex, $g_{\vt}(\vz)=f(\vt+\vz)+f(\vt-\vz)$ is convex and
$S(\vt,s,H)$ is a convex set. This is the only structural fact we use.
The reporter enumerates the grid points of this convex set by a
centerpoint search (Lemma~\ref{det:centerpoint}), using cutting
planes supplied by the atlas.

\begin{lemma}[Deterministic affordable-offset reporter]
\label{det:reporter}
After the atlas of Theorem~\ref{det:atlas} for the source has been
built, $Q$ adaptively chosen queries returning $O$ offsets in total can
be answered exactly and deterministically in time
\[
 \Ot_{r,q}\bigl(V^{3/2}+(Q+O)\sqrt V\,\bigr)\poly_{r,q}(\cB),
 \qquad V=V(K,I),
\]
including all preprocessing. No convolution is used.
\end{lemma}

\begin{proof}
We first cut the grid into buckets and prepare a separation structure
for the finite atlas entries in each bucket. A query is then split into
$O_r(1)$ search boxes. In each, every grid point either is an output or
yields a cut, and a pruned bisection lists the outputs.

\emph{Buckets.} If $C=\emptyset$ there are no queries and nothing is
built, so assume $C\ne\emptyset$. Write $\Bx_P=\{F\vz:|z_j|\le L_j\}$ as
in~\eqref{det:box} and use the corresponding integer coordinates of
$G$. Partition $G$ into boxes with $\lceil qH_bL_j\rceil$ consecutive
integers in coordinate $j$. For every box $U$ meeting $K$, let $m_U$ be
its number of grid points, including those outside $K$, and let
$\Sigma_U$ be the set of finite atlas entries in $U$. If $U$ meets $K$,
its real cell lies in $K+2H_b\Bx_P\subseteq\vt_0+3(K-\vt_0)$ for any
$\vt_0\in C$. Since the cells are disjoint, this gives
$\sum_Um_U=O_{r,q}(V)$. Since
\[
 \Sigma_U=G\cap(U\cap K\cap Z)\quad\text{and}\quad U\cap K\cap Z\ \text{is convex},
\]
we have $G\cap\conv(\Sigma_U)=\Sigma_U$. For every box with
$\Sigma_U\ne\emptyset$, build the hierarchy of Lemma~\ref{det:forest} on the sites $\Sigma_U$ with zero
heights and $b=\lceil\sqrt{m_U}\,\rceil$. A grid point of $U$ outside
$\Sigma_U$ can then be strictly separated from $\conv(\Sigma_U)$ in
$\Ot_r(\sqrt{m_U})\poly_r(\cB)$ time, using the support queries of that
lemma and exact fixed-dimensional separation--optimization for
well-described polytopes~\cite{GLS88}. The preprocessing costs
$\sum_U\Ot_r(m_U^{3/2})\le\Ot_r(V^{3/2})$, up to the bit factor.

\emph{Search boxes.} For a query, only $O_r(1)$ boxes can contain an
endpoint $\vt\pm\vz$ with $\vz\in H_bP$, and both endpoints lie in $K$.
For each ordered pair $(U_+,U_-)$ of them, the constraints
$\vt+\vz\in U_+$ and $\vt-\vz\in U_-$ define an integer search box for
$\vz$; every requested offset belongs to exactly one pair. Boxes with an
empty $\Sigma_{U_\pm}$ are skipped.

\emph{Cuts.} Let $\vz$ be a grid point of a search box. We check three
cases in turn. If $\vt+\vz\notin \Sigma_{U_+}$, which is a dictionary
lookup in the atlas, then the separator of $\vt+\vz$ from
$\conv(\Sigma_{U_+})$, written in $\vz$, separates $\vz$ from every
eligible offset; the same holds for $\vt-\vz$. If
$\vz\notin HP$, the column-body oracle of
Corollary~\ref{det:body-oracles}, applied to $\vz/H$, gives a separator.
Otherwise both endpoints carry exact values and supporting prices
$\vp_\pm$ from the atlas. For every eligible $\vu$ the subgradient
inequalities at $\vt\pm\vz$ give
\[
 (\vp_+-\vp_-)(\vu-\vz)\le g_{\vt}(\vu)-g_{\vt}(\vz)\le2s-g_{\vt}(\vz).
\]
If $g_{\vt}(\vz)>2s$, the right side is negative. In that case a zero
normal certifies that the search box contains no output, and a nonzero
normal separates $\vz$ strictly. If $g_{\vt}(\vz)\le2s$, which is an exact rational
comparison, $\vz$ is an output.

\emph{Enumeration.} An emptiness test for a subbox of a search box is
the search of Lemma~\ref{det:centerpoint}, in the integer coordinates of
$G$, with the cuts above as oracle: it stops at an output, at a zero
normal, or when no integer point is left. Every cut keeps every eligible
offset, so after $O_r(\log(V+2))$ oracle calls the test finds an output
or certifies that the subbox contains none. Every cut is computed from
the original atlas, the original oracles, and a grid point of the search
box, so bit lengths do not grow; the subboxes of the bisection below
have $\poly_r(\cB)$-bit corners, so $L_0=\poly_r(\cB)$ as well. To list all outputs, bisect the integer
coordinate intervals of the search box down to single points, pruning
each node with this emptiness test. Every expanded node is an ancestor of an output, so a
pair with $k$ outputs uses $O_r((1+k)\log^2(V+2))$ node
tests. Each test makes $O_r(\log(V+2))$ separation or oracle calls of
cost $\Ot_r(\sqrt V)\poly_r(\cB)$, since $m_U=O_{r,q}(V)$ and
$\Psi(I)=O_r(V)$. Summing over the $O_r(1)$ pairs and over the queries
proves the bound.
\end{proof}

Since $V_S\le\cV$ for every source, the computation in the proof of
Proposition~\ref{det:source-charge} shows that the reporters of all
sources cost $\Ot_r(\cV^{3/2})\poly_r(\cB)$ in total, the same order as
the atlases. This completes the deterministic local LP atlas and the
reporter of affordable offsets. Making the convolution and signed
transfer subroutines deterministic is a separate issue, treated next.
(A Las Vegas construction with exact output would not by itself have
established the result proved here.)

\section{Deterministic witnessed signed-box transfers}
\label{det:transfer-section}

This section computes the signed removed-item tables, the only part of
the algorithm that uses convolution. Throughout, we fix
$1<\rho\le2$ and assume that deterministic witnessed integer
$(\min,+)$-convolution of arrays of length $m$ takes
$\Ot(m^{\rho})\poly(\cB)$ time, where $\cB$ bounds the bit lengths of
the finite entries; entries $+\infty$ are permitted. Our goal is to
replace the color coding of~\cite[Lemma~6.4]{Zhou26} by a deterministic
construction. We start from the all-capacities reduction
of~\cite[Appendix~A, Theorem~A.1]{Chan26} and give its labelled version
with signed costs explicitly, including the treatment of repeated
columns.

The section proceeds in three steps. We first build canonical subsets
for repeated columns, then halve them deterministically with a rational
halver, and combine both into a recursion that computes the tables for
all exact scalar capacities. Finally, a scalar encoding of boxes and a
digit-sum penalty lift the scalar tables to signed boxes.

All constructions below are deterministic reductions to an exact
witnessed convolution oracle. In Theorem~\ref{det:main} this oracle may
be implemented by a randomized primitive. The correctness statements of
this section are then conditional on exact replies, and all oracle
errors are accounted for globally in
Section~\ref{det:probability-section}.

Storage bounds count retained table entries, canonical-subset records,
and witness pointers. We account for the temporary workspace of the
convolution primitive separately, because a time bound for the primitive
does not imply a linear workspace bound.

\begin{theorem}[Deterministic signed transfer]\label{det:transfer}
Fix $r$ and let
\[
 \Bx=\prod_{j=1}^r[-\Delta_j,\Delta_j],\qquad
 M=|\Bx\cap\Z^r|=\prod_{j=1}^r(2\Delta_j+1),
\]
where the $\Delta_j$ are nonnegative integers. Let $I$ be a labelled pool
with integer columns $\va_i$ and integer costs $c_i$, and define the
family of \emph{hereditarily clipped} subsets
\[
 \mathcal F_{\Bx}(I)=
 \Bigl\{Y\subseteq I:\ \sum_{i\in X}\va_i\in\Bx\text{ for every }X\subseteq Y\Bigr\}.
\]
There is a deterministic table $\mathsf D$ on $\Bx\cap\Z^r$ such that
every finite entry has an actual labelled witness and
\begin{equation}
 \min_{\substack{Y\subseteq I\\\sum_{i\in Y}\va_i=\vz}}c(Y)
 \ \le\ \mathsf D(\vz)\ \le\
 \min_{\substack{Y\in\mathcal F_{\Bx}(I)\\\sum_{i\in Y}\va_i=\vz}}c(Y),
 \label{det:transfer-dominance}
\end{equation}
where an empty minimum equals $+\infty$. For explicitly listed labels the
running time is $\Ot_r(|I|+M^{\rho})\poly_r(\cB)$.

For a compressed pool, suppose that its eligible nonzero vectors can be
enumerated in $\Ot_r(M)$ time, that the $k$ cheapest current labels of a
specified vector can be retrieved in $\Ot(k+1)$ time, and that the
optimal cost of the zero columns and a compressed attaining assignment
are available in $\Ot_r(M)$ time. Excluding the preparation of these
access structures, the running time is $\Ot_r(M^{\rho})\poly_r(\cB)$.
Only $\Ot_r(M)$ labels are materialized, and $\Ot_r(M)$ table entries
and witness pointers are retained.
\end{theorem}

\subsection{Canonical subsets for repeated columns}

The preprocessing in~\cite[Appendix~A]{Chan26} keeps a single best item
per weight. This fails for labelled $0$--$1$ items, since two labels of
equal weight may both be selected. Instead, we keep suitable cheapest
prefixes and build canonical subsets on a tree of \emph{sorted label
indices}. This approach also permits costs of arbitrary sign. The idea
is that every subset can be replaced, without increasing its cost, by a
union of few small precomputed optimal subsets.

\begin{lemma}[Canonical tables]\label{det:transfer-canonical}
Let $Y$ be a pool of labels with positive integer weights at most $u$
and arbitrary integer costs. For $b\ge1$ one can deterministically
construct a collection $\mathcal C$ of subsets of $Y$, each of size at
most $b$, in time
\[
 \Ot\bigl(b^{O(1)}(u^{\rho}+|Y|)\bigr)\poly(\cB),
\]
retaining $\Ot(b^{O(1)}(u+|Y|))$ entries and witness records, such that
the following holds. Every subset $S\subseteq Y$ has a replacement
$\widehat S$ of the same weight and cardinality, with
$c(\widehat S)\le c(S)$, that is a union of
\[
 O\bigl((|S|/b+1)\log(|Y|+2)\bigr)
\]
pairwise disjoint members of $\mathcal C$.
\end{lemma}

\begin{proof}
Sort the labels by weight and then by identifier, and build a balanced
binary tree on their indices. The labels of a node $v$ form a contiguous
interval $J_v$; write $a_v$ for its minimum weight and $d_v$ for its
maximum weight minus $a_v$. For each $0\le j\le b$, store the minimum
cost of a $j$-label subset of $J_v$ at each exact weight. These weights
lie in $[ja_v,j(a_v+d_v)]$, an interval of $jd_v+1$ integers.

Combine the tables of the two children over all $O(b^2)$ pairs of
cardinalities by witnessed convolution with the appropriate weight
offsets; each convolution has length $O(bd_v+1)$. Leaves store the empty
and singleton entries. At any fixed depth the node intervals are ordered
and disjoint in the sorted list, so
\[
 \sum_vd_v\le u-1,\qquad
 \sum_{v:\,d_v>0}d_v^{\rho}\le u^{\rho}.
\]
Nodes of zero width cost $b^{O(1)}$ each, and there are $O(|Y|)$ nodes.
Summing over the tree proves the time bound and, by counting array
lengths, the bound on retained entries.

For each finite entry, put its chosen subset into $\mathcal C$. A subset
of size at most $b$ can be reconstructed in $\Ot(b)$ time, so explicit
records of the canonical subsets cost only a further factor polynomial
in $b$.

For an arbitrary $S$, traverse the index tree and split a node exactly
when it contains more than $b$ labels of $S$. In each terminal node,
replace the part of $S$ by the stored optimum with the same cardinality
and weight. Terminal intervals are disjoint, so no label is used twice,
and the cost does not increase. At each depth at most $|S|/b$ nodes are
split, which proves the decomposition bound. Repeated weights cause no
difficulty because the leaves are labels.
\end{proof}

\subsection{A rational deterministic halver}

The recursion for scalar capacities splits a pool into two halves and
needs every canonical subset to be split almost evenly. The following
lemma achieves this deterministically by the method of conditional
expectations, using exact rational arithmetic.

\begin{lemma}[Halving canonical subsets]\label{det:transfer-halver}
Given $L$ subsets of a label set $Y$, each of size at most $b$, one can
deterministically partition $Y$ into $Y_1,Y_2$ such that every listed
subset $C$ satisfies
\[
 |C\cap Y_j|\le |C|/2+O\bigl(\sqrt b\log(L+2)\bigr),\qquad j=1,2.
\]
The construction uses $O(bL+|Y|)$ incidence updates and arithmetic on
numbers of bit length polynomial in $b+\log(L+2)$.
\end{lemma}

\begin{proof}
The case of an empty family is immediate. Otherwise put
$h=\lceil\sqrt b\,\rceil$, $q=1+1/h$, and
$D_0=h\lceil\log_2(4L+4)\rceil$. For independent uniform signs
$\epsilon_i\in\{-1,1\}$ consider
\[
 \Phi=\sum_C\Bigl(q^{\sum_{i\in C}\epsilon_i}
                         +q^{-\sum_{i\in C}\epsilon_i}\Bigr).
\]
Since
\[
 v_0:=\frac{q+q^{-1}}2=1+\frac1{2h(h+1)}
 \qquad\text{and}\qquad b\le h^2,
\]
we have $\mathbb E\Phi\le2Le^{1/2}<4L$, whereas $q^{D_0}\ge4L+4$
because $q^h\ge2$. Fix the signs one at a time, each time choosing
a sign that does not increase the conditional expectation; the average
of the two choices equals the current value. At the end
$\Phi<q^{D_0}$, so $\bigl|\sum_{i\in C}\epsilon_i\bigr|<D_0$ for
every $C$.

The computation uses exact rational arithmetic. If a set has sign sum $z$
on its assigned labels and $j$ unassigned labels, its conditional summand
is
\[
 (q^z+q^{-z})v_0^j,\qquad |z|+j\le b.
\]
The common denominator $[2h(h+1)]^b$ clears all these summands, and
their summed numerators have $O(b\log(b+2)+\log(L+2))$ bits. Precompute
the needed powers and update only the sets that contain the newly
assigned label; there are at most $bL$ such incidences. The two sign
classes give the partition.
\end{proof}

We apply Lemma~\ref{det:transfer-halver} to the collection of
Lemma~\ref{det:transfer-canonical}. Then every subset $S$ with
$|S|\le k$ has a replacement of the same weight and cardinality, and of
no greater cost, with at most
\begin{equation}
 |\widehat S|/2+
 O\bigl((k/b+1)\log(|Y|+2)\sqrt b\log(L+2)\bigr)
 \label{det:transfer-imbalance}
\end{equation}
labels in either part. We only need optimum-preserving replacements to
be balanced; the original subset itself need not be.

\subsection{All exact scalar capacities}

We now combine canonical tables and halvers into a recursion on the
cardinality of the subsets. Each level roughly halves the cardinality
bound, so the convolution lengths decrease geometrically and the total
cost is dominated by the root.

\begin{proposition}[Labelled scalar tables]\label{det:transfer-scalar}
For a pool $Y$ of labelled items with positive integer weights and
arbitrary integer costs, the minimum costs at all exact weights up to a
capacity $t$ can be computed deterministically in
$\Ot(t^{\rho}+|Y|)\poly(\cB)$ time. The construction retains
$\Ot(t+|Y|)$ table entries and witness pointers, and an attaining subset
can afterwards be recovered in time proportional to its number of labels,
up to logarithmic factors.
\end{proposition}

\begin{proof}
Discard weights exceeding $t$. We first solve a relaxed problem of low
cardinality: for weights at most $u$, produce witnessed costs on
$[0,ku]$ that dominate every subset of at most $k$ labels. A returned
witness may have more than $k$ labels, but it must be an actual subset of
the given pool.

\emph{Parameters.} Put
$\Theta=\max\{\lceil\log_2(|Y|+t+2)\rceil,C_\rho\}$, one value for the
whole recursion, and $b=(C\Theta)^8$, rounded
up, for sufficiently large constants $C$ and $C_\rho$, fixed in this
order ($C_\rho$ large compared with $C$). In every recursive
instance the canonical family satisfies $\log(L+2)=O(\Theta)$. Hence,
when $k>b$, the bound~\eqref{det:transfer-imbalance} is at most
$|\widehat S|/2+k/(100\Theta)$.

\emph{Recursion.} If $k>b$, construct the halver and recurse on its two
disjoint parts with
\[
 k'=\bigl\lceil k/2+k/(100\Theta)\bigr\rceil;
\]
then convolve the two child tables and clip the result to $[0,ku]$. If
$k\le b$, use the exact-cardinality tables of
Lemma~\ref{det:transfer-canonical} and take their minimum over
cardinalities at most $k$. Any cardinality bound exceeding the pool size
is replaced by that size.

\emph{Correctness.} Replace a required subset $S$ by $\widehat S$ from
the halver property. Each part of $\widehat S$ has at most $k'$ labels,
and by induction the child tables dominate these parts. Their witnesses
belong to disjoint pools, so they combine to an actual subset of the
same weight and no greater cost. Positivity of weights justifies the
scalar clips. No sign restriction on costs is used.

\emph{Running time.} Since $k>b\ge100\Theta$ at every nonterminal node,
integer rounding is absorbed and along every nonterminal edge
\[
 k_{\rm child}\le \varsigma k,\qquad \varsigma=\tfrac12+\tfrac1{50\Theta}.
\]
Up to logarithmic factors, a node with $|Y_v|$ labels costs
$|Y_v|+u^{\rho}+(k_vu)^{\rho}$. Choose $C_\rho$ so that
$2\varsigma^{\rho}\le c_\rho<1$. The convolution costs then sum to
\begin{equation}
 (ku)^{\rho}\sum_{j\ge0}(2\varsigma^{\rho})^j
   =O_\rho\bigl((ku)^{\rho}\bigr).
 \label{det:transfer-recurrence}
\end{equation}
The $u^{\rho}$ terms obey the same bound, because there are $O(k)$
nodes and $k\,u^{\rho}\le(ku)^{\rho}$. The pools at each level are
disjoint, so the label work sums to $\Ot(|Y|)$.

The depth is $O(\log(k+2))$. At level $j$ the scalar table lengths sum
to at most $ku(2\varsigma)^j=O(ku)$, since $\log k=O(\Theta)$. There are $O(k)$
nodes in total. Consequently the canonical tables, their explicit
witnesses, and the reconstruction records of the convolutions retain
$\Ot(ku+|Y|)$ entries and pointers.

\emph{All capacities.} Partition the weights into dyadic layers
$[a,2a)$. In one layer put $k=\lfloor t/a\rfloor$ and let $u<2a$ be its
maximum weight, so that $ku\le2t$. Every subset of weight at most $t$
has at most $k$ labels in this layer. The relaxed table clipped to
$[0,t]$ is therefore the exact minimum-cost table of the layer: it
dominates all feasible subsets, and every finite value is attained.
Merge the $O(\log(t+2))$ layer tables by convolution, clipping to
$[0,t]$ after each merge; positive weights preserve every required path.

All finite stored costs are costs of actual subsets and therefore have
$O(\cB+\log(|Y|+2))$ bits. The rational halver arithmetic is polynomial
in $\log(|Y|+t+2)$. Store attaining child indices at each gate; following
nonempty pointers recovers a subset in the claimed time. The record
count excludes the temporary workspace of the convolution primitive.
\end{proof}

\subsection{Signed boxes and hereditary clips}

We now prove the main theorem of this section. The plan is to split the
columns into coordinate orthants, encode each reflected orthant as a
scalar problem, solve it with the scalar tables of the previous
subsection, and merge the orthant tables by signed convolution.

\begin{proof}[Proof of Theorem~\ref{det:transfer}]
\emph{Reductions.} Discard columns outside $\Bx$: even a singleton
containing such a column is excluded from every hereditarily clipped
subset. Select the zero columns of negative cost and omit the other zero
columns, saving their cost and compressed witness separately. Partition
the nonzero columns into coordinate orthants, with a fixed convention for
zero coordinates, and reflect each orthant into
$\Bx_+=\prod_j[0,\Delta_j]$.

\emph{Scalar encoding.} Delete zero-width coordinates, put
$b_j=\Delta_j+1$, and define
\[
 E(\va)=\sum_ja_j\prod_{i<j}b_i,\qquad
 d(\va)=\sum_ja_j,\qquad T=\prod_jb_j-1<M.
\]
The map $E$ is linear and injective on $\Bx_+$, and every nonzero column
has a positive scalar weight. For each vector $\va$, keep at most
$\lfloor T/E(\va)\rfloor$ of its cheapest eligible labels: a subset of
scalar weight at most $T$ uses no more copies, and substituting cheaper
copies of the same vector does not increase cost. The number of retained
labels is
\begin{equation}
 |Y|\le\sum_{w=1}^T\lfloor T/w\rfloor
       =O\bigl(T\log(T+2)\bigr),
 \label{det:transfer-prefix}
\end{equation}
since $E$ is injective. This is the required multiplicity preprocessing.
If $T=0$, only the separately handled zero columns contribute.

\emph{Digit-sum penalty.} Let $C_*=\sum_{i\in Y}|c_i|$, put $K=2C_*+1$,
and assign the augmented cost $\widetilde c_i=Kd(\va_i)+c_i$. Apply
Proposition~\ref{det:transfer-scalar} with capacity $T$; each witness
also keeps its digit sum and its original cost. Fix $\vz\in\Bx_+$ and
consider a subset whose vector sum $\vz'$ has scalar weight $E(\vz')=E(\vz)$.
Then $\vz'$ can be carried to $\vz$, with no final overflow because
$E(\vz)\le T$, and each nontrivial carry decreases the digit sum by
$\Delta_j\ge1$. Hence
\[
 d(\vz')\ge d(\vz),\qquad d(\vz')=d(\vz)\iff\vz'=\vz .
\]
The penalty $K$ makes the digit sum the primary criterion and the
original cost the secondary one. Accept an output only when its witness
has digit sum $d(\vz)$. The resulting orthant table gives the exact
minimum cost at every attainable vector $\vz\in\Bx_+$.

The encoding also preserves vector sums along every required replacement
path. A subassignment of a required reflected orthant choice has a sum
$\vz''\in\Bx_+$. Any replacement with equal scalar weight and no greater
augmented cost has digit sum $d(\vz'')$: a larger digit sum would
increase the cost by at least $K-2C_*>0$, and a smaller one is
impossible. Hence its vector sum is also $\vz''$. Intermediate scalar
tables may contain other witnessed entries, but the digit test is
applied before an orthant table is exposed, so these entries do not
enlarge a source clip.

\emph{Merging orthants.} Reflect the orthant tables back and merge them
by witnessed signed convolution. Translate each coordinate by
$\Delta_j$, use radix $4\Delta_j+1$, and subtract $2\Delta_j$ from the
output digits. This is carry-free and uses $O_r(M)$ cells. Clip to $\Bx$
after \emph{every} orthant merge, and finally add the saved contribution
of the zero columns.

\emph{Dominance.} For $Y\in\mathcal F_{\Bx}(I)$, every orthant part lies
in the reflected box and every partial union of orthant parts has its
sum in $\Bx$. Replacing a part by a cheaper witness with the same vector
sum preserves all these partial sums, so the required path survives every
clip and is dominated. All finite entries are actual labelled subsets
because the orthants are disjoint. This proves both inequalities
in~\eqref{det:transfer-dominance}.

\emph{Cost.} For explicit input, grouping and selecting cheapest prefixes
cost $\Ot(|I|+M)$. Under the compressed interface,
\eqref{det:transfer-prefix} bounds materialization and sorting by
$\Ot_r(M)$. Proposition~\ref{det:transfer-scalar} and the constant
number of orthant merges give the stated time and record bounds. All
augmented costs and subset aggregates have $O_r(\cB+\log(M+2))$ bits.
\end{proof}

The endpoint translations, the reduced-cost normalization, and the
maximum-allowance convolution of~\cite[Lemma~6.5]{Zhou26} are
deterministic identities. Clearing their common rational denominator and
applying Theorem~\ref{det:transfer} therefore removes the color-coding
failures of the transfer component without changing the physical clips,
the witness interface, or the convolution exponent.

\section{Convolution primitives for arbitrary bit lengths}
\label{detmc:top}
\label{detmc:section}

This section shows that the convolution premise of
Section~\ref{det:transfer-section} can be met deterministically, with no
restriction on the magnitude of the entries. We use the deterministic
real-input result of Chan, Itakura, Ma, and Yu~\cite{CIMY26}. This is
simpler than deriving an arbitrary-bit algorithm from an integer 3SUM
theorem, which applies only to polynomially bounded inputs. The proof
has three parts: a bit-level simulation of the real algorithm, an
encoding of values, infinities, and witnesses into one integer, and a
blocking that reduces convolution to square matrix products.

\begin{lemma}[Deterministic full witnessed integer convolution]
\label{detmc:lemma}
Let $F$ and $G$ be integer arrays of length at most $m$, with entries
$+\infty$ allowed, whose finite entries have at most $\cB$ bits. All
values of their $(\min,+)$-convolution
\[
 (F\oplus G)[k]=\min_{i+j=k}\bigl(F[i]+G[j]\bigr),
\]
together with an attaining pair $(i,j)$ for every finite output, can be
computed deterministically with
\[
 \Ot(m^{1.99985})\poly\bigl(\cB+\log(m+2)\bigr)
\]
bit operations. The exponent is a conservative explicit bound, not a
claim of numerical optimality.
\end{lemma}

\begin{proof}
\emph{The external input.} Section~4 of~\cite{CIMY26} gives a
deterministic algorithm for real $(\min,+)$-matrix multiplication. It
first reduces rectangular products (Problem~4.1) to a halving procedure
that maintains an attaining inner index (Problem~4.3 and Lemma~4.4), and
then reduces this procedure to counts in a graph via Lemma~4.7.
Theorem~4.8, with the parameter choice of Corollary~4.9, yields the
explicit exponent $2.9997$. The corollary is stated for real APSP, but
its proof gives the same bound for square $(\min,+)$-products before
repeated squaring.

\emph{Bit simulation.} We check that the real algorithm runs on integer
inputs of arbitrary length with only polynomial bit overhead. In the
reduction of~\cite[Section~4]{CIMY26}, every input-dependent comparison
involves sums or differences of a constant number of input numbers.
After the differences are sorted, the
construction uses only their integer ranks, dyadic intervals, and counts
of pairs or of graph triangles. The ordinary matrix products with
selected entries act on these combinatorial graphs, not on input weights
of arbitrary precision. Consequently, for a square product of size $s$
with $\cB'$-bit integer inputs, the weight comparisons use numbers of
$O(\cB'+\log(s+2))$ bits and the counters are polynomially bounded. The
deterministic graph-counting subroutine uses polylogarithmic bit work per
word operation. Thus a square $(\min,+)$-product costs
\begin{equation}\label{detmc:eq-square}
 \Ot(s^{2.9997})\poly\bigl(\cB'+\log(s+2)\bigr).
\end{equation}
No separate tie-breaking transformation and no witness output of the
matrix product are needed: the encoding below makes all finite
candidates for an output entry distinct and carries the witness in the
value. Ties can remain only among sentinel sums, at entries that decode
to $+\infty$, and the source permits ties.

\emph{Encoding values, infinities, and witnesses.} We pack each value,
a witness index, and a sentinel for infinite entries into a single
nonnegative integer, so that one product returns all three. Let
\[
 U=\max\bigl(1,\max\{|F[i]|,|G[j]|:\ \text{the entry is finite}\}\bigr),
 \qquad \bar m=m+1,
 \qquad \Omega=8U\bar m+4m+1.
\]
Pad both arrays to length $m$. Replace finite entries by
\[
 F'[i]=\bar m\,(F[i]+U)+i,
 \qquad G'[j]=\bar m\,(G[j]+U),
\]
and replace every infinite or out-of-range entry by $\Omega$. All encoded
finite entries are nonnegative, every sum of two finite entries is less
than $\Omega$, and every sum involving an infinite entry is at least
$\Omega$. The encoded bit length is $O(\cB+\log(m+2))$. For a fixed
output index $k$, minimizing a finite encoded sum first minimizes
$F[i]+G[j]$ and then minimizes $i$, because $i<\bar m$. In particular,
distinct pairs $(i,j)$ with $i+j=k$ give distinct finite encoded sums.

\emph{Reduction to square products.} Put $s=\lceil\sqrt m\,\rceil$ and
form the $s\times s$ matrix and the $s\times O(m)$ matrix
\[
 \mathsf L_{q,r}=F'[qs+r]\quad(0\le q,r<s),\qquad
 \mathsf R_{r,j}=G'[j-r]\quad(0\le r<s,\ -s^2\le j<2m),
\]
with the value $\Omega$ at out-of-range indices. For $0\le k<2m-1$,
\begin{equation}\label{detmc:eq-block}
 (F'\oplus G')[k]=\min_{0\le q<s}(\mathsf L\otimes\mathsf R)_{q,k-qs},
\end{equation}
where $\otimes$ is the $(\min,+)$-matrix product. Indeed, each inner
index $r$ represents exactly the pair $i=qs+r$, $j=k-i$; every $i$
occurs, and invalid pairs have sentinel cost. Partition the columns of
$\mathsf R$ into $O(s)$ blocks of width at most $s$, padding the last
one. By~\eqref{detmc:eq-square} all these products cost
\[
 \Ot\bigl(s\cdot s^{2.9997}\bigr)\poly\bigl(\cB+\log(m+2)\bigr)
 =\Ot(m^{1.99985})\poly\bigl(\cB+\log(m+2)\bigr).
\]
Building the matrices and evaluating~\eqref{detmc:eq-block} take
$O(m^{3/2})$ further operations.

\emph{Decoding.} Write $D'[k]=(F'\oplus G')[k]$. If $D'[k]\ge\Omega$,
return $+\infty$. Otherwise return
\[
 (F\oplus G)[k]=\bigl\lfloor D'[k]/\bar m\bigr\rfloor-2U,
 \qquad i=D'[k]\bmod\bar m,
 \qquad j=k-i.
\]
The decoded $i$ is the least attaining first index, both entries are
finite, and one may verify $F[i]+G[j]=(F\oplus G)[k]$ directly. No
self-reduction, randomized witness search, or witness from the matrix
product is needed. For inputs of
unequal length, keep only their original output range. This gives the
full witnessed interface, including arbitrary bit lengths and infinite
entries.
\end{proof}

A faster randomized primitive is due to Zhou.

\begin{lemma}[Randomized full witnessed integer convolution~\cite{Zhou26}]
\label{detmc:random}
Witnessed $(\min,+)$-convolution of arrays of total length $m$, with
entries $+\infty$ allowed and finite entries of at most $\cB$ bits, can
be solved in worst-case time
\[
 \Ot(m^{\rho_0})\poly\bigl(\cB+\log(m+2)\bigr),\qquad
 \rho_0=\frac{784055881637}{400000000000}=1.9601397040925,
\]
with error probability at most $1/3$.
\end{lemma}

\begin{proof}
This is~\cite[Definition~3.4 and Lemma~3.6]{Zhou26} with joint error
probability $\eta=1/3$; we check that it provides our interface. Zhou's
definition covers integer arrays with $+\infty$ entries and $b$-bit
signed finite entries, measures cost in bit operations as
$\Ot(m^{\rho})\poly(b+\log(1/\eta))$, and bounds the probability
that any part of the output is wrong by $\eta$. Witnesses are added
by~\cite[Lemma~3.5]{Zhou26}, which encodes the first index into the
low-order digits at an extra cost of $O(\log(m+2))$ bits. Arbitrary bit
lengths are handled in~\cite[Appendix~B]{Zhou26} by recovering the
output one bit per round, using a random prime to reduce each round's
indexed-equality queries to Wang's 3SUM decision program on polynomially
bounded integers~\cite{Wang26}. The prime sampling is capped at
$O(\log L)$ trials, where $L$ bounds the integers of the reduced 3SUM
instances, so the running time is a worst-case bound. Infinite
entries are replaced by a sentinel exceeding every finite pair sum.
\end{proof}

\section{The bounded integer optimization framework}
\label{fw:frame}

This section and the next four combine the local LP atlas, the reporter
of affordable offsets, and the signed removed-item tables with a
recursion on actual labelled pools. The recursion is that
of~\cite{CMZ26b}, with the source-based accounting of
Zhou~\cite{Zhou26} extended to an arbitrary fixed rank. This section
sets up the root of the recursion: it reduces the input to a compressed
pool of signed binary units, fixes one perturbed objective used
throughout, and describes how pools and removal tables are accessed.

Throughout, $d$ is the number of input equations, $r$ is the current
rank, and $N$ is the coefficient bound of~\eqref{det:ip}. The variable
bounds are finite integers; costs, bounds, and right-hand sides may have
either sign. All dimension-dependent constants may depend on $d$ and on
the fixed parameters $\rho$ and $\eps$ of Theorem~\ref{det:main}, but
not on $N$, $n$, or the numerical values of capacities. In
Sections~\ref{fw:frame}--\ref{fw:charge} we assume that every
convolution reply is exact; Section~\ref{det:probability-section}
removes this assumption.

\paragraph{Bit model.}
We let $\cB$ bound the bit length of every input number together with
$\log(n+N+2)$. Every coordinate change, perturbation, comparison, and
witness reconstruction below uses exact finite arithmetic, and
Section~\ref{fw:arith} shows that all numbers that arise have
$\poly_{d,\rho,\eps}(\cB)$ bits. We do not identify an
arithmetic-operation bound with a bound independent of the cost
encoding. We use the following tools with their standard deterministic
bit-complexity bounds~\cite{Lenstra83,LLL82,GLS88}: fixed-dimensional
integer feasibility and optimization, Hermite and Smith normal forms,
LLL reduction, exact linear optimization from strong separation, volume
computation for explicit polytopes, and integer-point enumeration. Point
enumeration proceeds by coordinate bisection with integer-feasibility
pruning, with polynomial bit work per output.

\subsection{A compressed signed root}
\label{fw:root}

The root reduces the integer program to a short signed correction of an LP
optimum, expressed with binary units whose multiplicities stay
compressed. We begin with routine cleanup. Substitute fixed variables,
reject contradictory bounds, and settle zero-column variables at
cost-minimizing endpoints.

\emph{Inequality constraints.} A two-sided row
$b^-_j\le A_j\vx\le b^+_j$ can be turned into an equality as follows.
The variable box gives the finite row interval
\[
 L^{\rm box}_j=\sum_i\min(A_{ji}\ell_i,A_{ji}u_i),\qquad
 U^{\rm box}_j=\sum_i\max(A_{ji}\ell_i,A_{ji}u_i).
\]
Intersect it with the requested interval, obtaining $[L_j,U_j]$. If this
is empty, reject. Otherwise use $A_j\vx+s_j=U_j$ with
$0\le s_j\le U_j-L_j$ and a zero-cost slack. This adds at most $d$
variables, no row, and no coefficient larger than $N$. Hence
Theorem~\ref{det:main} also holds for $d$ inequality or two-sided
constraints.

\emph{Signed units.} Solve the continuous box LP of~\eqref{det:ip} once,
at cost $T_{\rm LP}$.
If it is infeasible, the integer problem is infeasible. Otherwise let
$\vx^\ast$ be an optimal vertex, let $\vpi$ be an equality price, and put
\[
 q_i=\lfloor x^\ast_i\rfloor,\qquad \theta_i=x^\ast_i-q_i,\qquad
 F=\{i:\theta_i>0\},\qquad \vb_0=\vb-A\vq .
\]
The columns of the interior variables of a vertex are independent, so
$|F|\le d$ and
\begin{equation}\label{fw:b0}
 \vb_0=\sum_{i\in F}\theta_i\va_i,\qquad \|\vb_0\|_\infty\le dN.
\end{equation}
For variable $i$ introduce conceptual \emph{forward units}
$(\va_i,c_i)$ with multiplicity $u_i-q_i$, and \emph{backward units}
$(-\va_i,-c_i)$ with multiplicity $q_i-\ell_i$. A unit is an actual
labelled binary choice; the multiplicities remain compressed. If
$y_i^+,y_i^-$ count the chosen units, then
\begin{equation}\label{fw:units}
 x_i=q_i+y_i^+-y_i^-,\qquad
 \ell_i\le q_i-y_i^-\le x_i\le q_i+y_i^+\le u_i .
\end{equation}
Thus units may be chosen independently in both orientations without
violating the original bounds. The equality and the objective change by
exactly the sum of the chosen signed columns and costs.

Every available unit has nonnegative reduced cost at $\vpi$. Indeed, a
nonempty forward pool implies $x^\ast_i<u_i$, a nonempty backward pool
implies $x^\ast_i>\ell_i$, and complementary slackness gives the required
signs. Mark one forward unit for every $i\in F$. The fractional
selection that assigns $\theta_i$ to these marked units represents
$\vx^\ast$ and has zero reduced cost.

\begin{lemma}[Short correction and compressed retention]\label{fw:kernel}
If~\eqref{det:ip} is feasible, an optimum has a signed-unit
representation of cardinality less than
\[
 K_{\rm px}=(d+1)(2dN+1)^d .
\]
Keeping the cheapest $K_{\rm px}$ available units of each nonzero signed
vector, or all of them if fewer exist, preserves feasibility and an
optimum; among cost ties the marked units have priority. The retained
universe has
\begin{equation}\label{fw:U}
 U\le K_{\rm px}\bigl((2N+1)^d-1\bigr)=O_d(N^{2d})
\end{equation}
virtual labels and can be stored and prepared with $\Ot_d(n+N^d)$
operations on input words, without expanding $U$.
\end{lemma}

\begin{proof}
We first bound the cardinality by a signed Steinitz argument, then show
that cheapest prefixes suffice, and finally explain how to retain them
without expanding multiplicities.

Choose a minimum-cardinality optimum among all feasible signed-unit
assignments. Its multiset $V$ of columns has no nonempty zero-sum
submultiset: deleting one keeps the target and all unit capacities, and
its cost is the sum of its nonnegative reduced costs. A positive cost
contradicts optimality and a zero cost contradicts minimum cardinality.

Append the at most $d$ vectors $-\theta_i\va_i$, $i\in F$, to $V$. The
resulting sequence sums to zero and each vector has $\ell_\infty$-norm at
most $N$. The Steinitz lemma orders it with all partial sums in
$[-dN,dN]^d$~\cite{EW19}. The appended vectors $-\theta_i\va_i$ cut the integral moves into
at most $d+1$ consecutive runs. Within a run the partial sums lie in one
translate of $\Z^d$ and are distinct, since a repetition would exhibit a
zero-sum submultiset of $V$. Each run therefore has fewer than
$(2dN+1)^d$ moves. This is the signed proximity argument
of~\cite[Lemmas~3.3--3.5]{CMZ26b}.

Within each signed-vector group, replace the selected units of this
optimum by the same number of cheapest available units. The target is
unchanged and the cost does not increase; \eqref{fw:units} proves
feasibility after replacement, even if the units come from both
orientations of one variable. Conversely every retained assignment
reconstructs an original assignment, so infeasibility cannot be lost.
The marked units have minimum cost within their signed-vector groups and
are retained by the tie rule.

It remains to perform the retention efficiently. After the marked units
are isolated, there are at most $2n+d$ initial runs of units; group them
by signed vector. Weighted selection on run multiplicities locates the
retained prefix of each group. A deterministic median-of-records pivot
discards a constant fraction of records per step, regardless of the
multiplicities. The retained objects are intervals
of the original runs, with at most one partial run per cutoff. Sort the
retained runs, assign consecutive ordinal intervals, and prepare prefix
sums and fixed-degree ordinal power sums. This costs $\Ot_d(n+N^d)$
rather than $O(U)$. Each run records its original variable and
orientation for reconstruction.
\end{proof}

The marked fractional selection is an optimal vertex of the kernel LP.
Indeed, a two-sided feasible direction would be supported on the marked
units and would induce one at $\vx^\ast$. Since we perturb the objective
once below, we do not require this vertex to remain optimal afterwards.
Instead, one grouped root LP for the perturbed objective costs
$\Ot_d(N^d)\poly(\cB)$ by Lemma~\ref{fw:grouped-lp}.

\subsection{Permanent types and one objective}
\label{fw:objective}

This subsection fixes two things once and for all: a canonical
orientation of every column type, and a single perturbed objective under
which tight labels always have independent columns. Every later pool,
table, and comparison uses this objective.

\emph{Permanent types.} Choose an integer encoding that is injective on
the original signed column box, for example
\[
 \mathrm{enc}_0(\va)=\sum_{j=1}^da_jB_0^{\,j-1},\qquad B_0\ge2N+1 .
\]
Complement every retained unit whose encoding is negative, recording the
exact target and cost translations, and merge equal oriented columns
into sorted cost profiles. Each \emph{permanent type} now has a distinct
positive readout. All $d$ vector equations are retained: the encoding
only identifies types and does not replace the vector constraint. Every
later restriction or lift stores a linear map back to the permanent
column, and hence to this readout. In particular, distinct permanent
types never have equal or opposite current columns.

\emph{The common objective.} Fix the prospective rank bound $D$ and a
coordinate bound $H_{\rm co}$,
specified in Sections~\ref{fw:scales} and~\ref{fw:arith}, respectively. Give the
virtual labels distinct identifiers $j\in[U]$, increasing within each
permanent type in cost order. The following is the uniform perturbation
of~\cite[Lemma~5.1]{Zhou26}. Put
\[
 L_0=D!(2H_{\rm co})^D,\quad S_0=(D+1)L_0U^D,\quad R_0=2S_0+2,\quad
 \chi_j=\sum_{h=0}^Dj^hR_0^h,
\]
choose an integer $A_0>2\bigl(U+(D+1)L_0\bigr)\chi_U$, and define the \emph{common
objective}
\begin{equation}\label{fw:cstar}
 c^\ast_j=A_0c_j+\chi_j .
\end{equation}
Every selection has total perturbation between $0$ and $U\chi_U<A_0/2$.
Consequently, for an integral strict threshold $s$,
\begin{equation}\label{fw:threshold}
 \vc^{\T}\vx<s\iff(\vc^\ast)^{\T}\vx<A_0s .
\end{equation}

The point of the perturbation is genericity. For every assignment of
integer columns of rank at most $D$ and coordinates bounded by
$H_{\rm co}$, the labels that are tight at any price have independent
columns. Indeed, take a minimal dependent set of
at most $D+1$ columns and a nonzero integral circuit $\vz$ with
$\|\vz\|_\infty\le L_0$. Vandermonde independence on the distinct
identifiers shows that some moment $\sum_jz_jj^h$ is nonzero. By the
choice of $R_0$, the highest nonzero moment dominates
$\sum_jz_j\chi_j$, which is therefore nonzero and has absolute value
below $A_0/2$. It cannot cancel $A_0\sum_jz_jc_j$, which is an integer
multiple of $A_0$. Thus the circuit is not tight. The same argument
covers zero columns by one-term circuits. In particular, the common
objective satisfies the genericity condition~\eqref{det:generic} of the
local LP atlas for every pool that arises. \emph{All} stored pools,
phase comparisons, source tables, and LP values use this same
$\vc^\ast$. Complementing a label or subtracting a rational price
internally does not change circuit independence.

The extra bit length is $O_D(\log(H_{\rm co}+U+2))$. Prefix sums on a
constant-cost run reduce to fixed-degree power sums of the ordinal.

\subsection{Immutable pool access}
\label{fw:pools}

Pools are never expanded into explicit label lists. Instead, a stored
pool is a versioned set of actual labels, accessed through immutable
views. Each permanent type has one current integral column and an
immutable view into its ordered original run list. A view consists of
ordinal intervals together with a persistent deletion tree. In
polylogarithmic time and with polynomial bit overhead, it returns the
live count, the $k$th live label, counts below a cost threshold, and
prefix costs. Disjoint interval restrictions, reversed views, and
deletion of an explicitly named label preserve this interface. These
operations are the compressed-pool construction
of~\cite[Lemmas~6.6--6.8]{Zhou26}. They follow from balanced
rank/select trees whose nodes retain counts and sums of the first $D$
ordinal powers. In particular, they provide the profile interface
assumed in Section~\ref{det:section}.

At a price $\vpi$, a type splits into a constant number of working views
according to the sign of $c^\ast_i-\vpi\va_i$; its negative side is read
in reverse cost order after complementation. A cheap view
$|c^\ast_i-\vpi\va_i|\le\vartheta$ is a contiguous interval before
persistent deletions. No operation expands a multiplicity, and source
preparation scans current type records, not the original input runs.

For a rank-$r$ pool $I$ we use $\Lambda_I$, $P_I$, $Z(I)$, and $\Psi(I)$
as in Section~\ref{det:overview}, and
\[
 f_I(\vt)=\min\{(\vc^\ast)^{\T}\vx:\ A_I\vx=\vt,\ \vzero\le\vx\le\vone\}.
\]
All geometry is relative to the column span. Rank-zero pools are empty
after zero columns are settled. For them we put $\Psi=1$, $f(\vzero)=0$,
and $f(\vt)=+\infty$ for $\vt\ne\vzero$; a retained rank-zero buffer
with its nonempty window has mass and construction charge one. For
$r>0$ we have $\Psi(I)\ge1$, and the number of distinct current types is
$O_r(\Psi(I))$ by Lemma~\ref{fw:counting}(1) below, since the type
columns are distinct lattice points of $P_I$.

The following two counting facts are used repeatedly.

\begin{lemma}[Lattice counting]\label{fw:counting}
Let $\Lambda$ be a lattice of rank $s\ge1$ and let $P'$ be a centrally
symmetric convex set containing $s$ independent vectors of $\Lambda$.
\begin{enumerate}
\item There is a measurable fundamental domain $\Phi$ of $\Lambda$ with
$\Phi\subseteq(s/2)P'$. Consequently, for every set $X$ in the span,
$|X\cap\Lambda|\le\vol_s(X+(s/2)P')/\covol\Lambda$ and
$\vol_s(X)/\covol\Lambda\le|(X+(s/2)P')\cap\Lambda|$.
\item For a centrally symmetric convex set $S$, an integer $w\ge1$, and
$C'>0$,
\begin{equation}\label{fw:residue}
 |C'S\cap\Lambda|\le w^s\,|(2C'/w)S\cap\Lambda| .
\end{equation}
\end{enumerate}
\end{lemma}

\begin{proof}
For item~1, let $\vb_1,\dots,\vb_s\in P'$ be independent lattice vectors.
Rounding the coefficients of a point in the basis $(\vb_l)$ moves it by
$\sum_l\delta_l\vb_l$ with $|\delta_l|\le1/2$, a vector of
$(s/2)\conv(\pm\vb_1,\dots,\pm\vb_s)\subseteq(s/2)P'$, to a point of the
sublattice they generate, which lies in $\Lambda$. A
translation-equivariant nearest-point rule in the gauge of $P'$, with a
fixed tie rule, therefore gives a measurable fundamental domain inside
$(s/2)P'$. The translates $\vx+\Phi$, $\vx\in X\cap\Lambda$, are disjoint
and lie in $X+(s/2)P'$; and the translates meeting $X$ cover $X$ and have
centres in $X+(s/2)P'$. For item~2, partition $C'S\cap\Lambda$ into
residue classes modulo $w\Lambda$, and choose one point $\vz$ in each
occupied class. The map $\vx\mapsto(\vx-\vz)/w$ injects that class into
$(2C'/w)S\cap\Lambda$.
\end{proof}

\begin{lemma}[Grouped LP and affordable images,
{\cite[Lemmas~5.2--5.3]{Zhou26}}]\label{fw:grouped-lp}
In fixed dimension, a pool of $m$ current type profiles with the preceding
access operations admits exact box-LP optimization at a rational target
in $\Ot(m+1)\poly(\cB)$ time. It returns infeasibility, or an optimal
price and a basic labelled witness represented by endpoint prefixes and
at most $r$ fractional labels. The same bound holds for optimizing a
linear function over
\[
 \{\mathsf U\vx:\ A\vx=\vy,\ (\vc^\ast)^{\T}\vx\le b,\ \vzero\le\vx\le\vone\}
\]
when the image dimension is fixed and the image column is constant on
each visited type; optimal equality and budget multipliers are
available.
\end{lemma}

We sketch why the lemma holds. The oracle works by exact strong
separation in a fixed-dimensional dual. For example, the dual objective
$g_{\vt}(\vpi)=\vpi\vt+\sum_i\min(0,c^\ast_i-\vpi\va_i)$ and a
supporting plane are obtained by one count and one prefix query per
type, and the support function of the target domain is
$\sum_{\va}u_{\va}\max(0,\vv\va)$. Exact separation--optimization and
fixed-size determinant bounds give polynomially many oracle
calls~\cite{GLS88}, also on lower-dimensional feasible sets. Ties in
temporary systems are broken by an auxiliary fixed-dimensional circuit
perturbation, after which the tied coefficients are solved directly; the
reported objective remains $\vc^\ast$.

\subsection{Witnessed transfers and allowance merges}
\label{fw:transfers}

This subsection packages the signed removed-item tables constructed
earlier in the form used by the recursion. A
removal table records, for each displacement, the cheapest way found to
remove items; it is then merged with the allowances of a source.

For a removal pool $X$ disjoint from the remaining pool, a \emph{removal
table} $\mathsf D(\vv)$ must satisfy two requirements. First, every
finite entry is the exact cost of an actual labelled assignment on $X$.
Second, for each assignment that the geometric proof needs to preserve,
its index is present with a cost no larger than that assignment's cost.
The table need not preserve that particular assignment: keeping a
cheaper attaining assignment at the same exact vector sum only increases
the residual allowance.

\begin{lemma}[Signed tables and allowance merges]\label{fw:transfer}
Let $\Bx=\prod_l[-\Delta_l,\Delta_l]$ be a symmetric integer box with $M$ cells in fixed dimension,
and consider the assignments of a signed labelled pool every
subassignment of which has its sum in $\Bx$. A removal table dominating
all of them can be constructed deterministically in
$\Ot(M^{\rho})\poly(\cB)$ time with exact convolution replies, provided
its active nonzero vectors can be enumerated in $\Ot(M)$ time and their
requested cheapest labels have output-sensitive access. It stores
$\Ot(M)$ entries and witness pointers, and every finite entry is
witnessed.

For a source allowance array $S(\vt)$, with absent entries equal to
$-\infty$, the maximum residual allowance
\begin{equation}\label{fw:merge}
 \mathsf A(\vu)=\max_{\vv}\bigl(S(\vu+\vv)-\mathsf D(\vv)\bigr)
\end{equation}
and one attaining source/removal pair at every finite entry are computed
by one further witnessed convolution of the same order, followed by
decoding and the physical crop, provided $S$ is read only on an integer
box of $O(M)$ cells containing $\vu+\Bx$ for every requested $\vu$.
\end{lemma}

\begin{proof}
The table is Theorem~\ref{det:transfer}; the hypotheses listed are its
compressed-pool interface, supplied by Section~\ref{fw:pools}.
Equation~\eqref{fw:merge} is the identity
$-\mathsf A=(-S)\oplus\mathsf D^{\leftarrow}$ with
$\mathsf D^{\leftarrow}(-\vv)=\mathsf D(\vv)$, evaluated by the carry-free
mixed-radix encoding used in the proof of Theorem~\ref{det:transfer}. A
common denominator of bounded bit length clears the rational normalized
costs and allowances; the sentinel for infinite entries has polynomial
bit length, and witnesses identify outputs that involve it.
\end{proof}

When the reference on $X$ is an endpoint assignment $\ve$, put
$\vb_X=A_X\ve$ and $d_X=(\vc^\ast_X)^{\T}\ve$. Toggling label $i$ uses
column $(1-2e_i)\va_i$ and reduced cost $(1-2e_i)(c^\ast_i-\vpi\va_i)$.
A toggle witness of displacement $\vv$ and reduced cost
$\widehat{\mathsf D}(\vv)$ has actual cost
\[
 d_X+\vpi\vv+\widehat{\mathsf D}(\vv),\qquad \vu=\vt-\vb_X-\vv .
\]
This is an exact affine conversion of~\eqref{fw:merge}. The integer
convolution input is obtained by clearing the price denominator; no
cardinality or cost coordinate is introduced. A coefficient box is only
a computational domain: every decoded output is intersected with the
proved physical crop before a child source is built.

\section{Affordable geometry on one rational grid}
\label{fw:geom}

This section studies the geometry of affordable offsets around a single
query. It shows how the anchors of a state are chosen, why their windows
are disjoint, and why the rank of the affordable set suffices to decide
how a query is dispatched. Everything is done on one fixed rational grid,
so that the reporter of affordable offsets returns exact finite lists.

Throughout this section the halo scale satisfies $H\ge1$. We fix one
immutable pool $I$ of rank $r\ge1$, with fractional cost function
$f=f_I$, column hull $P=P_I$, and lattice $\Lambda=\Lambda_I$. Every
query $(\vt,s)$ considered here has $\vt\in\Lambda$ and positive
\emph{gap} $g=s-f(\vt)>0$. Put
\[
 Q_{\vt,s}=\{\vz:\ f(\vt+\vz)+f(\vt-\vz)\le2s\},\qquad
 S_{\vt,s}=Q_{\vt,s}\cap HP ,
\]
so that $S_{\vt,s}=S(\vt,s,H)$ in the notation of
Section~\ref{det:reporter-section}. Intuitively, the first set collects
the symmetric offsets that the allowance can pay for, and the second
restricts them to the halo. These sets are compact,
symmetric, and convex, but they may have smaller dimension than the
pool. Put $A_r=3r+1$, and fix one grid denominator $q_0$ that is
divisible by $3A_r$ for every rank $r\le D$. With grid $\Lambda/q_0$,
the reporter of Lemma~\ref{det:reporter} returns exactly the list
$S_{\vt,s}\cap(\Lambda/q_0)$.

The first lemma turns this list into an explicit parallelotope that
approximates the section within its linear span up to a factor
depending only on the rank.

\begin{lemma}[Discrete section rounding]\label{fw:rounding}
From the list $S\cap(\Lambda/q_0)$, with $S=S_{\vt,s}$, one can obtain
\[
 U_3=3S\cap\Lambda,\qquad U_A=A_rS\cap\Lambda,\qquad
 \mathcal L=\Span U_3,\qquad k=\dim\mathcal L,
\]
and an explicit parallelotope $R\subseteq\mathcal L$ with
\[
 S\cap\mathcal L\subseteq R\subseteq2r(3r+1)S .
\]
The work is linear in the size of the list, up to a fixed-rank
polynomial bit factor. If $k=0$, take $R=\{\vzero\}$.
\end{lemma}

\begin{proof}
We first compute the two lattice lists, then show that the hull of the
larger one sandwiches the section, and finally extract a well-shaped
basis from it.

Multiply each listed point by three or by $A_r$, and retain exactly the
resulting lattice points. The divisibility of $q_0$ makes these two lists
exact. For $k>0$ choose independent $\vb_1,\dots,\vb_k\in U_3$. Every
$\vz\in S\cap\mathcal L$ lies in a cell of their sublattice, and every
vertex $\vv$ of that cell satisfies
\[
 \vv-\vz\in\sum_i[-\vb_i,\vb_i]\subseteq3kS,\qquad
 \vv\in(3k+1)S\cap\Lambda\cap\mathcal L\subseteq U_A\cap\mathcal L .
\]
Thus, with $C=\conv(U_A\cap\mathcal L)$, we have
$S\cap\mathcal L\subseteq C\subseteq A_r(S\cap\mathcal L)$.

To construct the parallelotope, apply Lemma~\ref{det:greedy-basis} to
$X=U_A\cap\mathcal L$, which is symmetric, so $P_X=C$. The initial
determinant is at least the covolume of the saturated lattice
$\Lambda\cap\mathcal L$, and $\vol_k(C)$ has polynomial bit length, so
only polynomially many scans occur. The resulting $R=B[-2,2]^k$
satisfies $C\subseteq R\subseteq2kC$. No facet enumeration of $C$ is
needed.
\end{proof}

\subsection{Locality and anchor packing}
\label{fw:anchors}

The queries of a state are grouped around a few \emph{anchors}. Each
query is assigned to a nearby anchor of at least the same gap, and the
anchors are spread out enough that their windows do not overlap. This
disjointness is what lets us charge the work at all anchors of a state
to the lattice points of a single buffer.

Put $C_\ast=\max_{1\le r\le D}2r(3r+1)$. Choose parameters
$0<\alpha\le1/(100C_\ast)$, $b=\alpha/4$, and $\gamma$ with
$C_\ast\alpha\le\gamma\le1/100$; their values are fixed in
Section~\ref{fw:scales}. Scan the distinct positive-gap queries of a
state in decreasing order of gap. Each query that is still unmarked
becomes an \emph{anchor} $v=(\vt_v,s_v)$. The anchor reports
$S_v=S_{\vt_v,s_v}$, constructs $R_v$ by Lemma~\ref{fw:rounding}, and
marks the unmarked query keys in $\vt_v+\alpha R_v$, including its own
key since $\vzero\in\alpha R_v$; these marked
queries are \emph{assigned} to the anchor. Write $Q_v=Q_{\vt_v,s_v}$ and $g_v$
for its gap.

\begin{lemma}[Locality and disjoint paid windows]\label{fw:locality}
Every query $q$ assigned to an anchor $v$ satisfies
\[
 \vt_q-\vt_v\in\gamma S_v,\qquad g_q\le g_v,\qquad
 Q_q\subseteq\tau_\gamma Q_v,\qquad \tau_\gamma=\frac{1+\gamma}{1-\gamma}.
\]
The windows $\vt_v+(b/2)S_v$ are pairwise disjoint. If a state buffer
contains all of them and has $M$ lattice points, then
\[
 \sum_v\mu_v\le M,\qquad \mu_v=|(\vt_v+(b/2)S_v)\cap\Lambda| .
\]
These statements also hold at targets on the boundary of the domain.
\end{lemma}

\begin{proof}
We work in the epigraph, where the symmetric bodies at nearby points
can be compared directly; locality and disjointness then follow by
projecting back to target space.

Let $\mathcal E=\epi f$ and, for $\vp=(\vt,s)\in\mathcal E$, let
$\mathcal M_{\vp}=(\mathcal E-\vp)\cap(\vp-\mathcal E)$; its horizontal
projection is $Q_{\vt,s}$. If $\vd\in a\mathcal M_{\vp}$ with $0<a<1$,
convex combinations in $\mathcal E$ give
$(1-a)\mathcal M_{\vp}\subseteq\mathcal M_{\vp+\vd}\subseteq(1+a)\mathcal M_{\vp}$.
The vertical section of $\mathcal M_{\vp}$ through zero is the interval
of radius $s-f(\vt)$, so the gaps at $\vp$ and $\vp+\vd$ differ by
factors between $1-a$ and $1+a$. Moreover, at a fixed horizontal target,
the body at gap $g'$ is contained in $\max(1,g'/g)$ times the body at gap
$g$. To see this, apply convexity to the function
$\vz\mapsto f(\vt+\vz)+f(\vt-\vz)-2f(\vt)$, which vanishes at zero.

We now prove locality. An assigned offset lies in $\gamma Q_v$, so it
lifts to $\gamma\mathcal M_v$. At its horizontal target the lifted gap is
at least $(1-\gamma)g_v$, while by the scan order the actual query gap is
at most $g_v$. The two comparisons above give
$Q_q\subseteq\tau_\gamma Q_v$. The other locality statements follow
directly from the marking rule and the scan order.

For disjointness, suppose that $\vt_v+bS_v$ and $\vt_q+bS_q$ intersect,
where $v$ was scanned earlier. Lift the common horizontal point into
$v+b\mathcal M_v$ and into $q+b\mathcal M_q$. Since $g_q\le g_v$, the two
lifted gaps have ratio at most $\tau_b$. Comparing the projected bodies
through that
point yields $Q_q\subseteq\tau_b^2Q_v$, hence $S_q\subseteq\tau_b^2S_v$,
and so $\vh=\vt_q-\vt_v\in b(1+\tau_b^2)S_v\subseteq3bS_v$. As
$\vh\in\Lambda$, it belongs to $3S_v\cap\Lambda\subseteq\mathcal L_v$,
and $\vh/\alpha\in(3/4)S_v\cap\mathcal L_v\subseteq R_v$. Thus $v$ would
have marked $q$, a contradiction. Hence the windows of radius $b/2$ are
disjoint, and the mass inequality follows by counting their lattice
points in the buffer. The argument nowhere uses that $Q_v$ or $S_v$ is
full-dimensional, so it also applies at boundary targets.
\end{proof}

The disjoint windows also pay for the lists returned by the reporter
and for the marking scans, since these lists are only a
polylogarithmic dilation of the windows.

\begin{lemma}[Output and marking charges]\label{fw:output}
For every fixed $C>0$, the total number of points of
$CS_v\cap(\Lambda/q_0)$, summed over the anchors of one state, is
$\Ot_D(M)$. Marking enumeration has the same bound.
\end{lemma}

\begin{proof}
We use the residue inequality~\eqref{fw:residue} of
Lemma~\ref{fw:counting}. Multiplication by $q_0$ identifies $CS\cap(\Lambda/q_0)$ with
$Cq_0S\cap\Lambda$. Take $w=\lceil4Cq_0/b\rceil$; translation by the
lattice point $\vt_v$ bounds the count by $w^r\mu_v$. The factor $w^r$
is polylogarithmic in the outer parameters, since $b$ is (see
Section~\ref{fw:scales}). Summing over the anchors with
Lemma~\ref{fw:locality} gives the first claim. For marking, use
$\alpha R_v\subseteq\alpha C_\ast S_v$ and the same inequality;
fixed-dimensional integer feasibility enumerates the explicit
parallelotope with polynomial bit work per point.
\end{proof}

\subsection{Discrete rank suffices for exact dispatch}
\label{fw:dispatch-geom}

Once an anchor has its parallelotope, it must decide how its assigned
queries are reduced. The next theorem shows that the dimension of the
span of the reported lattice points, the \emph{discrete rank} of the
anchor, is enough for this decision. Every changed nonbasic column of an
affordable witness lies in that span, the change of the basic
coordinates is confined to a dilate of the parallelotope, and only
boundedly many heavy endpoints change. A
drop in this discrete rank yields a child of smaller rank;
full rank yields either a deep reduction or a thinner child.

\begin{theorem}[Fixed-grid dispatch interface]\label{fw:dispatch}
Let $\vx^0$ be a complete basic optimum at an anchor $v$, with basic set
$B_0$, $|B_0|\le r$, and let $\mathcal L_v$, $k_v=\dim\mathcal L_v$, and
$R_v$ be obtained as above. For every integral affordable witness $\vx$
of an assigned query:
\begin{enumerate}
\item Every changed nonbasic endpoint column belongs to
$3S_v\cap\Lambda$, and hence to $\mathcal L_v$.
\item With $\boldsymbol\beta=\vx_{B_0}$ and
$\vd_B=A_{B_0}(\boldsymbol\beta-\vx^0_{B_0})$, one has
$\vd_B\in\Lambda\cap\mathcal L_v\cap\max(3,r)R_v$.
\item Call a nonbasic direction \emph{light} when $H\va_i\in R_v$, and
\emph{heavy} otherwise. At most $3k_vH$ heavy endpoints change, and every
subsum of their signed changes lies in $3k_vR_v$. The light pool
satisfies $HP_{\rm light}\subseteq R_v$.
\end{enumerate}
Branching on all basic labels and applying the test of item~2 to $\vd_B$
before creating a child source preserves every affordable witness. If
$k_v<r$, the light child loses rank. If $k_v=r$, the containment test
gives either $K_\ast P\subseteq S_v$, or a light child that either
loses rank or satisfies $\Psi_{\rm child}/\Psi(I)=O_D(K_\ast/H)$, for the constant $K_\ast$ of
Section~\ref{fw:scales}.
\end{theorem}

\begin{proof}
We first show that the whole deviation of a witness from the anchor
optimum lies in a dilate of the affordable body, which gives items~1
and~2. We then bound the heavy endpoints by
a determinant argument, and finally treat the full-rank case.

Normalize costs at a supporting price $\vpi$ of the anchor. The reduced
assignment cost is then nonnegative throughout the box and vanishes at
$\vx^0$. Write $\vu=\vt_q-\vt_v$. Convexity and locality bound the
assigned normalized allowance by $(1+2\gamma)g_v$, and there is a
fractional witness $\vy$ at $\vt_v-\vu/\gamma$ of normalized cost at
most $2g_v$.

For every $\boldsymbol\zeta$ with $|\zeta_i|\le|x_i-x^0_i|$, both
$\vx+\vx^0+\boldsymbol\zeta$ and $\vx+\vx^0-\boldsymbol\zeta$ lie in
$[0,2]^I$. Add $\gamma\vy$ and divide by $2+\gamma$. The two resulting
box-feasible vectors have targets $\vt_v\pm A\boldsymbol\zeta/(2+\gamma)$
and average normalized cost less than $\frac{1+4\gamma}{2+\gamma}g_v<g_v$.
Thus the whole deviation zonotope satisfies
\begin{equation}\label{fw:deviation}
 \{A\boldsymbol\zeta:\ |\zeta_i|\le|x_i-x^0_i|\}\subseteq3Q_v .
\end{equation}
A changed nonbasic endpoint contributes one entire column, so that
column lies in $3Q_v\cap P\cap\Lambda\subseteq3S_v\cap\Lambda$.

For item~2, the vector $\vd_B$ is integral because
$A_{B_0}\vx^0_{B_0}=\vt_v-\sum_{i\notin B_0}x^0_i\va_i\in\Lambda$. It
lies in $\mathcal L_v$, since the total target displacement and every
nonbasic deviation do; for the former,
$\vt_q-\vt_v\in\gamma S_v\cap\Lambda\subseteq3S_v\cap\Lambda$. Applying~\eqref{fw:deviation} only to the basic
coordinates gives $\vd_B\in3Q_v\cap rP\subseteq\max(3,r)S_v$. Since it
also lies in $\mathcal L_v$, section rounding gives the asserted
test.

For item~3, first let $k_v=0$. Then all changed nonbasic columns lie in
$\mathcal L_v=\{\vzero\}$, so no heavy endpoint changes. Now let
$k_v>0$, and choose a maximum-determinant basis of the symmetric body
$Q_v\cap\mathcal L_v$; this basis is used only in the analysis, not by
the algorithm. Its dual coordinate functionals
$\lambda_1,\dots,\lambda_{k_v}$ have magnitude at most one on that body,
since otherwise a replacement would increase the determinant. Moreover,
$\|\va\|_{Q_v\cap\mathcal L_v}\le\sum_l|\lambda_l(\va)|$.
Restricting~\eqref{fw:deviation} to the changed nonbasic endpoints shows
\[
 \sum_{i\ \text{changed, nonbasic}}\|\va_i\|_{Q_v\cap\mathcal L_v}
 \le\sum_l\sum_{i\ \text{changed, nonbasic}}|\lambda_l(\va_i)|\le3k_v .
\]
If $H\va_i\notin R_v$, its gauge exceeds $1/H$: otherwise
$H\va_i\in Q_v\cap HP\cap\mathcal L_v\subseteq R_v$. Hence fewer than
$3k_vH$ heavy endpoints change. Every subsum of their signed changes lies
in $3Q_v$ and in $3k_vHP$, hence in $3k_vS_v\cap\mathcal L_v\subseteq
3k_vR_v$.

Every light column lies in $R_v/H$, which proves the halo inclusion and
the rank loss when $k_v<r$. It remains to treat the full-rank case.
Test $C_\ast K_\ast P\subseteq R_v$ at the constantly many facets of
$R_v$, using the support oracle of Corollary~\ref{det:body-oracles}. If
the test passes, then $K_\ast P\subseteq S_v$. If it fails, some facet
functional $\lambda$ satisfies
$|\lambda\va_i|\le h_{R_v}(\lambda)/H<(C_\ast K_\ast/H)h_P(\lambda)$
for every light column, so the light hull lies in the corresponding
central slab of $P$. By Lemma~\ref{det:central}(2) the volume fraction of
this slab is at most $rC_\ast K_\ast/H=O_r(K_\ast/H)$. The same-rank child lattice has no smaller
covolume, which proves the normalized-volume bound.

Finally, boundary targets need no special treatment. If $\vt_v$ lies on a proper face of the fractional domain,
every symmetric feasible displacement is tangent to that face, so
$\Span Q_v$ is proper and $k_v<r$. Thus the rank test precedes the deep
test, and no minimal-face computation is required; conversely, $k_v=r$
supplies the full-dimensional interior hypothesis of the deep
reduction.
\end{proof}

\section{States, source windows, and local reductions}
\label{fw:states}

This section describes one step of the recursion. We first define
states, sources, and the buffers used for accounting. We then fix all
scales in advance, and finally show how an anchor is dispatched either
to a thin or rank-loss child or to the deep reductions.

A \emph{state} is a tuple $(I,j,\ell_\uparrow,\mathcal Q)$ consisting of
an immutable pool, an arithmetic depth, an accumulated number of lifts,
and a batch of strict queries. A query $q=(\vt_q,s_q)$ asks for a
labelled selection with target $\vt_q$ and common-objective cost below
$s_q$. We keep only queries with $\vt_q\in\Lambda_I$ and positive
fractional gap $g_q=s_q-f_I(\vt_q)>0$. At duplicate targets we retain
the largest allowance and one attaining incoming source.

A \emph{source} is the unit of work that we charge. It records a pool
version, its birth depth and lift count, an affine coordinate map, a
rational convex window $K$, and the predecessor and removal witnesses
used to extend a child answer to the parent. Different source windows
remain separate when batches merge. The window covers the required
offsets of every candidate target, enlarged by a fixed rank-dependent
multiple $c_{\rm box}H_{\rm birth}\Bx_{P_I}$ of the birth halo. This
enlargement ensures that the window hypothesis $K\supseteq C+H_b\Bx_P$
of Lemma~\ref{det:reporter} holds with $H_b=H_{\rm birth}$. The charge
of the source is $V(K,I)$ as in~\eqref{det:volume}. All windows
constructed below are explicit polytopes of rank-bounded description
complexity, namely parallelotopes, rounded sections, or slanted prisms.
A source continues to answer its queries after coalescence, since its
pool is unchanged and the depth-dependent halo only decreases.

The cost of a state is measured by its \emph{buffer}, the set of lattice
points near its query targets. For decreasing scales $H_j$ and
$0<\xi_j<1$, and with $Q_q$ as in Section~\ref{fw:geom}, put
\begin{equation}\label{fw:buffer}
 S_q=Q_q\cap H_jP_I,\qquad
 \mathcal D(I,j)=\Lambda_I\cap\bigcup_{q\in\mathcal Q}
 \bigl(\vt_q+(\xi_jQ_q\cap H_jP_I)\bigr),\qquad
 M(I,j)=|\mathcal D(I,j)| .
\end{equation}
Buffers are accounting objects only; the algorithm never enumerates
them. By convexity and $\xi_j<1$, every buffer point lies in the domain
of $f_I$, even when the target lies on the boundary of the domain.

\subsection{A priori scales}
\label{fw:scales}

All constants of the recursion are fixed before it starts. The two
schedules below, for the buffer radii and the halos, decrease with the
depth just fast enough to absorb the losses of one
step; the two displayed schedule inequalities at the end of this
subsection express this.

Fix $\delta>0$ and an accounting slack $\beta>0$, and let
$\nu=\lceil N^\delta\rceil$. Choose
\[
 T_\uparrow=\Bigl\lceil\frac{2(d+1)}{\delta}\Bigr\rceil+2,\qquad
 D=d+T_\uparrow+1,\qquad
 J\ge D+\lceil\log_2(C_dN^d)\rceil+2 .
\]
Here $C_d$ bounds the initial normalized volume. These constants are
chosen prospectively; the rank and lift argument of
Section~\ref{fw:rank} justifies them without assuming the rank bound in
advance. Fix a progression parameter
\begin{equation}\label{fw:dgap}
 0<\delta_{\rm GAP}<\frac{\delta}{2[d+(D-1)\delta]} .
\end{equation}
Let $k_0=C\nu\log_2^2(N+2)$, rounded up, where the constant is fixed
large enough to bound every reserve and outlier count in
Lemma~\ref{fw:progression}. (That lemma produces
$O(\nu\log^2(\nu L+2))$ outliers with $L$ polynomial in $N$, so a
single logarithm would not suffice.)
Put
\[
 \kappa=3k_0,\qquad E=(260+6J)k_0,\qquad
 C_\ast=\max_{1\le r\le D}2r(3r+1),\qquad \xi_\ast=2^{-10},
\]
and
\[
 \alpha=\frac{\xi_\ast}{10000C_\ast(J+1)},\quad \gamma=C_\ast\alpha,\quad
 b=\alpha/4,\quad \tau_\gamma=\frac{1+\gamma}{1-\gamma},\quad
 K_\ast=E/\gamma .
\]
Set
\[
 \Delta_\xi=\gamma+(\tau_\gamma-1)\xi_\ast+E/K_\ast,\qquad
 \xi_j=\xi_\ast/4+(J-j)\Delta_\xi .
\]
Then $\xi_\ast/4\le\xi_j<\xi_\ast/2$ and
\begin{equation}\label{fw:xi-schedule}
 \xi_j\ge\gamma+\tau_\gamma\xi_{j+1}+E/K_\ast .
\end{equation}
Choose $H_{\min}\ge\max\{K_\ast,JE,C_\ast/3\}$, with the additional
polylogarithmic enlargement~\eqref{fw:Hmin}, and put
\[
 \Delta_H=2E+2\gamma H_{\min},\qquad H_j=H_{\min}+(J-j)\Delta_H .
\]
Direct substitution gives
\begin{equation}\label{fw:H-schedule}
 H_j-H_{j+1}\ge\gamma H_j+E,\qquad H_0\le4H_{\min}.
\end{equation}
All these numbers are rational with polynomial bit length. In
particular $H_j=\Ot(\nu)$, $H_j/K_\ast$ is polylogarithmic, and
$bK_\ast\ge c_D\nu$.

\subsection{Uniform dispatch, including boundary queries}
\label{fw:dispatch-sec}

We now describe how a state is processed. The same procedure handles
interior and boundary targets: every feasible positive-gap query enters
the decreasing-gap anchor scan described above, and there is
no preliminary branch on minimal faces. All source atlases tabulate
$\Lambda_I/q_0$, with $q_0$ as in Section~\ref{fw:geom}.

For an unmarked anchor $v$, use the reporter to list
$S_v\cap(\Lambda_I/q_0)$ at halo $H_j$, form
$\mathcal L_v=\Span_{\R}(3S_v\cap\Lambda_I)$ and $k_v=\dim\mathcal L_v$,
and construct by Lemma~\ref{fw:rounding} an explicit parallelotope
$R_v\subseteq\mathcal L_v$ with
\begin{equation}\label{fw:Rv}
 S_v\cap\mathcal L_v\subseteq R_v\subseteq C_\ast S_v .
\end{equation}
Mark every available query in $\vt_v+\alpha R_v$.
Lemma~\ref{fw:locality} supplies
\begin{equation}\label{fw:packing}
 \sum_v\mu_v\le M(I,j),\qquad \mu_v=|(\vt_v+(b/2)S_v)\cap\Lambda_I|,
\end{equation}
and, for every assigned $q$,
$\vt_q-\vt_v\in\gamma S_v$, $g_q\le g_v$, and $Q_q\subseteq\tau_\gamma
Q_v$. By Lemma~\ref{fw:output}, the complete grid output and the marking
work are $\Ot_D(\mu_v)$.

The anchor is then dispatched by its discrete rank. If $k_v<r$,
apply the rank-loss transfer below. Only if $k_v=r$, test
$C_\ast K_\ast P_I\subseteq R_v$. A successful test implies $K_\ast
P_I\subseteq S_v$ and invokes the deep procedure of
Section~\ref{fw:deep}; otherwise apply the thin transfer below. A
boundary target automatically has $k_v<r$
(Theorem~\ref{fw:dispatch}). The dispatch therefore needs only finite LP
values, prices, and basic witnesses, and no minimal-face records. The
facet test uses a constant number of column-body support queries per
anchor (Corollary~\ref{det:body-oracles}). Since the reporter already
restricts to $H_jP_I$, no further clipping is needed. Both costs are
paid by Proposition~\ref{det:source-charge}, and no thin anchor scans
all current types.

\begin{lemma}[Operational thin and rank-loss source]\label{fw:thin}
At an anchor that is not sent to the deep procedure, branch on all at
most $r$ basic labels of its stored basic optimum. The remaining labels
can be transferred to a light child with the following properties.
\begin{enumerate}
\item Every affordable integral witness of every assigned query survives
some branch and a dominating removal-table entry.
\item Each generated child target, before LP filtering, belongs to one
explicit translate of $O_D(R_v)$, and $H_jP_{\rm child}\subseteq R_v$.
\item If $k_v<r$, the child loses rank. Otherwise every same-rank child
satisfies $\Psi_{\rm child}\le C_D(K_\ast/H_j)\Psi(I)$.
\item The computational boxes, compatible-candidate scans, source
volumes, caller queries, and reported outputs are $\Ot_D(\mu_v)$ per
anchor. Witnessed tables cost $\Ot_D(\mu_v^{\rho})$, and all other
structural work costs $\Ot_D(\nu\mu_v)$.
\end{enumerate}
\end{lemma}

\begin{proof}
We first restrict the basic branches and identify the light and heavy
labels, then compute the child target, and finally bound the costs.

Let $\vx^0$ be the basic optimum at $\vt_v$, with basic labels $B_0$. By
Theorem~\ref{fw:dispatch}, every changed nonbasic column of an affordable
integral witness lies in $3S_v\cap\Lambda_I\subseteq\mathcal L_v$. All
other nonbasic labels can be fixed at their anchor endpoints. For a
basic choice $\boldsymbol\varepsilon\in\{0,1\}^{B_0}$ put
$\vd_B=A_{B_0}(\boldsymbol\varepsilon-\vx^0_{B_0})\in\Lambda_I$, and
discard it before source construction unless
\begin{equation}\label{fw:basic-test}
 \vd_B\in\mathcal L_v\cap\max\{3,r\}R_v .
\end{equation}
By item~2 of the dispatch theorem, the basic displacement of every
actual witness passes this test. Declare a nonbasic label light when
$H_j\va_i\in R_v$. Since $R_v\subseteq C_\ast S_v$ and
$H_j\ge C_\ast/3$, the reported list $3S_v\cap\Lambda_I$ contains every
light direction. The same list supplies all relevant heavy directions
through current-type dictionary lookups, so the parent types are never
scanned in full.

By Theorem~\ref{fw:dispatch}, at most $3k_vH_j$ heavy nonbasic endpoints
change and every subsum of their signed changes lies in $3k_vR_v$. Let
$\vz_H$ denote the table displacement and
$\vb_{\rm light}=\sum_{i\ \text{light}}x^0_i\va_i\in\Lambda_I$. For
$\vu=\vt_q-\vt_v$, the exact residual target is
\begin{equation}\label{fw:child-target}
 \vt_{\rm child}=\vb_{\rm light}+\vu-\vd_B-\vz_H .
\end{equation}
This proves the asserted crop, including for branches that LP filtering
later rejects. The fixed remote endpoints are represented by the anchor
template together with its target and cost aggregates, so only the
reported active types and $B_0$ are visited.

The light hull satisfies $H_jP_{\rm child}\subseteq R_v$, and
Theorem~\ref{fw:dispatch} gives the rank loss and the same-rank volume
decrease. For the removal tables, cap the eligible heavy toggles by
their required count and apply Lemma~\ref{fw:transfer} in a symmetric
box
$\prod_l[-\Delta_l,\Delta_l]$, taken in lattice coordinates of an adapted
basis of $\Lambda_I\cap\mathcal L_v$ as in~\eqref{det:box}, that contains
a fixed dilation of $R_v$ and is contained in a larger fixed dilation
$c_DR_v\subseteq c_DC_\ast S_v$. Every required subassignment lies in
that physical crop, and by~\eqref{fw:residue} the box has $\Ot(\mu_v)$
cells. There are at most $2^r$ basic branches. The source-volume and mass
charges are proved in Lemma~\ref{fw:halo}; they use~\eqref{fw:basic-test}
before filtering. Basic-label deletions and at most $\Ot(\nu)$ relevant
labels per visited direction cost $\Ot(\nu\mu_v)$.
\end{proof}

\subsection{The deep structural inputs}
\label{fw:deep}

At a deep anchor the affordable body contains a large multiple of the
column hull, and we can no longer make progress by thinning. Instead we
use three structural inputs, in order: a plateau of cheap labels whose
canonical core has a stable lattice, an additive progression built from
that core, and a trial that either answers the queries, restricts to a
proper sublattice, or lifts the pool to one more dimension. The three
results below treat these steps in turn.

We state these results in the form used by the algorithm. The
affordable and cost-sensitive reductions are proved
in~\cite[Sections~8--10]{CMZ26b} and, in the compressed witnessed form
used here, in~\cite[Lemmas~6.10 and~7.1--7.10]{Zhou26}. Zhou's
statements already concern a current pool of arbitrary fixed rank
$r\ge1$, because his lifts raise the rank, and the point-or-thin
primitive~\cite[Lemma~10.1]{CMZ26b} is stated in every fixed dimension.
Only the root differs: Zhou's root has rank one and normalized volume
$O(w_{\max})$, while ours has rank $d$ and $\Psi_0\le C_dN^d$. Each place
where this enters is adapted below, namely the progression hypothesis
$\nu\ge L^{\delta_{\rm GAP}}$ through~\eqref{fw:dgap}, the global
choice of $k_0$, and the depth bound $J$. The underlying progression
result is the count version~\cite[Theorem~3.1 and
Corollary~8.8]{CMZ26}, used exactly as in~\cite[Lemma~7.3]{Zhou26}. We
record the quantitative statements, and we give the strip argument in
full.

At a deep anchor choose its supporting price $\vpi$ and write $g=g_v$,
$Q=Q_v$, and $P=P_I$. With the fixed strict convention
$e_i(\vpi)=[\vpi\va_i>c^\ast_i]$ and $\epsilon_i=1-2e_i(\vpi)$, the
working columns and costs are $\bar\va_i=\epsilon_i\va_i$ and
$\bar c_i=|c^\ast_i-\vpi\va_i|\ge0$. A query has working target
$\bar\vt=\vt-A_I\ve$ and normalized allowance
\[
 a_q=s_q-f_I(\vt_v)-\vpi(\vt_q-\vt_v)\le(1+2\gamma)g ,
\]
and the anchor minimum is zero. For $\mathcal A\subseteq I$ let
\[
 T_{\mathcal A}(b)=\{\bar A_{\mathcal A}\vx:\ \vzero\le\vx\le\vone,\
 \bar\vc_{\mathcal A}^{\T}\vx\le b\},\qquad
 L_{\vpi}(\vartheta)=\{i:\ |c^\ast_i-\vpi\va_i|\le\vartheta\}.
\]
Thus the first set consists of the working targets reachable
fractionally within the given budget, and the second consists of the
\emph{cheap} labels at the given threshold. A pool is
\emph{$\kappa$-resilient} if deleting any at most $\kappa$ labels leaves
its lattice unchanged. The \emph{canonical core} $\mathcal R_\kappa(Y)$
is the union of all resilient subpools of $Y$. This union is itself
resilient, so the core is unique and monotone. It is also closed under
adding labels of $Y$ whose columns lie in its lattice.

\begin{lemma}[Affordable collar and canonical plateau]\label{fw:collar}
Suppose $K_\ast P\subseteq Q$. If $P\subseteq P^\sharp\subseteq
Q/K_\ast$, $0\le z\le\xi\le1$, and $\mathcal A$ omits at most $e$ labels
of $L_{\vpi}(\vartheta)$, then, whenever the radius is nonnegative,
\begin{equation}\label{fw:collar-eq}
 \bar\vt_v+zQ+\bigl((\xi-z)K_\ast-2\xi g/\vartheta-e\bigr)P^\sharp
 \subseteq T_{\mathcal A}(2\xi g).
\end{equation}
Put $\vartheta_0=g/(128k_0)$, $\upsilon=g/32$, and
$\Delta=\vartheta_0/(J+1)$. Among consecutive grid thresholds from
$\vartheta_0$ to $2\vartheta_0$ with spacing $\Delta$ there are
$\vartheta_-<\vartheta_+$ with equal full-rank canonical-core lattice
$\Gamma$. They can be found on compressed profiles in
$\Ot(m+\kappa^2\Psi(I))\poly(\cB)$ time, where $m$ is the number of
current types. For $\mathcal A_c=\mathcal R_\kappa(L_{\vpi}(\vartheta_-))$
and $I_\Gamma=I\cap\Gamma$, one has $\Lambda_{I_\Gamma}=\Gamma$, and at
most $\kappa J$ labels of $L_{\vpi}(\vartheta_+)$ lie outside $I_\Gamma$.

For every retained restored query with attaining assignment $Y$ on
$X=I\setminus I_\Gamma$, every point $\vy$ in its depth-$(j+1)$ buffer,
and every $\vpi'\in\partial f_{I_\Gamma}(\vy)$,
\begin{equation}\label{fw:every-price}
 \Gamma=\Lambda_{\mathcal R_\kappa(L_{I,\vpi'}((\vartheta_-+\vartheta_+)/2))},
 \qquad A_X\bigl(Y-\ve_X(\vpi')\bigr)\in EP .
\end{equation}
The assignment $Y$ and its predecessor are fixed before $\vpi'$ is
chosen.
\end{lemma}

\begin{proof}
We first prove the collar inclusion, then find the plateau, and finally
derive the every-price certificate.

The paired-body inequality gives $\bar\vt_v+\xi Q\subseteq T_I(2\xi g)$.
Removing costs above $\vartheta$ deletes fractional mass at most
$2\xi g/\vartheta$, and the $e$ missing labels delete at most $e$ more.
Comparing support functions and using $h_Q\ge K_\ast h_{P^\sharp}$
gives~\eqref{fw:collar-eq}.

At $\xi=1/128$ this supplies a translate of $(K_\ast/128-2k_0)P$ in every
cheap zonotope used by the construction. Each of these zonotopes
therefore remains full rank after $\kappa(J+r)$ deletions. The
canonical-core algorithm
of~\cite[Lemma~6.10]{Zhou26} extracts $\kappa+1$ disjoint actual-column
bases, enumerates the characters of their finite-index quotients, and
deletes a nonzero character support of size at most $\kappa$ whenever
one is found. Throughout, its current pool is exactly
$Y_0\cap\Lambda_Y$. Every successful restriction at least doubles the
index, so there are at most $J$ of them, and the character searches
cost $\Ot(m+\kappa^2\Psi)$.
Monotonicity of the threshold cores and the same index bound give the
adjacent plateau and the bound on excluded cheap labels.

For the last assertion we use the saved assignment $Y$. Its normalized
cost is below $a_q$. Hence at most $128(1+2\gamma)k_0$ costly labels
change, and at most $\kappa J$ cheap excluded labels can change, so
$A_X(Y-\ve_X(\vpi))\in(E/2)P$. Extending opposite child witnesses by the
same $Y$ puts the buffer point in the anchor's affordable neighbourhood.
Applying~\eqref{fw:collar-eq} at $\xi=1/64$ gives a collar
$\vy\pm q_{\rm col}\va_i$ of normalized cost at most $\upsilon$, where
$q_{\rm col}=16(J+1)k_0$. Every supporting price therefore satisfies
$|(\vpi'-\vpi)\va_i|\le\upsilon/q_{\rm col}=\Delta/4$. Consequently the
cheap pool at the midpoint threshold and price $\vpi'$ is sandwiched
between the two plateau pools, and monotonicity of the core proves the
core-lattice identity.
Endpoints of labels above $\vartheta_+$ cannot switch, and at most
$\kappa J$ other excluded labels do. This enlarges the displacement bound
to $EP$ and proves~\eqref{fw:every-price}. These are the arguments
of~\cite[Lemmas~7.1--7.2]{Zhou26}; no minimal face is computed.
\end{proof}

\begin{lemma}[Compressed progression and actual reserve]
\label{fw:progression}
For the full-rank cheap core $\mathcal A_c$ above one can compute a
partition
\[
 \mathcal A_c=I_{\rm out}\,\dot\cup\,I_{\rm res}\,\dot\cup\,I_{\rm rem},
 \qquad |I_{\rm out}|,|I_{\rm res}|\le k_0,
\]
a symmetric integer box $\mathsf B\subset\R^k$, integral type
coordinates $\vu_i\in\mathsf B$, an integral linear map $\phi$ with
$\phi\vu_i=\bar\va_i$, and a reserve centre $\vz_0$, such that
\begin{equation}\label{fw:reserve}
 \vz_0+(c_{\rm GAP}\nu\mathsf B\cap\Z^k)\subseteq
 \Bigl\{\sum_{i\in Y}\vu_i:\ Y\subseteq I_{\rm res}\Bigr\}.
\end{equation}
The rank $k$ and $c_{\rm GAP}>0$ depend only on $\delta_{\rm GAP}$. The
remainder spans $\R^k$ after any $\nu$ labelled deletions,
$\phi(\Z^k)=\Gamma$, and the entire reserve has normalized cost at most
$g/64$. The construction costs $\Ot(m\nu+\nu^2L)\poly(\cB)$ with
$L=O_D(\nu^{r-1}\Psi(I))$, and returns actual outlier and reserve
identifiers. An attainable deficit in the old coordinates can be
recovered as an actual reserve subset in $\Ot(\nu^{r+1}\Psi(I))\poly(\cB)$
time.
\end{lemma}

\begin{proof}
This is~\cite[Lemma~7.3]{Zhou26}, which is stated for a pool of
arbitrary rank $r\ge1$. We repeat the argument because its hypothesis
$\nu\ge L^{\delta_{\rm GAP}}$ is verified there from the
knapsack-specific bound $\Psi=O(w_{\max})$, and here
from~\eqref{fw:dgap}. We encode the core as positive integers, apply the
common-progression theorem to them, and then translate its conclusion
back to vectors.

Rebox the old lattice so that $|a_l|\le q_l$ with integers $q_l\ge1$ and
$\prod_lq_l=O_D(\Psi)$. Mixed radices $2\lceil C\nu q_l\rceil+1$ are
injective on all $O(\nu)$-term signed sums used in the reserve argument,
while a single encoded column has magnitude at most $L$. A permanent
type has the positive encoded value $|\mathrm{enc}(\va)|$, independent of
its working sign. Distinct permanent types have distinct values, since
equal values would make their current columns equal up to sign,
contrary to their distinct permanent readouts.

The counting version of the common-progression
theorem~\cite[Theorem~3.1 and Corollary~8.8]{CMZ26} applies with
$h=\nu$. Its hypotheses are distinct positive integer values at most $L$,
arbitrary finite multiplicities, and $\nu\ge L^{\delta_{\rm GAP}}$; the
last follows from~\eqref{fw:dgap}, $\Psi\le C_dN^d$, and $r\le D$, for
all sufficiently large $N$. It supplies $O(\nu\log^2(\nu L+2))$
outliers, $O(\nu)$ reserve labels, the filling box, and a robust
remainder span in the stated time, and it is deterministic
by~\cite[Corollary~8.8]{CMZ26}. The all-outlier outcome is excluded by
full-rank $\kappa$-resilience.

The filling comparisons in~\cite[Lemma~3.4]{CMZ26} give an integral
old-vector readout. By mixed-radix injectivity, equal scalar sums of
the $O(\nu)$ reserve vectors have equal old-vector sums, and unit-square
and consecutive-triple comparisons make the resulting map affine on the
filled box. Compute its linear part on a basis of actual remainder types and
verify it in one scan. The core's resilience gives
$\Lambda_{I_{\rm rem}}=\Gamma$, and reserve filling at two neighbouring
integer points gives $\phi(\Z^k)\subseteq\Gamma$. Undoing temporary
encoding signs complements the corresponding reserve choices and
translates $\vz_0$. Every reserve label costs at most $g/(64k_0)$.
Finally, the $\Ot(\nu)$ actual reserve labels have all old-coordinate
sums in a box of $\Ot(\nu^r\Psi)$ cells, and an ordinary labelled
reachability dynamic program with predecessor pointers recovers a
prescribed attainable old target in $\Ot(\nu^{r+1}\Psi)$ time. This
recovery need not reproduce a prescribed progression representation.
\end{proof}

\begin{proposition}[Deep trial, restoration, and lift]\label{fw:trial}
Let $I^\circ=I_{\rm res}\cup I_{\rm rem}$ and use a fresh witnessed
transfer on $I\setminus I^\circ$, followed by lattice and positive-gap
filtering. Call a survivor \emph{high} when its normalized allowance is at
least $g/4$.
\begin{enumerate}
\item If the trial has no survivor, none of the assigned queries has an
integer witness.
\item If it has survivors but none is high, $\Gamma$ is a proper
full-rank sublattice of $\Lambda_I$. A fresh transfer on $I\setminus
I_\Gamma$ preserves the decision and gives the every-price
certificate~\eqref{fw:every-price}. Its arithmetic depth is $j+1$.
\item If a high entry exists, one fixed-dimensional affordable-fibre
computation either returns an actual integral witness for an assigned
parent query, or returns a lifted pool $I'$ of rank $r+1$ whose
old-coordinate readout and stored costs are preserved. With a constant
integer $F_D$ depending only on the bounded progression dimension,
\begin{equation}\label{fw:lift-volume}
 \Psi(I')\le\frac{2(r+1)F_D}{\nu}\Psi(I).
\end{equation}
A fresh transfer on $I\setminus I'$ gives an old residual $\vh$. Before
filtering, enumerate only heights $z$ in one computable common strip
\begin{equation}\label{fw:strip}
 b^\ast-256F_D\le z-\boldsymbol\lambda\vh\le b^\ast+255F_D .
\end{equation}
This strip retains every required completion after replacement by a
cheaper attaining removal witness, and contains every point of every
resulting actual buffer. There are at most $511F_D+1$ heights per old
residual. A lift preserves the arithmetic depth and increases the lift
count by one.
\end{enumerate}
Every required removal assignment in these operations has every subsum
in $O(E)P$. The structural work, including core tests, progression
construction, grouped image and fibre LPs, retrieval of exceptional
labels, and terminal reserve recovery, is $\Ot_D(\nu\mu_v)\poly(\cB)$,
even if all generated batches are rejected. The removal tables have
$\Ot_D(\mu_v)$ cells each and are charged separately.
\end{proposition}

\begin{proof}
We give the quantitative statements behind each outcome, as proved
in~\cite[Lemmas~7.5--7.10]{Zhou26}, treating items~1 and~2 first, then
the integer-point and thin outcomes of item~3, and finally the costs.

We begin with a displacement bound for all three items. Outside the
trial pool there are at
most $\kappa J+k_0$ cheap labels, and an affordable assignment changes
fewer than $128(1+2\gamma)k_0$ other labels, so all its selected
subassignments have displacement in $(E/2)P$. The dominating table
preserves every integral completion, which proves item~1. If
$\Gamma=\Lambda_I$, the zero working assignment of the anchor outside
$I^\circ$ has compatible residual allowance $g$,
and~\eqref{fw:collar-eq} supplies a trial fractional solution of cost at
most $\upsilon/2$, so a high entry must survive. This proves properness
in item~2. Restoration recomputes the table on $I\setminus I_\Gamma$; it
does not reuse a table for a different remaining pool. Any high restored
entry would, by the same affordable-depth bound, produce a high trial
entry. Lemma~\ref{fw:collar} supplies the every-price certificate for the
actual restored witnesses.

For item~3 choose one high residual $\vy_0$, and put
\[
 \widehat{\mathcal J}_b=\vz_0+\Bigl\{\sum_{i\in I_{\rm rem}}\vu_ix_i:\
 0\le x_i\le1,\ \sum_{i\in I_{\rm rem}}\bar c_ix_i\le b\Bigr\},\qquad
 \mathcal K=\widehat{\mathcal J}_\upsilon\cap\phi^{-1}(\vy_0).
\]
The affordable-depth bound permits us to fix the reserve to a witness
for $\vz_0$ and to prescribe either endpoint on any at most $\nu$
remainder labels, with remainder cost at most $\upsilon$. Without those
prescriptions, a neighbourhood of $\vy_0$ is feasible even at half
budget.
Compatibility gives the affine lattice $\phi^{-1}(\vy_0)\cap\Z^k$. If
$k=r$, its unique point is integral. Otherwise the fixed-dimensional
point-or-thin theorem, in the optimization-and-membership form
of~\cite[Lemma~10.1]{CMZ26b}, applied in the $k-r$ Smith coordinates of
this affine lattice, returns an integer point of $\mathcal K$ or
an integral functional $\psi$, nonzero on $\ker\phi$, of width at most
$F_D$ on $\mathcal K$. Its polynomially many oracle calls are grouped
affordable-image operations of Lemma~\ref{fw:grouped-lp}.

At an integer point, minimize the remainder cost in its image fibre and
round down at most $k$ fractional labels. The error is integral in
$k\mathsf B$, and $k\mathsf B\subseteq c_{\rm GAP}\nu\mathsf B$ for $N$
above the cutoff of Section~\ref{fw:rank}, so~\eqref{fw:reserve} and the reserve dynamic program
recover its old-coordinate deficit. The resulting cost is at most
$\upsilon+g/64<g/4$, and the remainder, reserve, and trial removal are
disjoint actual pools. This returns a witnessed parent solution.

In the thin outcome choose optimal equality and budget multipliers
$\boldsymbol\lambda,\tau$ for minimizing $\psi$ on $\mathcal K$, with
$\tau\ge0$, and put
$\omega_i=\psi(\vu_i)-\boldsymbol\lambda\bar\va_i+\tau\bar c_i$ and
$e^\omega_i=[\omega_i<0]$. The reduced-cost identity on the fibre is
\[
 \psi(\vz)-\min_{\mathcal K}\psi=\sum_i|\omega_i||x_i-e^\omega_i|
 +\tau\Bigl(\upsilon-\sum_i\bar c_ix_i\Bigr)\le F_D .
\]
Half-budget feasibility gives $\tau\le2F_D/\upsilon$, and prescribing
opposite endpoints on the $\nu$ largest absolute scores shows that their
sum is at most $F_D$. Remove those labels; each remaining label satisfies
$|\psi(\vu_i)-\boldsymbol\lambda\bar\va_i|\le2F_D/\nu$, because
$|\omega_i|\le F_D/\nu$ and
$\tau\bar c_i\le(2F_D/\upsilon)\cdot g/(64k_0)=F_D/k_0\le F_D/\nu$. Robust spanning
of the remainder and $\psi\notin\operatorname{rowspan}\phi$ give lifted
rank $r+1$. Absolute affine scores on a sorted type profile are retrieved
from its two ends, so the exceptional set costs $\Ot(m+\nu)$.

It remains to find the strip of heights; we follow the proof
of~\cite[Lemma~7.9]{Zhou26}. Use a fresh table on the
disjoint union of the old trial complement, the
reserve, and the exceptional labels, with reserve choices free. Its
affordable assignments have all subsums in $(E/2+2k_0)P$. Replacing a
removal assignment by a cheaper table witness of the same old weight does
not change the appended coordinate of a completion, which depends only
on the labels in $I'$.

\emph{A feasible neighbourhood.} Put $P_{\rm ref}=S_v/K_\ast$, so that
$P\subseteq P_{\rm ref}\subseteq Q/K_\ast$ at a deep anchor, and
\[
 \mu_0=(1/128-\gamma)K_\ast-2k_0-2E .
\]
The high residual satisfies $\vy_0\in\bar\vt_v+\gamma Q+(E/2)P$.
Apply~\eqref{fw:collar-eq} with $P^\sharp=P_{\rm ref}$, $\xi=1/128$,
$z=\gamma$ and $\vartheta=\vartheta_-$, so that $2\xi g/\vartheta_-\le2k_0$.
Erode by the at most $\kappa J+k_0\le E$ cheap labels missing from
$I^\circ$, and absorb the offset $E/2$. This gives
$\vy_0+\mu_0P_{\rm ref}\subseteq T_{I^\circ}(\upsilon/2)$. Fixing the
reserve to a witness for $\vz_0$ costs $k_0$ more units of radius, so
every $\vy\in\vy_0+(\mu_0-k_0)P_{\rm ref}$ is the image under $\phi$ of a
point of $\widehat{\mathcal J}_{\upsilon/2}$.

\emph{Every required point lies near $\vy_0$.} Consider a retained
residual query whose actual attaining assignment $Y$ has $d(Y)<a_q$, and
a point of its buffer at scale $0\le\xi\le\xi_\ast$. Mixing a paired
witness with a centre witness in proportion $\xi$ gives a fractional
completion $\vx$ on $I'$ with
\[
 d(Y)+\bar\vc_{I'}^{\T}\vx\le(1+\xi_\ast)(1+2\gamma)g<2g ;
\]
for a true integral completion the same holds at $\xi=0$. Combine $\vx$
with the exceptional labels of the removal assignment and use the fixed
reserve witness for $\vz_0$. The result is a point of
$\widehat{\mathcal J}_{4g}$ with old target $\vy$, and
\[
 \vy\in\vy_0+R_pP_{\rm ref},\qquad R_p=(1+2\gamma)K_\ast+E+2k_0 .
\]
The four contributions are:
\begin{itemize}
\item the assigned query targets, which differ by a vector of
$2\gamma S_v$;
\item the projected child-buffer displacement, which lies in $S_v$ by
extending its two defining LP witnesses with the same removal
assignment, using $Q_q\subseteq\tau_\gamma Q_v$ and
$\xi_\ast\tau_\gamma<1$;
\item the two trial-complement assignments, each with drift in
$(E/2)P$;
\item the replacement of the actual reserve part by the fixed reserve
witness, which moves the weight by at most $2k_0P$.
\end{itemize}

\emph{Concavity.} For feasible $(\vy,b)$ put
\[
 G(\vy,b)=\max\bigl\{\psi(\vz)-\boldsymbol\lambda\vy+\tau b-\alpha_0:\
 \vz\in\widehat{\mathcal J}_b,\ \phi\vz=\vy\bigr\},\qquad
 \alpha_0=\psi(\vz_0)-\boldsymbol\lambda\phi(\vz_0)
 +\sum_{i\in I_{\rm rem}}\min(0,\omega_i).
\]
For any representation $\vz=\vz_0+\sum_i\vu_ix_i$ with $\phi\vz=\vy$,
\[
 \psi(\vz)-\boldsymbol\lambda\vy+\tau b-\alpha_0
 =\sum_{i\in I_{\rm rem}}|\omega_i||x_i-e^\omega_i|
 +\tau\Bigl(b-\sum_i\bar c_ix_i\Bigr).
\]
Hence $G\ge0$, and $G$ is concave because representations can be mixed.
Moreover $G(\vy_0,\upsilon)=\max_{\mathcal K}\psi-\min_{\mathcal K}\psi
\le F_D$. By the scales of Section~\ref{fw:scales}, where $K_\ast=E/\gamma$
and $\gamma\le2^{-10}/10^4$, we have $\mu_0-k_0\ge\max\{\nu,R_p/254\}$.
Hence $\vy'=\vy_0-(\vy-\vy_0)/254$ is feasible at budget $\upsilon/2$. Since
$\upsilon=g/32$,
\[
 (\vy_0,\upsilon)=\tfrac1{255}(\vy,4g)+\tfrac{254}{255}(\vy',\upsilon/2),
\]
and concavity together with $G(\vy',\upsilon/2)\ge0$ gives
$G(\vy,4g)\le255F_D$. Therefore every such representation has
$\sum_{i\in I'}|\omega_i||x_i-e^\omega_i|\le255F_D$, while
$\tau\sum_{i\in I'}\bar c_ix_i\le(2F_D/\upsilon)\cdot4g=256F_D$. With
$\vh=\sum_{i\in I'}\bar\va_ix_i$, $z=\sum_{i\in I'}\psi(\vu_i)x_i$ and
$b^\ast=\sum_{i\in I'}\min(0,\omega_i)$, the identity
\[
 z-\boldsymbol\lambda\vh=b^\ast+\sum_{i\in I'}|\omega_i||x_i-e^\omega_i|
 -\tau\sum_{i\in I'}\bar c_ix_i
\]
gives~\eqref{fw:strip}. The bound covers every integral completion and
every actual buffer point, even when the table chooses a new, cheaper
attaining assignment, and it imposes no feasibility requirement on the
heights inside the strip.

Store the lifted column in its permanent orientation as
$(\va_i,\epsilon_i\psi(\vu_i))$. Its appended coordinate is constant on
each permanent type, and reboxing stores an inverse readout of all old
coordinates. The shear $(\vh,z)\mapsto(\vh,z-\boldsymbol\lambda\vh)$ has
determinant one and puts the symmetric child hull inside
$P\times[-2F_D/\nu,2F_D/\nu]$. The child lattice projects to a sublattice
of the old lattice and its vertical kernel has a positive integral
period, so its covolume is at least the old covolume.
This proves~\eqref{fw:lift-volume}.

Finally we bound the costs. Deepness and~\eqref{fw:packing} give
$\mu_v\ge c_D\nu^r\Psi(I)$,
by Lemma~\ref{fw:counting}(1) applied to $X=((bK_\ast-r)/2)P$, since $bK_\ast\ge2r$ for $N$ above that cutoff. By Lemma~\ref{fw:collar}, core work is
$\Ot(m+\kappa^2\Psi)=\Ot(m+\nu^2\Psi)$, because
$\kappa=3k_0=O(\nu\log^2 N)$; progression work is
$\Ot(m\nu+\nu^{r+1}\Psi)$; the other grouped operations and the reserve
dynamic program have the same bound. Since $m=O_D(\Psi)$ and $r\ge1$, all
are $\Ot(\nu\mu_v)$. The removal tables live in boxes covering $O(E)P$,
with $O_D(E^r\Psi)=\Ot(\nu^r\Psi)=\Ot(\mu_v)$ cells. The argument applies
before any trial filtering.
\end{proof}

\section{Charging sources and coalescing arithmetic phases}
\label{fw:charge}

This section settles two accounting questions left open by the local
reductions. First, every source window must be paid for when it is
built, before any LP test discards its candidates; otherwise a source
that is rejected in full would be free. Second, the buffers produced by
proper full-rank restorations must be merged so that their masses do
not multiply along a chain of restorations. The first two subsections
answer the first question; the last subsection answers the second.

\subsection{Windows are paid before filtering}

We first show that the window of every source, including its halo, is
paid by the anchor that generates it. The following physical-window
estimate is that of~\cite[Lemma~4.2]{Zhou26}, stated with possibly
different parent and child ranks.

\begin{lemma}[A physical halo charge]\label{fw:halo}
Let $\Gamma\subseteq\Lambda$ have rank $s\ge1$, and let $P'$ be the
symmetric hull of generators of $\Gamma$. Suppose a compact child window
$K$ satisfies, for some $\vt,\vb_0\in\Lambda$ and some centrally symmetric
convex $S$,
\[
 \vb_0+K+(6+s/2)P'\subseteq\vt+CS .
\]
For any $b>0$ and $k=\max\{1,\lceil4C/b\rceil\}$,
\begin{equation}\label{fw:halo-eq}
 1+\frac{\vol_s(K+6P')}{\covol\Gamma}\le
 \bigl(1+k^{\rank\Lambda}\bigr)\,|(\vt+(b/2)S)\cap\Lambda| .
\end{equation}
Consequently all nonlifted sources generated at an anchor have total
charge $\Ot_D(\mu_v)$, including those whose candidates all fail
subsequent LP tests. The same holds for lifted sources, by horizontal
folding and~\eqref{fw:strip}.
\end{lemma}

\begin{proof}
We first prove the displayed inequality and then verify its containment
hypothesis for each kind of source: thin or rank-loss, deep, and lifted.

By Lemma~\ref{fw:counting}(1) for $\Gamma$ and $P'$, the normalized
volume of $K+6P'$ is at most the number of points of $\Gamma$ in
$K+(6+s/2)P'$. After adding $\vb_0$, these become distinct
parent-lattice points of $\vt+CS$. By~\eqref{fw:residue} with residues
modulo $k\Lambda$, translated by the lattice point $\vt$,
$|(\vt+CS)\cap\Lambda|\le k^{\rank\Lambda}|(\vt+(2C/k)S)\cap\Lambda|$,
and the chosen $k$ gives~\eqref{fw:halo-eq}. The additional one on the
left is paid by $\vt$, which lies in the smaller set.

\emph{Thin and rank-loss sources.} Here \eqref{fw:child-target}
and~\eqref{fw:basic-test} put all candidates in
$\vb_{\rm light}+O_D(R_v)$, and $H_jP'\subseteq R_v$. Thus the full
source halo, its fixed bucket enlargement, and its fundamental-domain
padding remain in a fixed dilation of $S_v$, and the translation
$\vt_v-\vb_{\rm light}$ is integral. If the child has smaller span, we
apply its lattice test first. From any compatible candidate $\vt_0$, we
round the symmetric section of this physical crop in the child span and
translate it by $\vt_0$. This needs no LP-feasible candidate, and the
containment still holds with a different rank-dependent constant.

\emph{Deep trials and restorations.} The removal displacement is
$O(E)P$, the child hull is contained in $P$, and $K_\ast P\subseteq
S_v$. After the integral endpoint translation, the target crop and the
full halo lie in
$\vt_v+\bigl(\gamma+O_D((E+c_{\rm box}H_j)/K_\ast)\bigr)S_v$. The
dilation is polylogarithmic, so~\eqref{fw:halo-eq} pays the window
before any LP filter. Different incoming windows are never replaced by
their convex hull.

\emph{Lifts.} We first enumerate the constantly many integer heights
in~\eqref{fw:strip} and form a slanted prism whose old-coordinate
projection lies in the preceding deep crop. In the sheared coordinates of
Proposition~\ref{fw:trial}, adding the source halo increases the vertical
width by $O_D(H_j/\nu)$, and the fixed navigation padding has the same
form. The projected lattice is a sublattice of the parent and the
positive vertical period is at least one. The normalized prism volume is
therefore at most the normalized volume of the projected window times
$O_D(1+H_j/\nu)$, a polylogarithmic factor. We apply the same
fundamental-domain charge in the projection. This construction is paid
even if every enumerated height is later rejected.
\end{proof}

The bounds on child masses are stronger than the construction bounds in
one respect: they do not depend on the halo height. For thin and
rank-loss children, actual buffer displacements lie in
$H_jP'\subseteq R_v$, and~\eqref{fw:child-target} sends every child
buffer, under a fixed integral translation, into $\vt_v+O_D(S_v)$. Its
mass is therefore $\Ot_D(\mu_v)$ per basic branch, independently of the
ratio $H_j/K_\ast$. For a lift, extending the two defining child witnesses by
the same attaining removal assignment puts the projected $Q$-body inside
the parent's assigned $Q$-body; the horizontal clipping and the $O(E)P$
drift again put the projection in $\vt_v+O_D(S_v)$. The actual buffer
satisfies the constant-width strip~\eqref{fw:strip}, not the taller
construction halo, so each projected lattice point has at most
$511F_D+1$ possible heights. We denote the resulting polylogarithmic mass
factors by $A_{\rm geo}$ for each thin or rank-loss family and by
$A_{\rm lift}$ for the lift family. Both are independent of the absolute
choice of $H_{\min}$. Construction charges, by contrast, may depend
polylogarithmically on $H_j/K_\ast$, which is harmless.

\subsection{The source constructor}
\label{fw:constructor}

We now fix the order in which a source is built, so that its window is
charged before any of its candidates can be discarded. For each
transfer, the following operations are executed in this order.
\begin{enumerate}
\item Compute the witnessed removal table and the allowance merge of
Lemma~\ref{fw:transfer}, saving one attaining predecessor and removal
witness per finite target.
\item Decode and apply the physical crop. For a lift, enumerate the
common-strip heights. Apply the exact child-span and child-lattice tests.
\item If no compatible candidate remains, stop. A rank-zero child retains
the target zero exactly when its allowance is positive.
\item Otherwise form the explicit source window of Lemma~\ref{fw:halo},
including the halo $c_{\rm box}H_{\rm birth}\Bx_P$ required by the
reporter. Prepare current profile views and construct the source atlas
of Theorem~\ref{det:atlas}, the column-body structures of
Corollary~\ref{det:body-oracles}, and the reporter of
Lemma~\ref{det:reporter}.
\item Filter candidates by the exact test $f(\vt)<s$. Attach their
source pointers and attaining witnesses to all survivors, even when they
will join a larger arithmetic batch.
\end{enumerate}
The window is constructed and charged before the last step, so a source
rejected in full is still charged to its generating anchor. A source
that reaches the fourth step has a nonempty window, since the window
contains the candidates that survived the third step; hence
$m_I=O_D(\Psi(I))=O_D(V(K,I))$. The preliminary
operations of the local LP atlas are therefore paid as well; these are a
joint feasibility test $Z(I)\cap K\ne\emptyset$ by zonotope support
scans, and one grouped LP at a seed. The internal endpoint normalization
and lattice reboxing of a source are converted back to the stored
common-objective coordinates, and neither changes the pool or its phase
objective. Finally, the implementation never needs to construct a large
convex hull to obtain a volume parameter: it uses the estimate
$\widehat V(K,I)$ of Section~\ref{det:section}, which is within a
constant factor of $V(K,I)$.

\subsection{A certified arithmetic phase}
\label{fw:phase}

A chain of proper full-rank restorations passes to finer and finer
sublattices of one root pool. If each restoration paid separately for
its buffer, the masses could add up along the chain. We avoid this by
grouping such restorations into a phase and merging their buffers, and
we show that the merged buffers have total moment at most a logarithmic
factor times the moment of the root buffer.

A \emph{phase} has a fixed root pool $I_0$, fixed stored coordinates, and
root lattice $\Gamma_0=\Lambda_{I_0}$. A current lattice $\Gamma$
represents exactly
\[
 I_\Gamma=I_0\cap\Gamma=\{i\in I_0:\ \va_i\in\Gamma\},\qquad
 \Lambda_{I_\Gamma}=\Gamma .
\]
Only proper full-rank restorations remain inside a phase. Thin exits,
rank losses, and lifts start new phase roots, and their actual pool
identities are retained; equal lattices from different phase-root pools
are not identified. The endpoint extension below adds back, at a
supporting price, the items that a restriction has excluded. For
$R\subseteq I$ define it by
\begin{equation}\label{fw:extension}
 U_{I,R}(\vy;\vpi)=\vy+\sum_{i\in I\setminus R}\va_ie_i(\vpi),\qquad
 e_i(\vpi)=[\vpi\va_i>c^\ast_i].
\end{equation}
A phase is \emph{certified} if its buffers lie in their fractional
domains and, for every nonroot buffer point $\vy\in\mathcal D_\Gamma$,
one predecessor lattice $\Pi\supsetneq\Gamma$ satisfies, for every
$\vpi\in\partial f_{I_\Gamma}(\vy)$,
\begin{equation}\label{fw:certified}
 U_{I_\Pi,I_\Gamma}(\vy;\vpi)\in\mathcal D_\Pi,\qquad
 \Gamma=\Lambda_{\mathcal R_\kappa(L_{I_\Pi,\vpi}(\tau))}
 \ \text{for some }\tau\ge0 .
\end{equation}
The predecessor is independent of the price; the threshold may depend on
it.

\begin{lemma}[Certified phase moments]\label{fw:phase-moments}
Suppose all lattices in a certified phase have index at most $B_{\rm ind}$
in $\Gamma_0$, and write $M_\Gamma=|\mathcal D_\Gamma|$. For every
$\theta\ge1$,
\begin{equation}\label{fw:phase-eq}
 \sum_\Gamma M_\Gamma^{\theta}\le
 \bigl(1+\lfloor\log_2B_{\rm ind}\rfloor\bigr)M_{\Gamma_0}^{\theta}.
\end{equation}
\end{lemma}

\begin{proof}
We show that endpoint extension preserves supporting prices and is
injective, and then that all lattices reaching a common root point form
a single nested chain.

If $\vpi$ supports $f_R$ at $\vy$, an optimum at $\vy$ together with the
excluded endpoints $\ve_{I\setminus R}(\vpi)$ minimizes reduced cost on
the full box, so $\vpi$ supports $f_I$ at $U_{I,R}(\vy;\vpi)$.
Conversely, any supporting price at this extended target supports $f_R$
at $\vy$: compare arbitrary feasible points of $R$ after adding the same
fixed excluded endpoints. At one fixed price, extensions compose by
partitioning the excluded labels.

For any choices $\vpi_{\vy}\in\partial f_R(\vy)$, the map
$\vy\mapsto U_{I,R}(\vy;\vpi_{\vy})$ is injective. If two images
coincide, take the inner product of their difference with
$\vpi_{\vy_1}-\vpi_{\vy_2}$. The contribution of $\vy_1-\vy_2$ is
nonnegative by the two supporting inequalities, and every excluded-label
contribution is nonnegative and strictly positive if its endpoints
differ, by the strict convention. All endpoints therefore agree, and then
$\vy_1=\vy_2$.

For each $\vy\in\mathcal D_\Gamma$ choose a supporting price and keep it
fixed while extending along a certified predecessor chain to a root point
$\vu$. The root image is $U_{I_0,I_\Gamma}(\vy;\vpi_{\vy})$, independent
of the chain. Choose a common supporting price $\vpi^\ast$ at $\vu$; by
the reverse implication it supports every point on every predecessor
chain reaching $\vu$. Write $C(\tau)=\mathcal R_\kappa(L_{I_0,\vpi^\ast}(\tau))$
and $\Gamma(\tau)=\Lambda_{C(\tau)}$. These lattices are nested.
Restricting the root to $I_0\cap\Gamma(\sigma)$ leaves $C(\tau)$
unchanged when $\tau\le\sigma$; when $\tau\ge\sigma$, the restricted core
contains $C(\sigma)$ and all its columns lie in $\Gamma(\sigma)$, so its
lattice is $\Gamma(\sigma)$. Inductively, every lattice on a predecessor
chain reaching $\vu$ belongs to this one nested family. Strict inclusions
double the index, so at most $1+\lfloor\log_2B_{\rm ind}\rfloor$ such
lattices occur, and injectivity gives at most one owner of $\vu$ in each.
Summing over root points proves the case $\theta=1$. Injectivity also
gives $M_\Gamma\le M_{\Gamma_0}$; multiplying by $M_{\Gamma_0}^{\theta-1}$
proves~\eqref{fw:phase-eq}. This is the common-price argument
of~\cite[Lemma~4.5]{Zhou26}.
\end{proof}

Operationally, we process the phase lattices in order of increasing
index, using the Hermite normal form in the fixed root basis as a
canonical key. Since the index grows strictly along restorations, every
arrival at a lattice precedes its processing. Duplicate targets are
merged by largest allowance, retaining one attaining source, and the
merged target receives the maximum incoming arithmetic depth. Source
windows and their preprocessing remain separate.

\begin{lemma}[The merged buffers are certified]\label{fw:certified-lemma}
The preceding coalescence rule and the proper-restoration branch of
Proposition~\ref{fw:trial} satisfy~\eqref{fw:certified}. Source
construction and failed filters remain charged to their generating
anchors.
\end{lemma}

\begin{proof}
Take a point $\vy$ of a merged buffer. Choose a covering merged query and
its saved attaining incoming source before choosing any price. Since both
buffer scales decrease with depth, the maximum incoming depth places
$\vy$ in the birth buffer of this source. Let $q$ be its saved parent
query and $Y$ its saved assignment on $X=I_\Pi\setminus I_\Gamma$, and
write $\vy=\vt_q-A_XY+\widehat\vz$. Extending the two opposite-point
child witnesses by the same $Y$ gives $Q_{\rm child}\subseteq Q_q$, hence
$\widehat\vz\in\tau_\gamma\xi_{j+1}Q_v\cap H_{j+1}P_I$. For every
supporting price $\vpi'$ at $\vy$, Lemma~\ref{fw:collar} gives
$A_X(Y-\ve_X(\vpi'))\in EP_I$ and the canonical-core lattice identity.
The endpoint extension differs from $\vt_v$ by an element of
\[
 (\gamma+\tau_\gamma\xi_{j+1}+E/K_\ast)Q_v\cap(\gamma H_j+H_{j+1}+E)P_I ,
\]
and~\eqref{fw:xi-schedule}--\eqref{fw:H-schedule} put this in the parent
buffer. Its endpoint extension is an integral parent-lattice point, so
the same predecessor certifies both clauses of~\eqref{fw:certified} for
every price. This is the mechanism of~\cite[Lemma~4.6]{Zhou26}, applied
to the common objective~\eqref{fw:cstar}.
\end{proof}

\section{Global moments and the proof of Theorem~\ref{det:main}}
\label{det:composition}

We first bound the rank, the volume, and the depth along the recursion,
then prove the global source and moment bounds, and finally substitute
the local costs of Sections~\ref{det:section}--\ref{detmc:top}. Until
Section~\ref{det:probability-section} every convolution reply is exact.

\subsection{Rank, volume, and depth}
\label{fw:rank}

Normalized volume is the quantity that drives the recursion: every
reduction step either keeps it or shrinks it, and the steps that change
the structure shrink it by a definite factor. This gives bounds on the
rank and on the number of reduction edges along any path.

We begin with restrictions. Normalized volume is nonincreasing under a
pool restriction, even if the rank drops. In the lower-rank case, extend
an independent set of the
child span by parent columns that are independent modulo that span. The
join of the child symmetric hull with the crosspolytope on these added
columns lies in the parent hull, and integrating its sections gives
\[
 \frac{r!}{2^r}\vol_r(\mathrm{join})
 =\frac{s!}{2^s}\vol_s(P_{\rm child})\,
 \bigl|\det(\bar\va_1,\dots,\bar\va_{r-s})\bigr|,
\]
where bars denote projections to a complementary space. The child
covolume is at least that of the parent lattice intersected with its
span, and the projected determinant is at least the projected parent
covolume. Their product is therefore at least the parent covolume. A
proper full-rank restoration at least doubles the covolume, so it at
least halves $\Psi$.

We next bound the number of lifts. The root has $\Psi_0\le C_dN^d$. For
sufficiently large $N$, every lift made while the rank is at most $D$
decreases $\Psi$ by a factor at least $N^{\delta/2}$,
by~\eqref{fw:lift-volume}. Up to the first $T_\uparrow$ lifts the rank
is at most $d+T_\uparrow<D$, so this estimate applies without
presupposing the final rank bound. More than $T_\uparrow$ lifts would
then force $\Psi<1$. Thus
\begin{equation}\label{fw:rank-bound}
 r\le d+\ell_\uparrow\le D,\qquad
 \Psi(I)\le C_{d,\delta}\frac{N^d}{\nu^{\ell_\uparrow}} .
\end{equation}
All other changes are volume-nonincreasing. The choice~\eqref{fw:Hmin}
below makes every same-rank thin exit halve $\Psi$, and every lift also
at least halves it for large $N$. Hence proper restorations, same-rank
thin exits, and lifts number at most $\lceil\log_2\Psi_0\rceil$ on any
path. There are also at most $d+T_\uparrow$ rank losses, since each
reduces the rank and only a lift increases it. The prescribed $J$
therefore exceeds the number of reduction edges on a path. The buffer
index $j$ increases only on proper restorations. At a merge we use the
incoming path of maximum arithmetic depth, which preserves
$2^j\Psi(I)\le\Psi_0$. Within a phase,
\[
 [\Gamma_0:\Gamma]\le\Psi(I_0)/\Psi(I_\Gamma)\le\Psi(I_0),
\]
so the factor in Lemma~\ref{fw:phase-moments} is $O_d(\log(N+2))$.
Finally, the finitely many $N$ below the parameter-dependent cutoff are
handled directly after the compressed reduction. For them \eqref{fw:U}
bounds $U$ by a constant, so expansion and exhaustive optimization cost
$O(1)\poly(\cB)$ after the compressed reduction.

\subsection{The global source theorem}
\label{fw:global-sec}

We now combine the local charges into the two global bounds that drive
the running time: a bound on the moments of the state masses, which pays
for the signed removed-item tables, and a bound on the total charge of
all sources, which pays for the local LP atlases and the reporters.

Put $\Xi=(\nu N)^d$. The root mass and its source charge are
$\Ot_d(\Xi)$: the root buffer and source window lie in a translate of a
fixed enlargement of $H_0P_0$, and Lemma~\ref{fw:counting}(1) gives
$O_d(H_0^{r_0}\Psi_0)$ lattice points and normalized volume. This also
pays for an entirely rejected initial threshold query.

\begin{theorem}[Global source and moment bounds]\label{fw:global}
For every fixed $\beta>0$, the scales can be chosen as above so that, with
exact convolution replies, every strict decision invocation satisfies,
uniformly for $1\le\theta\le2$,
\begin{equation}\label{det:moments}
 \sum_{u\ \text{processed}}M_u^{\theta}
 =\Ot_{d,\delta,\beta}(\Xi^{\theta}N^\beta),
\end{equation}
where $M_u$ is the buffer mass of state $u$. If $V_S$ are the
construction charges of all created sources, $Q_S$ their query counts,
and $O_S$ the numbers of offsets reported by their reporters, then
\begin{equation}\label{det:ledger}
 \cV=1+\sum_S(V_S+Q_S+O_S)=\Ot_{d,\delta,\beta}(\Xi N^\beta).
\end{equation}
These sums include transient trial sources, tests of incompatible
candidates, source seed work, and batches that leave no surviving query.
Each anchor $v$ of a state $u$ generates $O_D(1)$ removal tables with
$\Ot_D(\mu_v)$ cells each, where $\sum_v\mu_v^{\rho}\le M_u^{\rho}$, and
all structural work outside
atlases, reporters, and removal tables is
\begin{equation}\label{det:structural}
 \Ot_{d,\delta,\beta}(\nu\,\Xi N^\beta)\poly_{d,\delta,\beta}(\cB).
\end{equation}
\end{theorem}

\begin{proof}
We prove the moment bound with a potential on a contracted recursion
tree, and then derive the ledger and structural bounds from it by
charging each source and each anchor to its state.

Contract every arithmetic phase to one macro node, and let
$L_{\rm ph}=O_d(\log(N+2))$ be the maximal phase factor. By
Lemma~\ref{fw:phase-moments} with $\theta=1$ and the child-mass bounds
after Lemma~\ref{fw:halo}, the total child mass of each thin or rank-loss
family is at most $GM$, and the lift family has total mass at most $FM$,
where
\[
 G=L_{\rm ph}(A_{\rm geo}+1)\ge1,\qquad F=L_{\rm ph}A_{\rm lift}\ge1 .
\]
These factors are polylogarithmic and independent of $H_{\min}$. Set
$\eps_0=\beta/d$, $\Omega=12\lceil G\rceil^2$, and
$Z=12\Omega\lceil F\rceil^2$, and choose the remaining halo enlargement
before processing any query:
\begin{equation}\label{fw:Hmin}
 H_{\min}\ge C_DK_\ast\max\{2,\Omega^{1/\eps_0}\} .
\end{equation}
Rounding the exponent upward if necessary keeps this a polylogarithmic
enlargement. Take $C_D$ at least the thin-slab constant of
Lemma~\ref{fw:thin}; the extra factor two makes every same-rank thin exit
halve $\Psi$. Since $G$ and $F$ do not depend on $H_{\min}$, this choice
is not circular.

For a macro node of rank $r$, lift count $\ell_\uparrow$, mass $M$, and
volume $\Psi$, we use the potential below. Its three factors reward,
respectively, a loss of rank, a lift, and a decrease of volume.
\[
 \Phi_\theta=\Omega^rZ^{T_\uparrow-\ell_\uparrow}M^{\theta}\Psi^{\eps_0}.
\]
A same-rank thin family loses at least a factor $\Omega$ through
$\Psi^{\eps_0}$, by Lemma~\ref{fw:thin} and~\eqref{fw:Hmin}. A rank-loss
family loses at least that factor through $\Omega^r$, while its volume
does not increase. Since $\sum_xM_x^\theta\le(\sum_xM_x)^\theta$ and
$\theta\le2$, either family has total potential at most
$(G^2/\Omega)\Phi_\theta\le\Phi_\theta/12$. A lift family gains at most
$\Omega$ from the rank, loses $Z$ from the lift count, and has
nonincreasing volume, so its potential is at most
$(\Omega F^2/Z)\Phi_\theta\le\Phi_\theta/12$. These three families have
total potential at most $\Phi_\theta/4$. The root factors
$\Omega^{r_0}Z^{T_\uparrow}$ are polylogarithmic, and
$\Psi_0^{\eps_0}=O_{d,\beta}(N^\beta)$. Summing the geometric decrease
over the macro tree gives $\Ot(\Xi^\theta N^\beta)$, and restoring the
internal arithmetic states adds only their phase factor. This
proves~\eqref{det:moments}. The argument relies on each phase having one
actual root pool; no coalescence between unequal pools was used.

We turn to the source ledger. At every processed state, anchor
packing~\eqref{fw:packing} charges $\sum_v\mu_v$ to $M_u$.
Lemma~\ref{fw:halo} pays all newly generated source volumes and
candidate scans by $\Ot(M_u)$, including failed batches. Each query is
assigned to the immutable source that answers it. The queries of a
state have distinct targets in its buffer, so there are at most $M_u$ of
them. Each becomes at most one anchor, with one reporter call and
$O_D(1)$ column-body calls; this pays $\sum_SQ_S$.
Lemma~\ref{fw:output} pays all reported grid offsets by $\Ot(\mu_v)$.
Adding the initial source charge and using~\eqref{det:moments} with
$\theta=1$ proves~\eqref{det:ledger}.

By Lemma~\ref{fw:thin} and Proposition~\ref{fw:trial}, an anchor $v$
generates $O_D(1)$ removal tables, each with $\Ot_D(\mu_v)$ cells, and
since $\rho\ge1$ we have
$\sum_v\mu_v^\rho\le(\sum_v\mu_v)^\rho\le M_u^\rho$. Finally, thin
extraction and the grouped deep work cost $\Ot(\nu\mu_v)$ per anchor,
hence $\Ot(\nu\sum_uM_u)=\Ot(\nu\,\Xi N^\beta)$ in total.

We note that this potential yields a moment and source-volume bound,
not merely a bound on $\sum M\Psi$. The adjustable polylogarithmic thin
contraction in~\eqref{fw:Hmin} absorbs the family factors that recur at
every level, and so avoids a loss of the form $C^{\Theta(\log N)}$.
\end{proof}

\paragraph{Deterministic structural routines.}
We record here why everything outside the convolution primitive is
deterministic. Integer normal forms, rational LLL, fixed-dimensional integer
feasibility, and exact LP separation--optimization are
deterministic~\cite{LLL82,Lenstra83,GLS88}. The integer centerpoint
routine is the deterministic fixed-dimensional statement
of~\cite[Theorem~30]{BasuOertel17}, together with deterministic
integer-point counting in fixed dimension~\cite{Barvinok94}. For the multiset progression
construction we use the deterministic implementation
of~\cite[Corollary~8.8]{CMZ26}, which replaces the randomized
Freiman-homomorphism routines in Theorem~3.1 of that paper without
changing its asymptotic bound; this is also the construction invoked
in~\cite[proof of Lemma~7.3]{Zhou26}. The compressed profiles, rank and
select operations, source merging, Hermite-normal-form lattice keys, and
the common objective are deterministic. Balanced search trees are used
for keys and persistent records, so no randomized hashing is needed. The
local atlas (Theorem~\ref{det:atlas}), the reporter
(Lemma~\ref{det:reporter}), and the removal tables
(Theorem~\ref{det:transfer}) are deterministic. Hence the numerical
convolution primitive is the only possible source of randomness.

\subsection{Paying for construction, queries, and transfers}
\label{det:paying}

We now substitute the costs of the local implementations into the
global bounds. The local LP atlases and the reporters are paid by the
source ledger, and the signed removed-item tables by the moment bound.

\emph{Atlases and column-body queries.} By Theorem~\ref{det:atlas},
deterministic atlas construction costs
\[
 \Ot_D\Bigl(\sum_SV_S^{3/2}\Bigr)\poly_D(\cB)
 \le \Ot_D(\cV^{3/2})\poly_D(\cB).
\]
The column-body structures with zero heights have preprocessing
$\Ot_D(\Psi(I_S)^{3/2})$ and cost $\Ot_D(\sqrt{\Psi(I_S)})$ per call, with
the same polynomial bit factors. The support calls for facet tests of
rounded bodies and the membership tests for emitted offsets are covered by
Proposition~\ref{det:source-charge}, whose hypothesis is supplied by
Section~\ref{fw:dispatch-sec}: every source makes $\Ot(Q_S+O_S+V_S)$
such calls. Since $\Psi(I_S)=O_D(V_S)$ for every source that builds
these structures, all of which have nonempty windows
(Proposition~\ref{det:source-charge}), their entire additional charge is
again $\Ot_D(\cV^{3/2})\poly_D(\cB)$.

\emph{Reporting.} By Lemma~\ref{det:reporter} and the estimate following
it, all queries to all sources are answered in
\[
 \Ot_D\Bigl(\sum_S\bigl(V_S^{3/2}+(Q_S+O_S)\sqrt{V_S}\bigr)\Bigr)\poly_D(\cB)
 \le\Ot_D(\cV^{3/2})\poly_D(\cB)
\]
time, deterministically and without convolution calls. Thus atlases,
column-body queries, and reporting together cost
$\Ot_D(\cV^{3/2})\poly_D(\cB)$.

\emph{Removed-item tables.} By Lemma~\ref{fw:transfer}, a removal table
with $\Ot_D(\mu_v)$ cells and its allowance merge cost
$\Ot_D(\mu_v^{\rho})\poly_D(\cB)$; the compressed pools of
Section~\ref{fw:pools} supply the access to eligible vectors and cheapest
labels assumed there. By Theorem~\ref{fw:global} and~\eqref{det:moments}
with $\theta=\rho$,
\[
 \sum_u\sum_{v\text{ in }u}\mu_v^{\rho}
 \le\Ot_D\Bigl(\sum_uM_u^{\rho}\Bigr)
 =\Ot_{d,\delta,\beta}(\Xi^{\rho}N^\beta).
\]
The endpoint-cost translations and the allowance merge are deterministic
given exact replies, and preserve actual disjoint labelled witnesses.

Together with~\eqref{det:structural}, these bounds give the total time
\begin{equation}\label{det:three-terms}
 T_{\rm LP}+\Ot_{d,\rho,\delta,\beta}
 \bigl(n+\cV^{3/2}+\Xi^{\rho}N^\beta+\nu\,\Xi N^\beta\bigr)
 \poly_{d,\rho,\delta,\beta}(\cB).
\end{equation}

\subsection{Slack and the explicit exponent}

Apart from the input and the initial LP, the three powers of $N$
in~\eqref{det:three-terms} have exponents at most
\[
 \tfrac32\bigl[d(1+\delta)+\beta\bigr],\qquad
 \rho\,d(1+\delta)+\beta,\qquad
 d+(d+1)\delta+\beta .
\]
Recall $\rhos=\max\{3/2,\rho\}\ge3/2$. For a requested $\eps>0$ choose
\[
 \delta=\min\Bigl\{\frac18,\ \frac{\eps}{8\rhos}\Bigr\},
 \qquad \beta=\frac{\eps d}{8\rhos}.
\]
Since both $3/2$ and $\rho$ are at most $\rhos$, the first two
exponents are at most $\rhos d(1+\delta)+\rhos\beta\le
(\rhos+\eps/4)d$. The third is at most $[1+3\eps/(8\rhos)]d$, which is also below
$(\rhos+\eps)d$. All additional logarithmic factors and polynomial bit
costs fit into~\eqref{det:main-bound}.

\subsection{Decision correctness and reconstruction}
\label{fw:correct}

We now check that the algorithm decides each strict threshold
correctly, that a witness can be reconstructed within the same time
bound, and that binary search turns decisions into optimization.

For one strict threshold, we initialize the root source, lattice-test
and LP-filter its query, and process its phase. Within a phase we merge
by largest allowance and dispatch every surviving query as in
Section~\ref{fw:dispatch-sec}. A state returns YES if a rank-zero
completion or a rounded high-fibre completion yields a witness, and NO
after all anchors and all recursively generated exits return NO.
Arithmetic descendants are coalesced in the current phase; other exits
start fresh phase roots.

With exact convolution replies this is an exact decision procedure.
Every atlas entry is exact (Theorem~\ref{det:atlas}), every required
affordable offset is reported (Lemma~\ref{det:reporter}), and every
removal table dominates its required assignments
(Lemma~\ref{fw:transfer}). Lemma~\ref{fw:thin} proves completeness of the
rank-loss and thin branches, and Proposition~\ref{fw:trial} proves
completeness of trial rejection, proper restoration, and the high-fibre
terminal or lift alternative. Each finite table entry is an actual
assignment on the removed pool, so every child witness extends through a
disjoint labelled assignment to a parent witness
(compare~\cite[Lemma~4.1]{CMZ26b}). Taking a maximum allowance retains an
attaining source and therefore preserves the disjunction over incoming
queries. Rank, normalized volume, and arithmetic index are the progress
measures of Section~\ref{fw:rank}, so the recursion terminates.

\emph{Reconstruction.} Witnesses are stored in compressed form: each
transfer saves an endpoint template, signed toggle witnesses, and one
predecessor. The local LP atlas supplies target and cost aggregates in
constant time, basic labels in $O_D(1)$ time, and prefix information for
a requested type in polylogarithmic time. Remote fixed endpoints
therefore never require a scan of a tile's selected types. Along a
successful path the removed pools and the terminal active pool are
disjoint. Table traceback visits only nonempty witness branches, so its
work is bounded by the materialized labels and pointers already charged
to construction. To recover the original variable counts, we aggregate
the selected counts by permanent type, undo the endpoint and complement
translations, and sweep the original run lists. Equivalently, within a
permanent type one may select the cheapest prefix of the recovered
count; this preserves the vector sum and cannot increase the cost.
Formula~\eqref{fw:units} then recovers the original bounded vector. A
successful chain has only
$O_{d,\delta}(\log(N+2))$ phase, rank, and volume steps, so this costs
$\Ot(n+N^d)$ beyond the paid witness work. The algorithm checks the
original equalities, bounds, and threshold cost of the result.

\emph{Optimization.} Put $c_{\max}=\max\{1,\|\vc\|_\infty\}$. Every
selection of the compressed root has cost in $[-Uc_{\max},Uc_{\max}]$.
The strict threshold $Uc_{\max}+1$ first decides feasibility. Binary
search over this integer interval then locates the minimum with
$O(1+\log(U(c_{\max}+1)))$ strict decisions, each translated
by~\eqref{fw:threshold}. Finally, the fixed objective offset of the root
is added. The large values of $\vb,\vell,\vu$ enter only the initial LP
and the reconstruction. Each decision runs within the
bound~\eqref{det:three-terms}, and the $O_d(\log N+\cB)$ decisions
multiply it by a factor absorbed into $\Ot(\cdot)\poly(\cB)$; the
envelope $W_\star$ of Section~\ref{det:probability-section} bounds the
work summed over all decisions. This establishes the exponent bound and
correctness for every execution whose convolution replies are exact, and
for every permitted choice of attaining witnesses.

\subsection{Exact arithmetic}
\label{fw:arith}

We check that every number in the algorithm has polynomially bounded
bit length. All stored labels have the one integer cost $c^\ast_i$, and
every stored equality and lift height is integral in its lattice
coordinates.

\emph{The coordinate bound, without circularity.} Put
$H_{\rm co}=\max\{N,\,c_DC_{d,\delta}N^d\}$, where $C_{d,\delta}$ is the
constant of~\eqref{fw:rank-bound} and $c_D$ bounds the constants
$O_r(1)$ in~\eqref{det:box} for $r\le D$. This value depends only on
$d$, $\delta$, $D$, and $N$, so it is fixed before $\vc^\ast$ is chosen.
Genericity is a property of the column system alone: integral circuits
do not depend on the coordinates in which the columns are written. The
argument of Section~\ref{fw:objective} therefore applies to a pool as
soon as its rank is at most $D$ and \emph{some} basis of its lattice
gives all its columns integer coordinates of absolute value at most
$H_{\rm co}$; nothing needs to be stored in that basis. We show by
induction along the recursion that every pool that arises has this
property. The root has rank $d\le D$ and columns with entries at most
$N\le H_{\rm co}$. Suppose the property holds for every pool on the path
from the root to a node. Then $\vc^\ast$ is generic for each of them, so
the lemmas of Sections~\ref{det:section} and~\ref{fw:geom}--\ref{fw:charge}
apply along this path. The derivation of~\eqref{fw:rank-bound} in
Section~\ref{fw:rank} uses only these lemmas along the path, and it
bounds the rank before the final rank bound is known. It gives every
child rank at most $D$ and $\Psi\le C_{d,\delta}N^d$. In the lattice
basis $F$ of~\eqref{det:box}, every column lies in $P_I\subseteq\Bx_P$
and so has integer coordinates with $|z_j|\le L_j\le\prod_lL_l\le
c_D\Psi\le H_{\rm co}$. This completes the induction. Hence the
perturbation~\eqref{fw:cstar}, chosen before the recursion, covers every
later column system and adds only $O_{d,D}(\log(N+2))$ bits. Together
with~\eqref{fw:U} and fixed-dimensional determinant bounds, this gives a
uniform polynomial bound on LP values, prices, source facets, lift
multipliers, and basic coefficients. A working price normalization uses
one bounded-bit common denominator for its batch, and stored-coordinate
conversion returns to the common objective. Temporary progression images
have bounded dimension and retain their full old-vector readout; no scalar
encoding replaces the vector equation. Hence every number in the
algorithm has $\poly_{d,\rho,\eps}(\cB)$ bits.

\subsection{Amplification and the end of the proof}
\label{det:probability-section}

It remains to remove the assumption that every convolution reply is
exact. If the primitive is deterministic, or more generally has error
probability $p=0$, every reply is exact, the algorithm above has the
stated running time, and Theorem~\ref{det:main} is proved. Otherwise let
$0<p<1$ be the error probability of the primitive, as in the hypothesis
of Theorem~\ref{det:main}. Number the logical convolution calls
$c=1,2,\dots$ in the order in which the execution makes them, including
the calls of all decisions of the objective search, and replace the
$c$th call by $\ell_c$ independent witnessed repetitions, where
\begin{equation}\label{det:repetitions}
 \ell_c=\Bigl\lceil\frac{\log(2c^2/p)}{\log(1/p)}\Bigr\rceil
     =1+\Bigl\lceil\frac{\log(2c^2)}{\log(1/p)}\Bigr\rceil .
\end{equation}
The index $c$ is known when the call is made, so the number of
repetitions does not depend on any bound fixed in advance. If $p$ is not rational, fix
once a rational $\bar p$ with $p\le\bar p\le\tfrac65p$ and $\bar p<1$,
and use $\bar p$ in place of $p$ in~\eqref{det:repetitions}; then
$\ell_c$ is the least integer with $\bar p^{\ell_c-1}\le1/(2c^2)$, which
is decided by exact rational comparisons. All bounds below hold with
$\bar p$ in place of $p$, and the failure probability is at most
$\pi^2\bar p/12\le\tfrac{\pi^2}{10}p<p$.

\paragraph{Amplifying one witnessed call.}
For arrays $F,G$ let $H=F\oplus G$. For the $c$th call, run the primitive $\ell_c$ times with
fresh independent randomness. At each output index $k$, check every
finite candidate and its witness: the indices must be in range and
satisfy $i+j=k$, they must refer to finite inputs, and the candidate
value $\widehat H[k]$ must equal $F[i]+G[j]$. Reject outputs of the wrong
shape and fields exceeding the prescribed bit bounds, which are derived
from the input, before using them in arithmetic. Discard invalid
candidates and malformed or oversized trial outputs, and treat missing
entries as $+\infty$. A discarded trial does not by itself cause a
global failure report. Among the valid candidates, keep one of minimum
value, with a fixed tie rule. For properly bounded outputs these checks
take time linear in the array lengths, up to polynomial bit factors.
Parsing any other output takes at most its emitted size, which is
bounded by the worst-case time of the primitive. All validation costs
are therefore absorbed by that time bound.

Every retained candidate is attained and hence is at least $H[k]$. If
any one trial returns its entire witnessed output correctly, the combined
output equals $H$ at every index and has valid attaining witnesses. For
every fixed input, the $c$th combined call therefore fails with
probability at most $p^{\ell_c}\le p/(2c^2)$. This also holds
conditionally on any previous history of the computation, because the
guarantee is uniform over inputs and the new randomness is independent.
Witness checking alone does not certify optimality; the probability
bound relies on the possibility of an entirely correct trial.

\paragraph{Error probability.}
Consider the first logical call that returns an inexact combined reply.
By the conditional bound and a union bound over its possible positions,
\[
 \Pr[\text{some combined reply is inexact}]
 \le\sum_{c\ge1}\frac{p}{2c^2}=\frac{\pi^2}{12}\,p<p .
\]
This holds however many calls the execution makes, so correctness needs
no bound on the number of calls. If every combined reply is exact, the
reduction and the objective search are correct
(Section~\ref{fw:correct}), and neither the work guard nor the trial
clocks below ever stop the computation.

\paragraph{Worst-case work.}
A guard on the work is still needed. Checked replies have one-sided
error: every retained candidate is attained, so an inexact reply can only
make a table entry too large, and every finite entry remains witnessed by
an actual assignment. This does not keep the analysis of
Sections~\ref{fw:frame}--\ref{fw:charge} valid. For example, properness
of a restoration in Proposition~\ref{fw:trial}(2) relies on the trial
table dominating the zero working assignment; an inexact table can make
the recursion restore to the same lattice, which breaks the depth and
index bounds of Section~\ref{fw:rank}. An inexact reply can also change
the lengths of later convolution calls, so a guard that counts only the
calls and the surrounding work does not bound their aggregate cost: a
linear budget on array lengths does not bound $\sum_cm_c^{\rho}$ by the
same budget. We therefore charge every logical call for its own cost.

\emph{Call charges.} By hypothesis there are constants $C_\star\ge1$ and
$\kappa_\star\ge0$, fixed together with the primitive, such that one run of
the primitive on arrays of total length $m$ whose finite entries have at
most $b$ bits takes at most
\[
 \tau(m,b)=C_\star\,(m+2)^{\rho}\,\bigl(b+\log(m+2)\bigr)^{\kappa_\star}
\]
steps. For the $c$th logical call let $m_c$ be its total array length
and $b_c$ the bit bound prescribed for its entries, both known before the
call is made, and put $\tau_c=\tau(m_c,b_c)$. The call is charged
$\tau_c$ once, independently of its number $\ell_c$ of repetitions. In
addition, every trial of the $c$th call is stopped after $\tau_c$ steps
and then discarded. By hypothesis this never interrupts a trial, but it
makes the bounds below independent of the behaviour of the primitive on
any particular input.

The constants $C_\star$ and $\kappa_\star$ are therefore an input to the
construction, not merely a fact used in its analysis. If $C_\star$ were
underestimated, the clock could cut off correct trials and the guard
could stop executions with exact replies, so the error bound would fail.
Accordingly, for $p>0$ the algorithm of Theorem~\ref{det:main} depends on
the primitive together with these constants; for $p=0$ no guard is used
and the constants are not needed. For the primitives of
Corollary~\ref{det:explicit} they are fixed numbers determined by the
cited algorithms, so the corollary is unaffected.

\emph{The guard.} The \emph{charged work} after input preparation is the
logical work, that is, all work except the runs of the primitive and
their checks, plus the sum of the charges $\tau_c$ of the calls made so
far. In an execution with exact replies, convolution is called only by
the removed-item tables (Theorem~\ref{det:transfer} and
Lemma~\ref{fw:transfer}), and their cost is bounded by charging every
call the hypothesis bound and summing. Inside one table this summation
is carried out in the proofs of Lemma~\ref{det:transfer-canonical}
(via $\sum_vd_v^{\rho}\le u^{\rho}$),
Proposition~\ref{det:transfer-scalar}
(via~\eqref{det:transfer-recurrence} and the merge of the dyadic layers),
and Theorem~\ref{det:transfer}. Over all tables it is the paragraph on
removed-item tables in Section~\ref{det:paying}, which sums
$\mu_v^{\rho}$ by Theorem~\ref{fw:global} and~\eqref{det:moments}. These
arguments use only the form of the per-call bound: the power $m^{\rho}$,
superadditivity of $m\mapsto m^{\rho}$, and a factor that is
polylogarithmic in $m$ and polynomial in the entry size. The charge
$\tau_c$ has the same form with explicit constants. Every call has
$m_c\ge1$, and $(m+2)^{\rho}\le3^{\rho}m^{\rho}$ for $m\ge1$, so the shift
from $m$ to $m+2$ costs only the constant factor $3^{\rho}$. Moreover, in such executions
$b_c=\poly_{d,\rho,\eps}(\cB)$ by Section~\ref{fw:arith}, and
$\log(m_c+2)=O(\log N+\cB)$. The same summation therefore bounds
$\sum_c\tau_c$ by the convolution term of~\eqref{det:three-terms}, up to
a factor $\poly_{d,\rho,\eps}(\cB)$, and Section~\ref{fw:charge} together
with~\eqref{det:three-terms} bounds the logical work. We therefore fix, from these uniform bounds for
executions with exact replies, a computable upper envelope
\begin{equation}\label{det:call-bound}
 W_\star=\Ot_{d,\rho,\eps}\bigl(N^{(\rhos+\eps)d}\bigr)
 \poly_{d,\rho,\eps}(\cB)
\end{equation}
on the charged work, together with the prescribed domain and
arithmetic-size bounds. These bounds hold uniformly over all permitted
attaining witnesses. The computation stops and reports failure if its
charged work would exceed $W_\star$, in particular before issuing a call
whose charge would do so, or if a size bound is violated; by the
preceding paragraph this happens only after an inexact reply.

\emph{Time.} In every execution, exact or not, the charges satisfy
$\sum_c\tau_c\le W_\star$. Since $\tau_c\ge1$, at most $W_\star$ calls
are made, so every call made has index $c\le W_\star$ and, $\ell_c$ being
nondecreasing in $c$, uses at most $\ell_{W_\star}$ repetitions. Each
trial of the $c$th call takes at most $\tau_c$ steps, and so emits at
most $\tau_c$ symbols; its validation takes $O(\tau_c)$ steps up to a
factor $\poly(b_c)$. The runs of the primitive and their checks therefore
take at most
\[
 \sum_c\ell_c\,\tau_c\,\poly(b_c)
 \le\ell_{W_\star}\,W_\star\,\poly_{d,\rho,\eps}(\cB)
\]
time in total, and the remaining work after input preparation is at
most $W_\star$. Hence the worst-case running time is
\begin{equation}\label{det:amplified-time}
 T_{\rm LP}+\Ot_{d,\rho,\eps}
 \bigl(n+\ell_{W_\star}N^{(\rhos+\eps)d}\bigr)
 \poly_{d,\rho,\eps}(\cB),
\end{equation}
in which, for fixed $C_\star$ and $\kappa_\star$, the hidden constants and
degrees do not depend on $p$; the dependence on $p$ enters only through
$\ell_{W_\star}$.

\paragraph{The number of repetitions.}
By~\eqref{det:repetitions} and~\eqref{det:call-bound},
\[
 \ell_{W_\star}=O_{d,\rho,\eps}\Bigl(1+\frac{\log N+\log(\cB+2)}{\log(1/p)}\Bigr).
\]
The individual counts $\ell_c$ depend on $c$ and $p$. The budget
$W_\star$ does not depend on $p$: it is set by the uniform bounds for
exact executions and by the primitive's running-time constants
$C_\star,\kappa_\star$. The clock rule $\tau_c=\tau(m_c,b_c)$ likewise has no
explicit dependence on $p$, although the limit for a particular call
depends on its size $m_c$ and bit bound $b_c$, and hence on the adaptive
input of that call. For fixed $p$ the bound is
$O_{d,\rho,\eps,p}(\log N+\log(\cB+2))$. For $p\le1/2$ we have
$\log(1/p)\ge\log 2$, so, with $C_\star$ and $\kappa_\star$ held fixed, the
amplification overhead is uniform in $p$. Its constant grows only as
$p\to1$, and then like $1/(1-p)$. The factor
$\log N$ is absorbed by $\Ot$ and the factor $\log(\cB+2)$ by
$\poly(\cB)$, so~\eqref{det:amplified-time} becomes~\eqref{det:main-bound}
with error probability at most $p$. This completes the proof of
Theorem~\ref{det:main}. Any smaller target error $\eps_{\rm fail}$ is
obtained in the same way, replacing $p$ by $\eps_{\rm fail}$ in the
numerator of~\eqref{det:repetitions}, at the price of an additional
factor $O(1+\log(1/\eps_{\rm fail})/\log(1/p))$.

\subsection{Explicit instantiations}
\label{det:instantiation}

We prove Corollary~\ref{det:explicit}. In both cases we apply
Theorem~\ref{det:main} with a fixed $\eps>0$, so the hidden dependence on
$\rho$ and $\eps$ is absorbed into $\Ot_d$ and $\poly_d$.

\emph{Deterministic bound.} The primitive of Lemma~\ref{detmc:lemma} is
deterministic and gives $\rho=1.99985$, hence
\[
 \rhos+10^{-5}=\max\{\rho,3/2\}+10^{-5}=1.99985+10^{-5}=1.99986 .
\]
Taking $\eps=10^{-5}$, Theorem~\ref{det:main} gives the
deterministic bound~\eqref{det:explicit-det}.

\emph{Randomized bound.} The primitive of Lemma~\ref{detmc:random} has
error probability $p=1/3$ and gives $\rho=\rho_0=1.9601397\ldots$, hence
\[
 \rhos+10^{-5}=\max\{\rho_0,3/2\}+10^{-5}<1.96015<1.9602 .
\]
Taking $\eps=10^{-5}$ and $p=1/3$, Theorem~\ref{det:main} gives
the randomized bound~\eqref{det:explicit-rand} with error probability at most $1/3$.

The numerical choices are conservative. What matters is that both
savings are absolute positive constants per equation, so they do not
decrease as $d$ grows. In both cases the bottleneck is the exponent
$\rho$ of the signed removed-item tables; the exponent $3/2$ of the
local LP atlas would become the bottleneck only for a
convolution algorithm with exponent below $3/2$.

\end{document}